\documentclass[a4paper,11pt]{article}

\usepackage[mathlines]{lineno}
\usepackage{jcappub}

\usepackage[table,svgnames,dvipsnames]{xcolor}
\usepackage[normalem]{ulem}
\usepackage{aas_macros}
\usepackage{hyperref}
\usepackage{verbatim}
\usepackage{tabularx}
\usepackage{orcidlink} % for author list
\usepackage{hanging} % for affiliations 
\usepackage{breqn}
\usepackage{multirow}
\usepackage{makecell}
\usepackage{siunitx}
\usepackage{dcolumn} % Align table columns on decimal point
\usepackage{bm} % bold math
\usepackage{xspace}
\usepackage{graphicx} % Include figure files
\usepackage{subfigure}
\usepackage{cases} % for dealing with mathematics,
\usepackage{booktabs}
\graphicspath{{figures/}}

\usepackage[nameinlink,noabbrev]{cleveref}
\crefname{equation}{Eq.}{Eqs.}
\crefname{section}{Section}{Sections}
\crefname{figure}{Figure}{Figures}
\crefname{table}{Table}{Tables}
\crefname{appendix}{Appendix}{Appendices}
\Crefname{figure}{Figure}{Figures}
\Crefname{equation}{Equation}{Equations}
\Crefname{section}{Section}{Sections}
\Crefname{table}{Table}{Tables}

\newcommand{\iron}{\texttt{iron}}
\newcommand{\redrock}{\textsc{Redrock}}
\newcommand{\hMpc}{\ h\,\text{Mpc}^{-1}}
\newcommand{\Mpch}{\ h^{-1}\,\text{Mpc}}

\renewcommand*\vec[1]{\ensuremath{\boldsymbol{#1}}}
\newcommand*\tens[1]{\ensuremath{\mathsf{#1}}}

\title{DESI 2024 II: Sample Definitions, Characteristics, and Two-point Clustering Statistics }
\author{{DESI Collaboration}:}
\emailAdd{spokespersons@desi.lbl.gov}
\affiliation{Affiliations are in Appendix \ref{sec:affiliations}}

\author[1]{{A.~G.~Adame},}
\author[2]{{J.~Aguilar},}
\author[3]{{S.~Ahlen}\orcidlink{0000-0001-6098-7247},}
\author[4]{{S.~Alam}\orcidlink{0000-0002-3757-6359},}
\author[5,6]{{D.~M.~Alexander}\orcidlink{0000-0002-5896-6313},}
\author[2]{{M.~Alvarez},}
\author[7]{{O.~Alves},}
\author[2]{{A.~Anand}\orcidlink{0000-0003-2923-1585},}
\author[8,7]{{U.~Andrade}\orcidlink{0000-0002-4118-8236},}
\author[9]{{E.~Armengaud}\orcidlink{0000-0001-7600-5148},}
% ...

\abstract{We present the samples of galaxies and quasars used for DESI 2024 cosmological analyses, drawn from the DESI Data Release 1 (DR1). We describe the construction of large-scale structure (LSS) catalogs from these samples, which include matched sets of synthetic reference `randoms' and weights that account for variations in the observed density of the samples due to experimental design and varying instrument performance. We detail how we correct for variations in observational completeness, the input `target' densities due to imaging systematics, and the ability to confidently measure redshifts from DESI spectra. We then summarize how remaining uncertainties in the corrections can be translated to systematic uncertainties for particular analyses. 
We describe the weights added to maximize the signal-to-noise of DESI DR1 2-point clustering measurements.
We detail measurement pipelines applied to the LSS catalogs that obtain 2-point clustering measurements in configuration and Fourier space.
The resulting 2-point measurements depend on window functions and normalization constraints particular to each sample, and we present the corrections required to match models to the data.
We compare the configuration- and Fourier-space 2-point clustering of the data samples to that recovered from simulations of DESI DR1 and find they are, generally, in statistical agreement to within 2\% in the inferred real-space over-density field.  
The LSS catalogs, 2-point measurements, and their covariance matrices will be released publicly with DESI DR1.}

\begin{document}

\maketitle

\flushbottom

\section{Introduction}
\label{sec:intro}

The large-scale structure (LSS) of the Universe, which can be measured by the clustering of galaxy and quasar tracers, provides a means to test cosmological models. Galaxy redshift surveys measure the angular coordinates and redshift distances of many galaxies and thus enable measurement of their clustering in 3D, from which cosmological information can be inferred. 
The Dark Energy Spectroscopic Instrument (DESI; \cite{Snowmass2013.Levi,DESI2016a.Science,DESI2016b.Instr,DESI2022.KP1.Instr}) is carrying out a Stage-IV redshift survey aiming to significantly improve the cosmological constraints derived from clustering measurements made with samples of galaxies, quasars and the Lyman-$\alpha$ forest.

DESI is a robotic, fibre-fed, highly multiplexed spectroscopic instrument that operates on the Nicholas U. Mayall 4-meter telescope at Kitt Peak National Observatory (KPNO) in Arizona. DESI is conducting a five-year survey over $14\,200$ square degrees, which will measure the spectra of $\sim40$ million galaxies and quasars in the redshift range  $0.1<z<4.2$, covering several different target classes \cite{TS.Pipeline.Myers.2023}. During bright time of telescope operation, DESI conducts the bright galaxy survey (BGS) at low redshifts, $0.1<z<0.4$. During dark time, DESI targets luminous red galaxies (LRGs) in the redshift range $0.4<z<1.1$, emission-line galaxies (ELGs) in $0.8<z<1.6$, and quasars (QSOs) over $0.8<z<2.1$. The Lyman-$\alpha$ forest spectral absorption in a further population of high-redshift quasars at redshifts $2.1<z<4.2$ is used to trace the distribution of neutral hydrogen, and a sample of stellar objects is also observed in the overlapping Milky Way Survey (MWS; \cite{MWS.TS.Cooper.2023}). During the first 13 months of main survey operation, DESI successfully observed spectra of over 18 million unique objects, more than 75\% of which are extragalactic. The key cosmological goals of clustering analyses using these data that form DESI Data Release 1 (DR1; \cite{DESI2024.I.DR1}) include:
\begin{itemize}
    \item localization of baryon acoustic oscillation (BAO) feature to probe the nature of dark energy through measuring distance scales as a function of redshift \cite{DESI2024.III.KP4,DESI2024.IV.KP6}, 
    \item analysis of the redshift-space distortion (RSD) signature that alters the clustering amplitude as a function of the angle to the line of sight and allows the rate of structure growth to be measured \cite{DESI2024.V.KP5},
    \item and measurement of the scale-dependent `bias' signature imprinted by squeezed primordial non-Gaussianity ($f^{\rm local}_{\rm NL}$) on the clustering on the largest scales \cite{ChaussidonY1fnl}.
\end{itemize}

These cosmological goals are achieved through measurement of the 2-point clustering signal of the different galaxy and quasar tracers. This signal is captured in the 2-point correlation function (2PCF) or its Fourier-space analogue, the power spectrum. These statistics encode the clustering of cosmological density fluctuations: however, survey operations, target selection effects, instrumental effects and astrophysical foregrounds all produce additional non-cosmological fluctuations in the observed galaxy density, and unless corrected for these contribute spurious correlations to the measured clustering. Accurately characterising the survey selection function is a key requirement for DESI. This is achieved in multiple steps, the first of which is through the creation of random catalogs of unclustered distributions of points covering the observed survey region. Then, the effects of known non-cosmological sources of density fluctuations in the data can be incorporated into this random catalog by adjusting the density or through the additional use of weights. Rather than apply weights to randoms to match data we can alternatively apply inverse weights to the data to remove effects. Our adopted approach depends on the particular nature of each effect and is detailed in this work. In the final catalogs, the ratio of weighted galaxy counts to weighted random counts is intended to produce a density field that is free from non-cosmological fluctuations.
 These weighted data and random catalogs are used to obtain measurements of the clustering signal that can be accurately modeled, including additional correction terms for observational effects. One purpose of this work is to identify the aspects of the analysis that require these correction terms and how they can be modeled.

This paper describes the selection of the galaxy and quasar catalogs used for the cosmological analyses and released as part of DESI DR1, the creation of the random catalogs and correction of survey-specific effects and foregrounds, and measurement and validation of the 2-point clustering statistics. We summarize here the work of many supporting studies, including a technical overview of the DESI LSS catalog creation \cite{KP3s15-Ross}, the pipeline for simulating DESI fiber assignment \cite{KP3s7-Lasker}, the catalog blinding scheme and its validation \cite{KP3s9-Andrade,KP3s10-Chaussidon}, and the creation and use of a new map of Galactic extinction based on spectra DESI has measured of stars \cite{KP3s14-Zhou}. The impact of imaging survey systematics on target selection is studied by \cite{KP3s13-Kong} for LRGs and \cite{KP3s2-Rosado} for ELGs, and the impact of this for full-shape clustering measurements is presented in \cite{KP5s6-Zhao}, for primordial non-Gaussianity measurements in \cite{ChaussidonY1fnl},  and for BAO in \cite{KP3s2-Rosado}. Systematic variations in the DESI spectroscopic success rate and our approach to modelling and removing the trends from the DR1 data are described in \cite{KP3s3-Krolewski}, and the ELG spectroscopic success rate and effects of catastrophic redshift errors are studied in \cite{KP3s4-Yu}. A general overview of the effects of the DESI fiber assignment algorithm on the DR1 sample and a method to quickly emulate fiber assignment effects in simulations is presented in \cite{KP3s6-Bianchi}, while the method for mitigating fiber assignment effects in our clustering analyses is described and validated in \cite{KP3s5-Pinon}. \cite{KP3s8-Zhao} presents an overview of all DESI DR1 simulations; all of these are based on measurements \cite{abacushod2, abacusHODELG, bgs_hod} of the clustering signal in DESI Early Data Release \cite{DESI2023b.KP1.EDR}. Finally, the methods for determining the covariance of the measured 2-point clustering statistics are described in \cite{KP4s7-Rashkovetskyi,KP4s8-Alves} and validated in \cite{KP4s6-Forero-Sanchez}.

\begin{table}
    \centering
    \small
    \resizebox{\columnwidth}{!}{%
        \begin{tabular}{|l|r|r|}
            \hline
             Ref.  &  Topic & Section\\
            \hline
\cite{KP3s15-Ross} & DESI LSS catalogs & \cref{sec:LSScat,sec:footprint,sec:compdef,sec:weights} \\
            \cite{KP3s9-Andrade} & Catalog-level blinding  & \cref{sec:blinding}\\
            \cite{KP3s10-Chaussidon} & Catalog-level blinding method for $f_{\rm NL}$ measurements & \cref{sec:blinding}\\
\cite{KP3s6-Bianchi} & Incompleteness due to fiber assignment  & \cref{sec:fiberassigncomp}\\
\cite{KP3s5-Pinon} & Removing scales affected by fiber 
assignment incompleteness & \cref{sec:fiberassigncomp}\\
\cite{KP3s7-Lasker} & Alternative realizations of DESI fiber assignment & \cref{sec:altmtl} \\
\cite{KP3s14-Zhou} & Improved Galactic extinction maps from DESI Observations of stars& \cref{sec:imsys}\\
\cite{KP3s13-Kong} & Forward modelling imaging systematics for DESI LRGs& \cref{sec:imsys}\\
\cite{KP3s2-Rosado} & Correcting for imaging systematics in DESI ELGs  & \cref{sec:imsys}\\
\cite{KP3s3-Krolewski} & DESI spectroscopic systematics & \cref{sec:specsys}\\
            \cite{KP3s4-Yu} & Correcting for spectroscopic systematics in DESI ELGs & \cref{sec:specsys}\\
          \cite{KP4s6-Forero-Sanchez} &	Comparison between analytical and mock-based covariance matrices  & 	\cref{sec:cov}\\
            \cite{KP4s7-Rashkovetskyi}  &	 Analytic covariance matrices for correlation functions& \cref{sec:cov}\\
             \cite{KP4s8-Alves} &	Analytic covariance matrices for power spectra 	& \cref{sec:cov}\\
\cite{KP3s8-Zhao} & Simulations of DESI LSS & \cref{sec:mocks}\\
            \hline
             \end{tabular}
    }
    \caption{\label{tab:supportingpapers}
    The list of the papers supporting this paper and the corresponding sections 
    where their results are discussed.}
\end{table}

The results presented here are part of a wider series of key papers based on the DESI DR1. These include measurement of BAO in galaxies and quasars \cite{DESI2024.III.KP4} and in the Lyman-$\alpha$ forest \cite{DESI2024.IV.KP6}, cosmological model constraints derived from BAO \cite{DESI2024.VI.KP7A}, analysis of the full-shape of the 2-point clustering power spectrum including redshift-space distortions \cite{DESI2024.V.KP5}, and cosmological implications of these full-shape measurements \cite{DESI2024.VII.KP7B}.

This paper is structured as follows: In \cref{sec:data}, we summarize the DESI DR1 data and how it is transformed into LSS catalogs. In \cref{sec:zcut}, we describe the spectroscopic selection criteria applied to DESI DR1 LSS catalogs and present the resulting redshift distributions and sample sizes. In \cref{sec:footprint}, we present the sky geometry of the DESI DR1 LSS catalogs and the various veto masks applied within the area. In \cref{sec:fiberassigncomp}, we summarize the details of fiber assignment incompleteness in DR1 and how its effects are mitigated in both the construction and analysis of the DR1 LSS catalogs. In \cref{sec:imsys}, we present how properties of the imaging used to select DESI samples impart spurious density variation into the DR1 LSS catalogs and how we correct for this. In \cref{sec:specsys}, we summarize trends in the DESI spectroscopic success rates with DESI observing properties and how we conclude they have a negligible effect on DR1 2-point clustering measurements. In \cref{sec:weights}, we present how weights are applied to the LSS catalogs, drawing on the previous three sections, and the normalizations of the DR1 samples. \cref{sec:compare_to_sdss} compares DESI DR1 to the Sloan Digital Sky Survey (SDSS) in footprint and redshift coverage and consistency in redshift measurements for the more than 400,000 objects with both SDSS and DESI spectra. \cref{sec:2pt}, we described how 2-point statistics are measured from the LSS catalogs in both configuration- and Fourier-space, how window functions are estimated to allow comparison between the 2-point statistics and cosmological models, and how covariance matrices that allow the consistency between the measurements and models are estimated. In \cref{sec:mocks}, we describe how simulations of the DR1 data were produced. In \cref{sec:clus}, we present comparisons between the 2-point clustering of the DR1 data and our simulations of it. Finally, we conclude in \cref{sec:conclusions}.

Throughout this work, for the calculation of the distance-redshift relation and to set the initial conditions of any simulations we use a fiducial flat $\Lambda$CDM cosmological model with:
$\omega_{\mathrm{b}} = 0.02237, \; \omega_{\mathrm{cdm}} = 0.12, \; h = 0.6736, \;  A_{\mathrm{s}} = 2.083 \cdot 10^{-9}, \; n_{\mathrm{s}} = 0.9649, \; N_{\mathrm{eff}} = 3.044, \; \sum m_{\nu} = 0.06 \; \mathrm{eV}$ (with a single massive neutrino eigenstate). This model matches the mean of the posterior from fitting to the CMB temperature, polarisation and lensing power spectra as measured by Planck \cite{planck2018}.

\section{Data}
\label{sec:data}

The DESI instrument \cite{DESI2022.KP1.Instr} on the Nicholas U. Mayall
Telescope at Kitt Peak, Arizona measures the spectra of 5,000 `targets' \cite{DESItarget} at once, using robotic positioners to place optical fibers in the 7 square degree field of view of the focal plane \cite{Corrector.Miller.2023} at the celestial coordinates of the targets \cite{DESIfocalplane,DESIfa}. The fibers are divided into ten `petals' and carry the light to a corresponding ten climate-controlled spectrographs. Each set of  
targets assigned to a set of 5,000 fibers is represented by a specific central sky position and denoted as a `tile'. 

The DESI main survey started observations on May
14, 2021, after a period of survey validation \cite{DESIsv}.
We analyze the main survey data to be released with DESI DR1 \cite{DESI2024.I.DR1}; this includes observations through to June 14, 2022. The DESI spectroscopic pipeline \cite{DESIpipe} first processed these data the morning following observations for immediate quality checks, and then reprocessed them in a homogeneous processing run denoted as `\iron'.\footnote{It was processed with version 23.1 of the DESI software, available on NERSC via source /global/common/software/desi/desi_environment.sh 23.1.} We use the redshift catalogs produced with the \iron\ spectroscopic processing, which will be released in DR1. Full details of what we use are presented in \cref{sec:dataspec}.

DESI has two distinct observing programs for large-scale structure observations, referred to as `bright' and `dark' time \cite{surveyops}. The decision on which type of tile to observe is determined prior to every exposure, depending on the observing conditions \cite{surveyops}. As described in \cite{DESItarget}, dark and bright time each have their own set of target samples, with independent `Merged Target Ledgers' (MTL). Each MTL is used to track the observation history of the targets. The states of the successfully observed targets are updated in the MTL after every tile is observed and validated so that the completed targets will no longer compete with unobserved targets for fibers. The updated MTL is then used as an input to determine which targets are assigned to what fibers on every tile, using DESI's \textsc{fiberassign} software \cite{FBA.Raichoor.2024}.\footnote{\url{https://github.com/desihub/fiberassign}} DR1 contains 2744 tiles observed in `dark' time and 2275 in `bright' time.  Completeness, in terms of the ratio of observed spectra to total targets, is built up by overlapping tiles, nominally up to four times in bright and seven in dark time. The main survey strategy prioritizes observing tiles that overlap at any given area of the sky, after validating the quality of observations of any underlying tile \cite{surveyops}, rather than covering new area. 

In the following two subsections, we describe the input target samples used for these observations and then the outputs from the analysis of observed spectra. 

\subsection{Target Samples used for DR1 LSS Catalogs}
\label{sec:target}

DESI observes four classes of extra-galactic targets: quasars (QSO; \cite{QSOtarget}), luminous red galaxies (LRG; \cite{LRGtarget}), emission line galaxies (ELG; \cite{ELGtarget}), and a bright galaxy sample (BGS; \cite{BGStarget}). All four of these classes of DESI targets were selected based on photometry from Data Release 9 (DR9) of Legacy Survey (LS) 
\cite{LS.Overview.Dey.2019,LS.dr9.Schegel.2024} imaging. The LS data combines photometric data from multiple sources. DESI targeting in the North Galactic Cap (NGC) at declination $>32.375^\circ$ uses $g$ and $r$ band photometry obtained by the Beijing-Arizona Sky Survey (BASS; \cite{BASS.Zou.2017}) and the $z$ band photometry obtained by the Mayall z-band Legacy Survey (MzLS). At low declination and in the South Galactic Cap (SGC), all of the $g$, $r$, $z$ bands were observed using the Dark Energy Camera (DECam; \cite{DECam}), as part of the Dark Energy Camera Legacy Survey (DECaLS) and the Dark Energy Survey (DES; \cite{DES}). Further details on all of these imaging programs are available in \cite{LS.Overview.Dey.2019}.
These regions are denoted, respectively, as the `North' and `South' photometric regions. Infrared photometry in the $W1$ and $W2$ bands from the WISE satellite \cite{wright2010wide,lang2014unwise} is used over the entire sky. 

In this paper, we describe the samples used in the DESI DR1 cosmological analyses. The samples are first defined by their target bits, encoded in the \texttt{DESI\_TARGET} column of the target catalogs, which map directly to the priority with which targets are assigned fibers. When science targets compete for fibers, the one with the highest priority receives the assignment. Any given target can pass the selection cuts of multiple target classes and in such cases, the target is always assigned the highest of the potential priority values.

We create DR1 LSS catalogs for three (nearly) distinct target samples observed exclusively in dark time (LRG, ELG, QSO) and one observed exclusively in bright time (BGS). Below, we describe the target properties of the dark time tracers in the order of greatest to least priority, then describe the bright time sample. We finally discuss the associated random samples created to enable clustering measurements. In all cases, we describe any cuts applied at the level of targeting (i.e., without any information from spectroscopic observations) that produce the samples considered for DR1 LSS catalogs.

\paragraph{QSO:} QSO are assigned the highest priority (\texttt{PRIORITY} value 3400 in the target catalogs). They have the lowest sky density at 310 deg$^{-2}$. They were given a high priority to ensure high completeness. This is important given the low density of the sample, which means that measurements are shot-noise-limited. Further, each target determined to have a redshift $z>2.1$ is given three additional observations at high priority (\texttt{PRIORITY} value 3350 in the target catalogs). These additional observations are meant to increase the signal-to-noise of spectra with Lyman-$\alpha$ forest absorption features. The full details of the QSO target selection are provided in \cite{QSOtarget}. Imaging in all of the $g$, $r$, $z$, $W1$, and $W2$ bands is used, from which a random forest algorithm selects likely quasars, restricting to data with $r<23$. This algorithm was trained and applied separately in the BASS/MzLS region, the DES region, and the DECaLS region\footnote{See the beginning of \cref{sec:target} for more details on these regions.}. DES and DECaLS used the same instrument, but the DES region typically contains data with greater imaging depth than the DECaLS data. There are thus three distinct photometric selections for QSO applied to three distinct regions on the sky, and we correct for imaging systematics and assign redshifts to the randoms (see \cref{sec:imsys,sec:ranz}) for QSO separately in each of these three regions. However, for the results we present, we will typically show the combined DECam (DES + DECaLS) dataset.

\paragraph{LRG:} DESI LRG targets have a sky density of just over 600 deg$^{-2}$ and are given an intermediate priority (\texttt{PRIORITY} value 3200 in the target catalogs). They are selected as described in \cite{LRGtarget} using $g$, $r$, $z$, and $W1$ flux measurements. The specific selection is tuned separately in the BASS/MzLS and DECam regions to obtain a sample of passively evolving galaxies with an approximately constant number density $5\times10^{-4}\;h^3$Mpc$^{-3}$ in the redshift range $0.4<z<0.8$. Above this redshift the density falls, to less than $1\times10^{-4}\;h^3$Mpc$^{-3}$ by $z=1.1$ (see \cref{fig:nzall}), due to a $z$-band fiber magnitude threshold (see \cite{LRGtarget} for full details). 

\paragraph{ELG:} The total sky density of DESI ELG targets is $\sim$ 2400 deg$^{-2}$, but the sample is somewhat complicated as it is split into three groups of different priorities. The targets are initially assigned either lower priority (`\texttt{ELG\_VLO}': \texttt{PRIORITY} value 3000, 25\% of the total) or higher priority (`\texttt{ELG\_LOP}': \texttt{PRIORITY} value 3100, 75\% of the total) based on the photometric cuts described in \cite{ELGtarget}. The same selection cuts are applied to the photometry in the BASS/MzLS and DECam regions, with a $g_{\rm fiber}<24.1$ threshold. A random selection of 10\% of both priority groups are promoted to the same priority (3200) as LRG targets, and are given an additional targeting bit `\texttt{ELG\_HIP}'. This boosting of priority increases the chance that pairs of LRG and ELG at small angular separations will be observed. Any \texttt{ELG\_HIP} is always also either \texttt{VLO} or \texttt{LOP}. For the DR1 cosmological analyses, we select only \texttt{ELG\_LOP} targets for the final sample, 10\% of which are also \texttt{ELG\_HIP}. 
For our analyses of DR1 data, the \texttt{VLO} is omitted simply due to the complexity it added to an already complicated analysis, but we plan to include it in analyses of future DESI data releases.
 Additionally, any of these targets that are classified as QSO by the target selection pipeline, and thus included in the QSO sample, are rejected. This removes duplicates from the DR1 analysis and simplifies the priority masking (see \cref{sec:prioritymask}). We refer to this final selected sample simply as `ELG' from here on. 

\paragraph{BGS:} Of the targets observed in bright time, we use only the \texttt{BGS\_BRIGHT} sample \cite{BGStarget} for cosmological analysis. This sample is defined by a simple magnitude threshold of $r<19.5$, which provides a target density of 864 deg$^{-2}$ and is selected via the \texttt{BGS\_TARGET} column in the DESI target catalogs. In \cref{sec:zcut} below we describe a further absolute magnitude cut that is later applied to this target sample, with the resultant clustering sample denoted simply as `BGS'.

\paragraph{Random samples:} In addition to the DESI target samples, the DESI targeting team provide samples of uniform random sky positions occupying the same area as the DESI targets (covering the full DESI footprint), as described in Section 4.5 of \cite{DESItarget}. Conveniently, each entry in these `randoms' includes the most relevant metadata associated with the imaging data, at the given celestial coordinates. These data are processed in a manner that matches the processing of the target samples defined above, and they thus define a reference sample that matches the sky geometry of the observed DESI samples.

The randoms are divided into many distinct files, each with a density of 2500 deg$^{-2}$. The constant density is convenient for quick calculations of sky area. For DR1 LSS catalogs, we provide up to 18 of them. They can be used independently and the total number used depends on the density needed for a particular analysis. The combination of all 18 provides a sky density that is more than 100 times that of all of the DR1 LSS catalogs.
 
We use the DESI \textsc{fiberassign} software together with the details on all of the individual observed tiles and positioners to determine all of the targets and randoms that could have been reached by a DESI positioner and were thus a `potential assignment'. The collection of all potential assignments of targets or randoms forms the potential assignment galaxy and random catalogs. Since in the fiducial tiling, a given area on the sky is observable up to seven times and much of the focal plane can be reached by two positioners, all targets and randoms are (typically) assigned multiple tile and fiber values, and thus each unique \texttt{TARGETID} is likely to have multiple entries in the potential assignment catalogs.  Section 3.1 of \cite{KP3s15-Ross} describes this process in more detail.

\subsection{Adding Spectroscopic Information}
\label{sec:dataspec}

We use the `cumulative' tile-based redshift catalogs and associated data products from the \iron\ version of the DESI spectroscopic reduction pipeline \redrock\ \cite{Redrock.Bailey.2024,Anand24redrock}, which are released with DR1.\footnote{These are pairs of FITS files containing information on the redshift fits for all of the spectra observed on DR1 tiles.} Targets observed on multiple tiles get multiple entries and these are matched to the potential target catalogs via the target ID, tile ID, and fiber ID. The metadata associated with the particular coadded spectrum is matched via tile and fiber to the potential target and random catalogs. 

An exception to using the \iron\ reductions is for data taken on the night of December 12th, 2021. When investigating trends between the spectroscopic success and the observation date, \cite{KP3s3-Krolewski} found this night to have unusually low spectroscopic success. It was found that during the \iron\ processing, a bug caused incorrect calibration data to be used, only for this night's results. 
This issue was identified after the iron data was frozen and all the data taken for the night were reprocessed separately. We substitute the reprocessed data for the original data on the affected 8 dark and 9 bright tiles before constructing the LSS catalogs. These data will be released as a supplementary value-added catalog to the DR1 release.

To define the samples ultimately used for clustering analysis, we also use various additional data from the \iron\ spectroscopic reductions that are produced for every tile (and will be released with DR1) but are not included in the redshift catalogs.

For BGS, we use the information obtained from \textsc{fastspecfit} \cite{fastspecfit} for $k$-corrections, which are used to define the absolute magnitude threshold used for the DR1 cosmology sample described in \cref{sec:zcut}.
For ELG samples, we use the [OII] emission line flux measurements and uncertainties produced in the \texttt{emlin} files, which are used in defining the ELG spectroscopic success criteria, as detailed in \cref{sec:zcut}. We concatenate the information over all tiles and join it to the ELG potential target catalog via a match to the target ID, tile, and fiber. 

For the QSO samples, in addition to \redrock\ redshifts we use the results produced per tile by the machine learning-based classifier QuasarNET and the MgII `afterburner' \cite{QSOtarget}, which are used to define a `good' QSO.
The observations of QSO targets that pass the QSO selection are evaluated per tile and are then concatenated into a QSO catalog that is later joined to the QSO potential target catalog via a match to the target ID, tile, and fiber. A similar process concatenates the QSO information that is determined from spectra that are coadded across tiles (when observations on multiple tiles exist) and separated into \textsc{Healpix} \cite{Healpix} pixels. These separate `\textsc{Healpix}' QSO catalogs are used for Lyman-$\alpha$ forest analyses \cite{DESI2024.IV.KP6}, but are not used for the LSS analyses except for some comparisons.

Finally, for all samples, we use the information in the `zmtl' files, which include flags that indicate whether the DESI instrument was performing properly in terms of positioning, CCD wavelength coverage, calibrations, etc. The region of data flagged can be as large as a petal or as small as an individual fiber. We concatenate this information across all tiles, match it to the redshift catalog via the tile, fiber, and target ID, and store it in the column \texttt{ZWARN\_MTL}. This information is used for the hardware veto described in  \cref{sec:hardware_veto}.

\subsection{Transforming Data into LSS Catalogs}
\label{sec:LSScat}

The combination of all the data described in the previous two subsections provides information associated with every instance in which a DESI target or random could be reached by a DESI fiber positioner, which is recorded in the `combined information' catalogs described in Section 3 of \cite{KP3s15-Ross}. We define the DESI footprint\footnote{The footprint of DESI DR1 LSS samples can be seen in \cref{fig:allcomp}, which is discussed further in \cref{sec:comp} in the context of completeness variations.} as the area containing such reachable targets, and subsequently apply a series of veto masks to this, as described below. The area of this footprint, and coverage properties within it, can be matched to a resolution of less than one arcsecond\footnote{This accuracy is assessed by comparing the physical position determined by the DESI \textsc{fiberassign} software to the actual physical position on the focal plane that a fiber positioner was instructed to move to at the time of observation. These differ, e.g., due to dynamic changes in the DESI optics.} by randoms through the use of the DESI \textsc{fiberassign} software, as described in \cite{KP3s15-Ross}. 

To create what we denote as the `full' LSS catalogs, the data and random potential assignment catalogs are reduced to a sample with unique entries for each target ID. This process includes careful sorting to ensure that only the most relevant instance for each target is retained---most importantly, keeping good observations over non-observations---and is vital for obtaining accurate completeness corrections. The process is described fully in Section 4 of \cite{KP3s15-Ross}. The resulting catalogs are split by target class and include one entry per target. No cuts are applied to these full catalogs based on analysis of the observed spectra, but all of the relevant information is included in order to enable quality cuts as desired. For instance, one can apply simple criteria to the columns provided in the catalogs to obtain a sample with `good' redshifts within some desired redshift range. The fact that we do not apply cuts based on the spectra also means that we can quickly determine completeness statistics, in terms of both assignment and spectroscopic success. 

Several additional processing steps are applied to the LSS catalogs to produce the final `clustering' catalogs. First, the DR1 full catalogs are output in three stages: before any veto masks (`\texttt{full\_noveto}'), after fiducial veto masks (`\texttt{full}'), and after applying vetoes based on imaging properties recorded in \textsc{Healpix} maps (`\texttt{full\_HPmapcut}'). These veto masks are described in \cref{sec:footprint}. The full catalogs are used to determine corrections for variations in completeness (\cref{sec:comp}), imaging data properties (\cref{sec:imsys}), and spectroscopic data properties (\cref{sec:specsys}). Cuts on the spectroscopic information are then applied to the full catalogs to produce clustering catalogs, as described in \cref{sec:zcut}. We summarize and define new weights included in the clustering catalogs in \cref{sec:weights}. These clustering catalogs are then used as the inputs for all results presented in the sections after \cref{sec:weights}. Versions v1.2 (used in \cite{DESI2024.III.KP4,DESI2024.VI.KP7A}) and v1.5 (used in \cite{DESI2024.V.KP5,DESI2024.VII.KP7B}) will be released publicly with DR1. The differences between the versions are detailed in \cref{app:catver}. All of the results presented in this work are based on version v1.5 of the DESI DR1 LSS catalogs unless otherwise noted.

\subsection{Catalog Blinding}
\label{sec:blinding}

To protect against confirmation bias in our DESI DR1 cosmological inference, we applied a \emph{blinding} scheme to obscure the true cosmology during early analyses until the full large-scale structure analysis pipeline was finalised. This blinding scheme was applied at the catalog level, to produce blinded clustering catalogs for further analysis. The blinding was meant to alter three distinct pieces of information that can be extracted from the DESI 2-point measurement: 1) the location of the BAO feature; 2) the anisotropy in the clustering imparted via redshift-space distortions (RSD) due to structure growth; and 3) the large-scale scale-dependent bias that is generated by local primordial non-Gaussianity, $f_{\rm NL}$. The specific blinding method applied, and its validation using DESI simulations, is presented in \cite{KP3s9-Andrade}. We summarize the procedure here. 

The BAO and RSD signatures were blinded by shifting the measured DESI redshifts, following the methods proposed in \cite{brieden2020}. The measured redshifts were first altered in a way that would mimic a change in the dark energy equation of state parameters $w_0$ and $w_a$. To do this, the measured true redshifts were converted to comoving distances using the DESI fiducial cosmology, and then coherently shifted based on the expected difference in redshifts between an object at that same comoving distance from an observer in a cosmological model with hidden values of $w_0$ and $w_a$, and in the fiducial cosmological model. A further shift was applied to blind the RSD structure growth measurements. To do this, RSD effects present in the measured redshifts were approximately subtracted based on an estimate of the local displacement field and the fiducial growth rate, and then new RSD shifts were applied to match the effect of a blinded growth rate value $f$ (full details of the method can be found in \cite{KP3s9-Andrade}). The shift in $f_{\rm NL}$ was implemented by altering the weight column of the LSS catalogs, using the methodology described in \cite{KP3s10-Chaussidon}.

To choose a $(w_0,w_a)$ pair for blinding, we produced a list of 1000 randomly sampled pairs, with the range of possible values bounded to keep the expected shift to the isotropic BAO scale measurement relative to its value in the fiducial cosmology to within 3\% over the redshift range $0.4<z<2.1$. The order of the pairs was randomized and they were written to a file on disk. The first time that the DESI DR1 LSS `clustering' catalogs were generated, a random integer was chosen as the row to select the $(w_0,w_a)$ pair used for DR1 blinding. The value of the integer was stored in a separate file, which was then read every subsequent time the LSS catalog production was run (following iterative improvements to the pipeline while the analysis remained blinded), so that the same blinding was consistently applied. 

Rather than being drawn randomly, the relative shift in $f$ was automatically calculated, using a linear RSD model as described in \cite{KP3s9-Andrade}, in order to approximately compensate for the expected change due to the $w_0,w_a$ blinding in the monopole of the redshift-space clustering, but with a maximum allowed shift of up to 10\% relative to its fiducial value. The procedure left the expected amplitude of the clustering monopole approximately unchanged by the blinding, which meant that, e.g., the clustering amplitude of the (blinded) data monopole would still be expected to match that of DESI mocks. However, the procedure imparted a shift in the amplitude of the higher-order multipoles determined by the unknown values of the $(w_0,w_a)$ pair, effectively blinding the true structure growth information in the data. The relative shift in $f_{\rm NL}$ was randomly chosen to be between $(-15,15)$, and the value applied to DR1 blinding was held fixed for all blinded catalogs by generating the value via a random seed determined from the random integer described above.

The application of the blinding scheme took the `full' catalogs described in the previous subsection as inputs. These catalogs contain all of the selection function details that are described in \cref{sec:comp,sec:imsys,sec:specsys}, but the blinding procedure itself changed this $n(z)$. The completeness-corrected $n(z)$ was determined from the number density in the full catalogs and the redshifts were shifted as described above. The $n(z)$ was then re-measured and a weighting was applied to the blinded clustering catalog to make the blinded $n(z)$ match the original $n(z)$. The steps of adding radial information to the randoms and adding an `FKP' \cite{FKP} weight to optimize the expected signal to noise given the number density variations then proceeded as described in \cref{sec:weights}. 

Six versions of the blinded DR1 LSS catalogs were produced and their clustering measurements analysed while the LSS pipeline was iteratively improved before the first unblinded version of the catalog was created. Changes to the LSS catalogs that occurred after unblinding are described in \cref{app:catver}.

\section{Redshift Selection for DR1 LSS Catalogs}
\label{sec:zcut}

\begin{figure*}
    \centering 
    \includegraphics[width=.49\columnwidth]{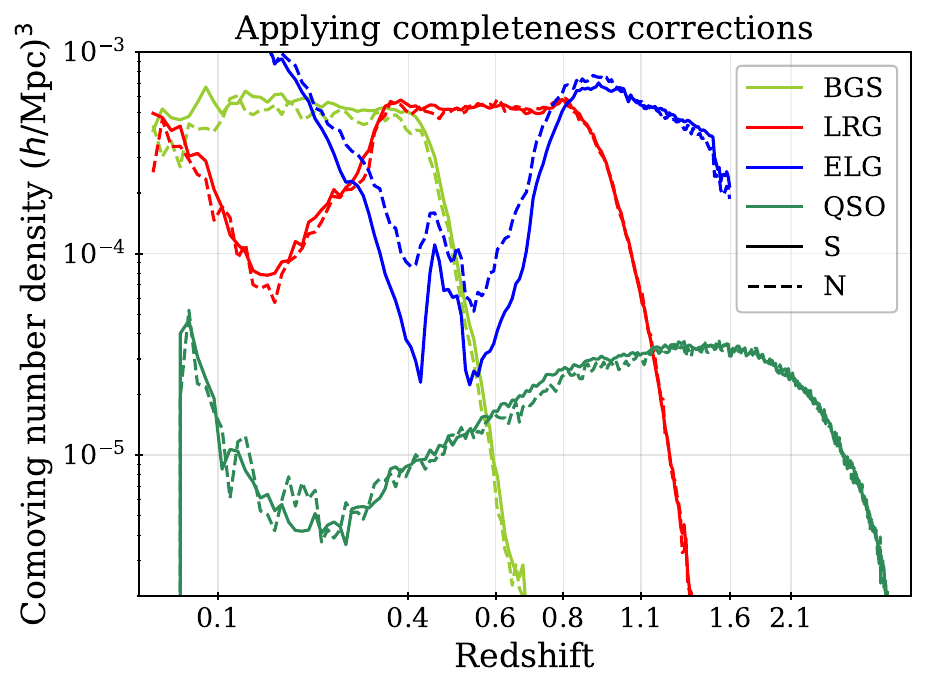} 
    \includegraphics[width=.49\columnwidth]{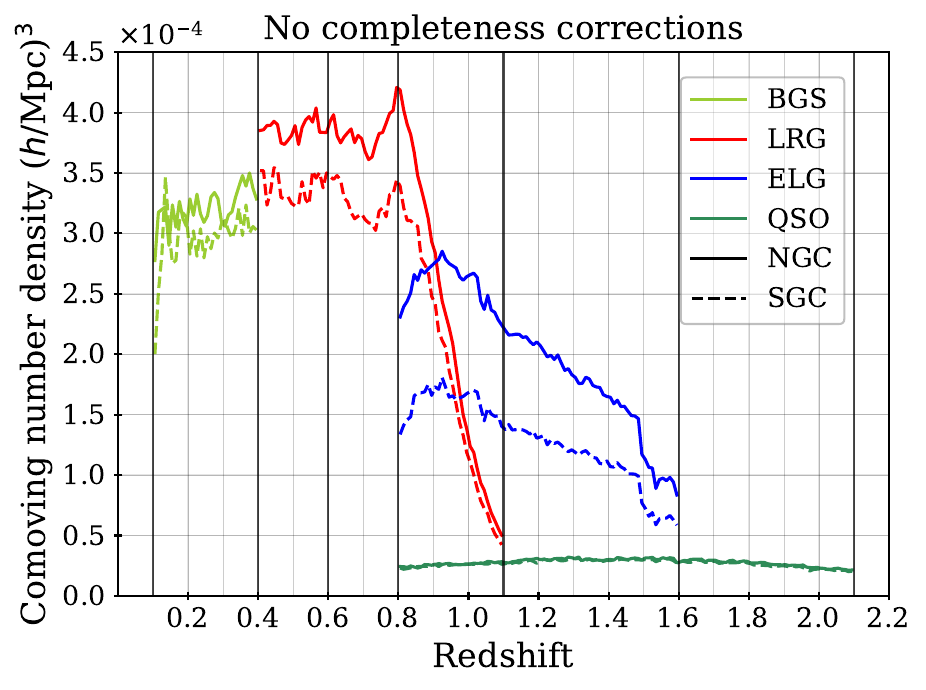}
    \caption{The left-hand panel displays the comoving number density for the four discrete galaxy and quasar tracers in the DESI DR1, estimated for a sample with no targeting incompleteness. Solid curves show the results for data targeted with DECam photometry (`S') and dashed curves for the target data that used BASS/MzLS photometry (`N'). The right-hand panel displays the same information, but without completeness corrections and with data limited to the redshift ranges adopted in DR1 cosmological analysis. In this panel the solid and dashed curves show results for the North and South galactic cap regions (NGC and SGC) respectively: the differences are driven primarily by the relative assignment completeness in each region.  
    Note that the left and right panels use logarithmic and linear axis scaling, respectively, which allow different distinctive features to be seen. In both panels, verticle grid lines appear at the limits of the redshift bins used to define the final DESI DR1 samples; on the right, they appear at greater thickness than those otherwise at every 0.1 in redshift. 
    }
    \label{fig:nzall}
\end{figure*}

For some fraction of the observed DESI spectra, secure redshifts could not be measured. We thus require spectroscopic success criteria that can be applied to the outputs of the redshift fitting pipeline that recover samples that maximize the sample size while maintaining sufficiently high purity and sufficiently low catastrophic failure rates.
For each DR1 sample, we apply the same spectroscopic success criteria as used for the DESI SV3 LSS catalogs \cite{DESI2023b.KP1.EDR,BGStarget,LRGtarget,ELGtarget,QSOtarget}, and we describe these below, together with the redshift cuts and the redshift binning that is applied within those cuts. For the galaxy samples, the spectral type from the redshift fit (which can be \texttt{QSO}, \texttt{GALAXY}, or \texttt{STAR}) does not enter in the success criteria; e.g., an observed LRG target that is classified by \redrock\ as a quasar or star but passes the success criteria defined below will be counted as a success, as we believe DESI has properly classified the observation based on the spectrum. \Cref{tab:Y1lss} summarizes the statistics of the final samples, while \cref{tab:zbins} provides the numbers of good redshifts in each of the redshift bins used for clustering measurements, and the redshift distributions are shown in \cref{fig:nzall}. 
\paragraph{BGS:}
For BGS, we apply the same spectroscopic success criteria as originally suggested in \cite{BGStarget}:
\begin{itemize}
    \item Spectroscopic success: \texttt{ZWARN}==0, $\Delta \chi^2 > 40$
\end{itemize}
Here $\Delta \chi^2$ is the difference in the fit $\chi^2$ for the best-fit and second-best fit redshift solutions from the \redrock\ pipeline \cite{Redrock.Bailey.2024}, and \texttt{ZWARN} != 0 is a \redrock\ output flag indicating any known problems with the data or the fit. Using this definition, and after applying the vetoes described in the following section, 98.9\% of observed \texttt{BGS\_BRIGHT} targets are classified as a success. A redshift selection $0.1<z<0.4$ is applied to the BGS sample,
and we apply an absolute magnitude cut $M_r < -21.5$, which provides a sample with an approximately constant number density, matching the number density of the LRG sample at redshift 0.4. The $M_r$ value is determined using the SDSS $r$-band $k$-corrected absolute magnitude determined using \textsc{Fastspecfit} \cite{fastspecfit_code}, $M_{r,\mathrm{fsf}}$, and an $e$ correction:
\begin{equation}
    M_r = M_{r,\mathrm{fsf}} +0.97*z-0.095\,.
\end{equation}
The redshift dependence of the $e$ correction matches that applied to the SV3 sample, and the constant 0.095 produces a sample that is a close match to the SV3 characteristics for any given $M_r$ cut. The $M_r < -21.5$ cut reduces the total number of successful redshifts from 4,036,190 (for the \texttt{BGS\_BRIGHT} sample within the area defined in the following section) to 485,331. Although it removes much of the data, the $M_r < -21.5$ selection produces a sample with an approximately constant number density around 5$\times 10^{-4}\;(h/{\rm Mpc})^3$, after completeness corrections. Applying redshift bounds $0.1<z<0.4$ further reduces the number of redshifts to the 300,043 that we use in the final DESI DR1 cosmological analysis. The upper bound in redshift separates the BGS and LRG samples, while the lower bound of $z>0.1$ was chosen as the effects of bright limits on the fiber magnitudes becoming increasingly important at lower redshifts, while this cut removes only a small fraction of the available volume.
The left-hand panel of \cref{fig:nzall} shows the comoving number density, $n(z)$, calculated based on the completeness-weighted counts of observed redshifts within redshift shells (the area that enters the volume calculation is described in the following section). One can see that the BGS number density is a close match to that of the LRG sample at the separation redshift $z=0.4$, and that it decreases sharply just above this. The right-hand panel of \cref{fig:nzall} shows the raw density, without completeness corrections, which represents the density one should use to estimate shot-noise contributions.
 This is just greater than 3$\times 10^{-4}\;(h/{\rm Mpc})^3$ for the BGS sample, which is dense enough to make shot-noise a minor contribution to the BGS statistical uncertainty in the DR1 2-point measurements (as $nP\sim 3$). Thus, despite removing nearly 90\% of the BGS sample, we expect the sample with the $M_r < -21.5$ cut applied to contain most of the clustering information useful to the cosmological analyses of \cite{DESI2024.III.KP4, DESI2024.V.KP5} and to have a nearly constant galaxy population that is simpler to model and simulate. LSS catalogs have been produced for the full DESI \texttt{BGS\_BRIGHT} (and \texttt{BGS\_ANY} that includes a selection to a fainter flux limit) DR1 samples and will be released publicly with DR1. However, they were not subject to the same scrutiny applied to the $M_r<-21.5$ sample that we refer to as `BGS' from here on.

\begin{table}
\centering
\begin{tabular}{|l|ccccc|}
\hline%\hline
Tracer & \# of good z & $z$ range & Area [deg$^2$] & $C_{\rm assign}$ & z succ. \% \\\hline
BGS ($M_r<-21.5$) & 300,043  & $0.1 < z < 0.4$ & 7473 & 63.6\% & 98.9\%\\
LRG  & 2,138,627 & $0.4 < z < 1.1$ & 5740 & 69.3\% & 99.1\%\\
ELG  & 2,432,072  & $0.8 < z < 1.6$ & 5924 & 35.2\% & 72.7\% \\
QSO & 1,223,391 & $0.8 < z < 3.5$ & 7249 & 87.4\% & 66.8\% \\
QSO & 856,831 & $0.8 < z < 2.1$ & 7249 & 87.4\% & 66.8\% \\
\hline
\end{tabular}
\caption{Statistics for each of DESI tracer types used for DESI DR1 cosmological analysis. We list the number of good redshifts included, the redshift range we included them from, the sky area occupied, and the observational completeness within that area. The criteria for selecting good redshifts and deciding redshift bounds are discussed in \cref{sec:zcut}. Determinations of the footprint area and completeness are presented in \cref{sec:footprint}. The spectroscopic success rates (`z succ \%'), area, and assignment completeness ($C_{\rm assign}$) are determined for the sample without any cuts on redshift, and these are thus the same for QSO $0.8<z<3.5$ and QSO $0.8<z<2.1$. The area is different for different tracer classes due to priority vetoes (e.g., a QSO target can remove sky area from lower priority samples) and small differences in the imaging vetoes applied. The assignment completeness is the percentage of targets within the DR1 footprint that were observed.
\label{tab:Y1lss}}
\end{table}

\begin{table}
\centering
\begin{tabular}{|l|cc|}
\hline%\hline
Tracer(bin) & \# of good z & $z$ range  \\\hline
BGS  & 300,043  & $0.1 < z < 0.4$ \\
LRG1  & 506,911 & $0.4 < z < 0.6$ \\
LRG2  & 771,894 & $0.6 < z < 0.8$ \\
LRG3  & 859,822 & $0.8 < z < 1.1$ \\
ELG1  & 1,016,365  & $0.8 < z < 1.1$ \\
ELG2  & 1,415,707  & $1.1 < z < 1.6$ \\
QSO & 856,831 & $0.8 < z < 2.1$  \\
\hline
\end{tabular}
\caption{The redshift ranges and number of good redshifts for each of the redshift bins that will be used for clustering measurements.
\label{tab:zbins}}
\end{table}

\paragraph{LRG:}
For LRGs, we apply the same spectroscopic success criteria as originally suggested in \cite{LRGtarget}:
\begin{itemize}
    \item Spectroscopic success: \texttt{ZWARN}==0, $\Delta \chi^2 > 15$
\end{itemize}
The LRG target selection was optimized for $z>0.4$ and the BGS sample covers $z<0.4$ at a higher number density. We therefore apply the redshift cuts $0.4<z<1.1$ for the LRG sample, using three redshift bins of $0.4<z<0.6$, $0.6<z<0.8$, and $0.8<z<1.1$. These provide samples with sufficient signal-to-noise for BAO measurements \cite{DESI2024.III.KP4} and match choices applied to previous SDSS studies. The split at $z=0.8$ was chosen to match the choice of the lower bound on the ELG sample (described next). \Cref{fig:nzall} shows that the LRG number density is nearly constant for $0.4<z<0.8$ and begins to drop for $z>0.8$: the redshift upper limit was chosen as the number density falls to less than 1$\times 10^{-4}\;(h/{\rm Mpc})^3$ above it. The redshift efficiency of the LRG sample is the highest of the DESI DR1 targets: 99.1\% of observed LRG targets (within the footprint defined in the following section) have a good redshift and 90\% are also within $0.4<z<1.1$. The DR1 LRG sample is the most efficient in terms of the fraction of observed spectra included in the clustering measurements.

\paragraph{ELG:}
For ELGs, we apply the same spectroscopic success criteria as originally suggested in \cite{ELGtarget}:
\begin{itemize}
    \item spectroscopic success: $\log_{\rm 10}(S_{\rm [O II]}) + 0.2\log_{\rm 10}(\Delta \chi^2)  > 0.9$,
\end{itemize}
where $S_{\rm [O II]}$ is the signal-to-noise ratio of the [OII] emission line doublet. We select ELGs in the range $0.8<z<1.6$. 72.7\% of ELG observations yield a successful redshift, 86\% of which are also within $0.8<z<1.6$. Below redshift 0.8, the expected signal-to-noise is dominated by LRGs and target density fluctuations become more severe, including strong variations in the redshift distribution with the imaging depth, detailed in \cite{KP3s2-Rosado}. Above redshift 1.6, the [OII] doublet cannot be observed with the DESI spectrograph; a significant fraction of the redshift failures are presumed to be $z>1.6$ galaxies. \Cref{fig:nzall} shows that the number density of ELGs decreases sharply for $z>1.5$ as at these redshifts the [OII] doublet falls at wavelengths that overlap highly with sky lines, increasing the noise entering the $S_{\rm [O II]}$ determination and thus lowering the success fraction. 
While this reduces the number density in the $1.5<z<1.6$ range, we are able to account for trends in the success with effective observing time (see \cref{sec:specsys}) and any impact of catastrophic redshift failures, e.g., due to misidentified sky lines, is found to be negligible in \cite{KP3s4-Yu}.
One can further observe that the difference between the raw number densities in the NGC and SGC is greater for the ELG sample than for any other. This is due to the difference in completeness between the NGC and SGC and will be discussed further in the following section; the ELG sample is most affected because it has the lowest priority. The ELG sample is split into two redshift bins, $0.8<z<1.1$ and $1.1<z<1.6$, with the split at $z=1.1$ motivated by it being the maximum redshift used in the LRG analysis.

\paragraph{QSO:}
We apply the same spectroscopic success criteria as originally suggested in \cite{QSOtarget}, and treat all instances where an object has either \redrock, MgII, or QuasarNET spectral identification as a QSO as successes. Thus the criterion is simply:
\begin{itemize}
    \item Spectroscopic success: Not rejected by the quasar catalog.
\end{itemize}
The QSO sample is the only sample in DR1 that uses the spectral type as a factor in the spectroscopic success criteria. This is a complicating factor in the modeling of redshift systematics for the sample, discussed further in \cite{KP3s3-Krolewski}. 
We consider a broad redshift selection $0.8<z<3.5$ for QSOs, although a smaller subset with $0.8<z<2.1$ is used for primary analyses of the quasar clustering. 66.8 \% of observed QSO targets (within the region defined in the following section) yield a successful redshift, with 93\% of them within $0.8<z<3.5$ and 65\% within $0.8<z<2.1$. For the DR1 LSS catalogs, we use the QSO redshift measurement based on the first tile a QSO is observed on. We find that doing so has a negligible effect on the overall spectroscopic success rate (changes are $\sim0.1\%$) and simplifies the modeling of the spectroscopic success rate.

The QSO sample is significantly less dense than the other DESI tracer samples, as can be seen in  \cref{fig:nzall}. In the $0.8<z<1.6$ range, the completeness-corrected number density is no more than 15\% of that of ELGs at any redshift, and is typically below 10\%. However, since the QSO completeness is approximately $3\times$ that of the ELG sample in DR1, the number of redshifts is just under a factor of 5 smaller (2,432,027 for ELGs and 502,462 for QSO). The 354,190 QSOs with $1.6<z<2.1$ provide the only tracers in that redshift range for DESI DR1. These QSO numbers can be compared to the 454,452 QSOs with $0.8<z<2.2$ used for SDSS DR16 clustering analyses \cite{ebosslss}.

\section{DESI DR1 Geometry and Veto Masks}
\label{sec:footprint}

We define the DESI footprint as the locations on the sky where it was possible to assign a fiber to a target and obtain a `good' DESI observation, which we fully define below. This definition is applied equally to data targets and random points, as described in \cite{KP3s15-Ross}. The randoms thus trace our footprint definition and, given that the input density of randoms is 2500 deg$^{-2}$ per random catalog, the area of the footprint can be trivially determined by counting the number of random points and dividing by 2500 deg$^{-2}$.
In what follows, we will present the total covered area in DR1 and then step through the area removed by each type of veto mask. The details for how these veto masks are applied in the LSS pipeline can be found in \cite{KP3s15-Ross}. Here, we describe the specific choices for the veto masks that are applied to the DR1 LSS catalogs.

\begin{figure*}
    \centering 
            \includegraphics[width=0.49\columnwidth]{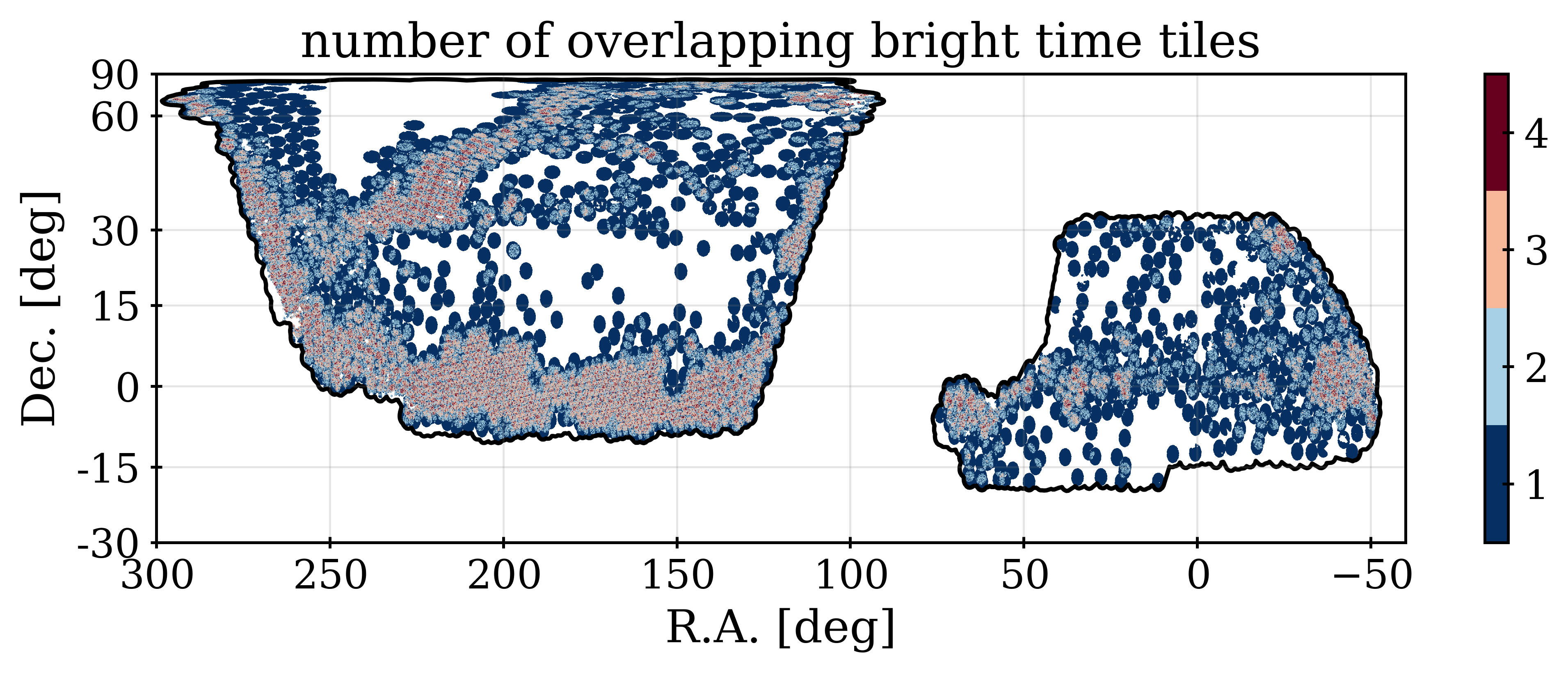}  
            \includegraphics[width=0.49\columnwidth]{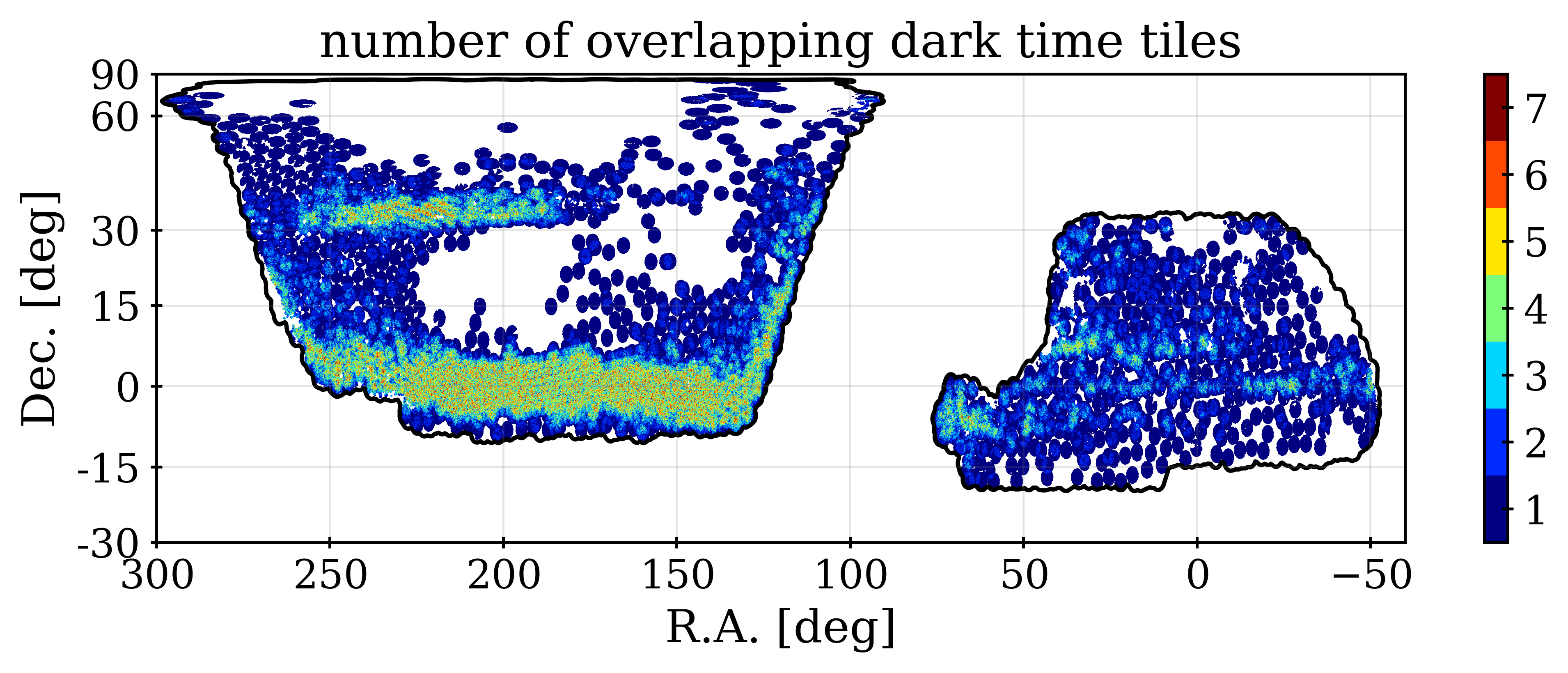} 
            \includegraphics[width=0.49\columnwidth]{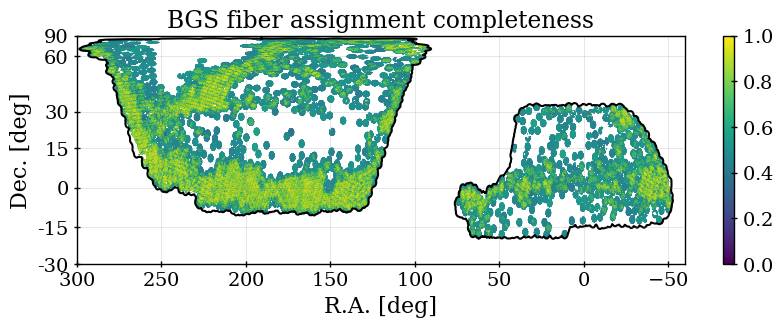}
            \includegraphics[width=0.49\columnwidth]{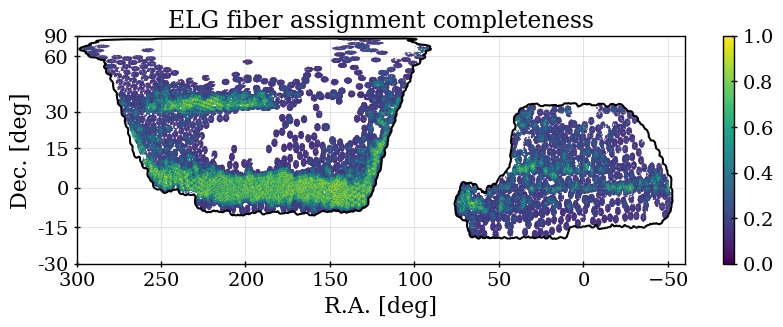}
    \includegraphics[width=0.49\columnwidth]{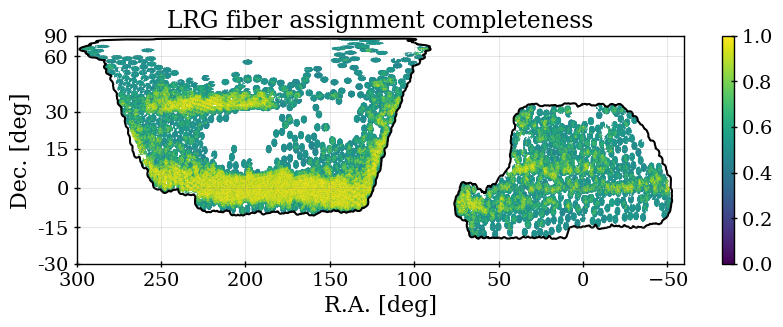}
    \includegraphics[width=0.49\columnwidth]{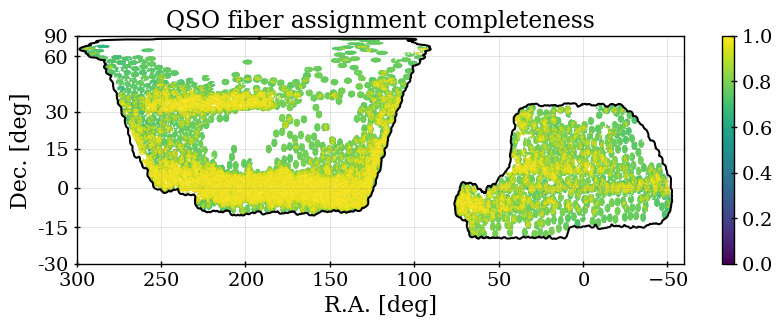}

    \caption{The top two panels show the number of overlapping tiles in DR1 bright and dark time. The lower four panels show the assignment completeness of the four DESI samples, within unique tile groupings. One can observe that the patterns in the completeness correspond to the number of overlapping tiles. The black outline shows the edges of the area within which DESI dark or bright tiles have been defined, as of May 17th, 2024. The completeness maps can be compared to the values in \cref{tab:statsvntile}.}
    \label{fig:allcomp}
\end{figure*}

\Cref{tab:Y1lss} includes details of the sky area for each tracer used in the DESI DR1 cosmological analyses. The footprint for these samples is shown in \cref{fig:allcomp}. The dark time tracers (LRG, ELG, and QSO) include the same tiles, and thus at a coarse level (at scales larger than an individual tile) their footprints are the same, as can be seen from the distribution of circular tiles in the figure. The differences in area \cref{tab:Y1lss} are thus only due to the differences in the veto masks at smaller scales, with the biggest effect coming from the priority veto, which is primarily due to QSOs. This veto removes more than 1300 deg$^2$ of the DR1 footprint, as can be seen by comparing the ELG and QSO areas. The LRG footprint is 187 deg$^2$ smaller than that of the ELGs because the bright star mask applied to LRGs removes more area than that applied to ELGs (see \cref{sec:imvetos}). Finally, the BGS sample was observed with a different set of tiles than the dark tracers, and thus has a different footprint (although from \cref{fig:allcomp} one can see that it is similar, by design \cite{surveyops}).

\begin{table*}
\centering
\begin{tabular}{|l|l|l|}
\hline%\hline
Mask/region &  Area [deg$^2$] & fraction of total   \\\hline
Total, dark time & 8,194.8 & 1\\
Total, bright time & 8,319.9 & 1\\
\hline
{\bf Regions (no vetos)} & &\\
\hline
DECam, dark time & 6574.8 & 0.802 \\
BASS/MzLS, dark time & 1620.0 & 0.198 \\
NGC, dark time & 5213.7 & 0.636 \\
SGC, dark time & 2981.1 & 0.364 \\
DES, dark time & 745.6 & 0.091 \\
DECam, bright time & 5760.5 & 0.692\\
BASS/MzLS, bright time & 2559.5 & 0.308\\
NGC, bright time & 5856.5 & 0.704 \\
SGC, bright time & 2463.5 & 0.296 \\
DES, bright time & 601.8 & 0.072 \\
\hline
{\bf Vetos} & & \\
\hline
Hardware, dark time & 254.2 & 0.031\\
Hardware, bright time & 193.3 & 0.023\\
Priority, LRG \& ELG & 1666.7 & 0.203\\
Priority, QSO & 38.5 & 0.005\\
Priority, BGS & 25.9 & 0.003\\
LRG imaging & 631.9 & 0.077 \\
QSO imaging & 488.4 & 0.060  \\
ELG imaging & 362.5 & 0.044 \\
BGS imaging & 373.0 & 0.045  \\
\hline
\end{tabular}
\label{tab:vetos}
\caption{The area in the DR1 footprint, covered by at least one DESI tile and determined by the number of random points (divided by their density of 2500 deg$^{-2}$), in different regions and within each kind of veto mask applied. Note that these veto masks can overlap and thus the total vetoed area is not the sum.}
\end{table*}

\subsection{Hardware Veto Masks}
\label{sec:hardware_veto}

All veto masks associated with individual components of the DESI instrument are grouped together to define `bad hardware' regions that are to be masked from the LSS catalogs. When producing the LSS catalogs with a unique entry per target (data and randoms), we prioritize the cases that were reachable by good hardware over the cases that were not, as detailed in \cite{KP3s15-Ross}. Data and randoms are flagged as bad if they can only be reached by a fiber defined as having bad hardware in the initial compilation.
 This minimizes the area lost to bad hardware (as such areas are likely to be recovered as good when a tile overlaps the area on a subsequent pass) and also allows us to determine the area lost. One can see in \cref{tab:vetos} that the area lost to bad hardware is only 2.3\% in bright time and 3.1\% in dark time.

\begin{figure*}
    \centering 
    \includegraphics[width=\columnwidth]{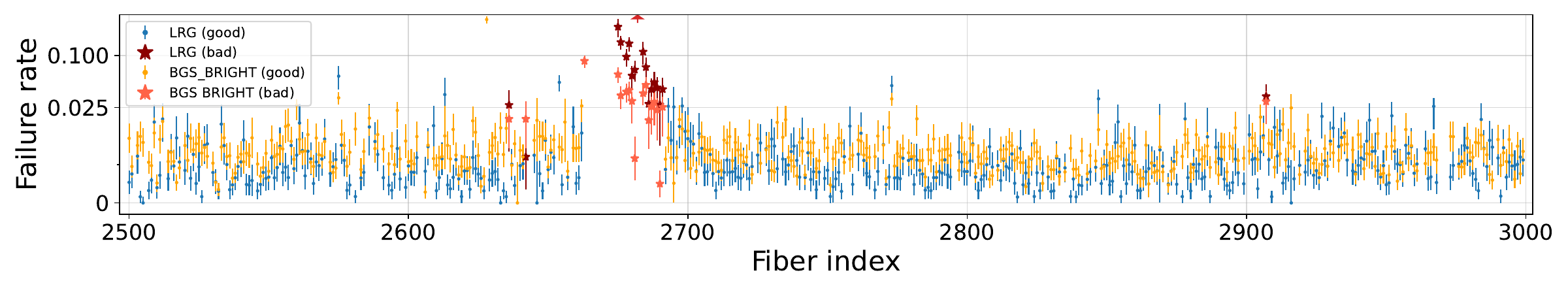}
    \caption{Failure rate for each fiber on petal 5 for LRG (blue) and BGS (orange; labeled \texttt{BGS\_BRIGHT} as that is the sample presented herein). Vetoed fibers, with a significantly higher failure rate, are colored dark red (LRG) or light red (BGS).}
    \label{fig:perfiber_success}
\end{figure*}

Three distinct sources of information define the DR1 hardware veto: 

\begin{itemize}
    \item The \texttt{ZWARN\_MTL} information compiled from the spectroscopic pipeline outputs (see \cref{sec:dataspec}) contains flags that indicate whether an observation passes the cuts to count as observed in the MTL. We apply the same definition as part of the hardware mask.\footnote{A difference between what we use and what was used for MTL decisions, however, is that we are using the information as determined during the \iron\ spectroscopic reductions, and the determination for the MTL is based on the `daily' version of the spectroscopic pipeline.}
    
    \item We require a minimum template signal-to-noise ratio, \texttt{TSNR2}. These values are determined for each tracer type and are proportional to the effective observing time. They are determined per coadded spectrum, but are independent of the target observed to produce the spectrum; they use a fixed template and the estimated noise. Each is defined in \cite{DESIpipe}. 
    In dark time, we apply a threshold $\texttt{TSNR2\_ELG} > 80$ for all samples, while in bright time, we apply a threshold $\texttt{TSNR2\_BGS} > 1000$. The spectroscopic success rates for spectra with \texttt{TSNR2} values below these thresholds decline dramatically and these cuts remove less than 1\% of the observed data. All tile and fiber combinations below these thresholds are marked as bad hardware. 

    \item We identify 60 poor-performing fibers, as defined in \cite{KP3s3-Krolewski}. All data from those fibers are flagged as bad hardware.
    \Cref{fig:perfiber_success} shows the LRG and BGS failure rate for each fiber on petal 5, highlighting this petal as it has the largest concentration of bad fibers, largely lying between \texttt{FIBER} 2675 and 2691. We describe the process used to identify bad fibers in \cite{KP3s3-Krolewski} and show failure rates for the other petals and tracers there.
\end{itemize}

\subsection{Priority Mask}
\label{sec:prioritymask}
To account for the areas on the sky where a given target type could not be observed, we apply a priority veto mask. Similar to the hardware veto, the priority veto is applied to the initial compilation of reachable data and random targets and is based on the metadata associated with the potential assignments; i.e., it is determined purely from the fiber assignment information. For both data and randoms, before cutting to unique objects, the priority of every target assigned on the given tile and fiber is known and stored in the catalogs as the column \texttt{PRIORITY\_ASSIGNED}. If the \texttt{PRIORITY\_ASSIGNED} is greater than that of the sample under consideration, the object's particular occurrence (associated with the given tile and fiber) is vetoed. Note that the given object can still be included (i.e., not ultimately vetoed) in the final catalog if it was a potential assignment on a different tile or fiber. For our dark time DESI DR1 samples, only QSO and rare high-priority strong lens candidates with priority 4000 cause priority vetoes. We choose not to have LRG targets cause priority vetos on ELGs, as 10\% of ELGs have the same priority as LRGs and there is significant overlap between the samples in redshift. For BGS, only white dwarf candidates have a higher priority (2998) than our BGS sample and cause a priority veto.

The priority mask for the LRG and ELG samples is caused by the same QSO and strong lens candidates and it is thus the same area of 1666.7 deg$^{2}$ for both, which is 20\% of the DR1 footprint. In the completed DESI survey, the impact of the priority mask will be much smaller, as the high-priority targets will already have been observed when a given area of the sky is revisited. One can compare coverage areas in \cref{tab:statsvntile} and observe that the area covered by dark time tracers matches to within 7\% in areas covered by more than one tile, e.g., the area of the LRG sample that is covered by two or more tiles is 3390.6 deg$^2$ and for QSO it is 3634.0 deg$^2$. 
For BGS, the white dwarfs remove 25.9 deg$^{2}$ and for QSO, the strong lens candidates remove 38.5 deg$^{2}$. Full details of the implementation of the priority mask in the LSS pipeline can be found in \cite{KP3s15-Ross}. 

An implicit assumption in the application of the priority mask is that the higher priority sample is not correlated with the sample being masked. This is not strictly true for QSO and LRG/ELG, as the samples overlap in redshift, though the angular correlations are strongly diluted due to the breadth of the QSO redshift distribution. This makes simulations that include all three tracers and are processed to produce LSS catalogs in the same way as for the real DESI data important. These are described in \cref{sec:mocks}.

\subsection{Imaging Veto Masks}
\label{sec:imvetos}
We apply two types of veto masks that are related to imaging conditions. One set, which we denote `bright object', masks areas around bright stars, galaxies, and defects in the imaging. The bright object masks are different for each tracer type. The other, which we denote as `property', masks data in regions with bad imaging conditions and we use the mask for all tracers. We provide the details for each below.

\subsubsection{Bright Object Masks}

All tracers had masks defined by the maskbits in the Legacy Survey DR9 imaging\footnote{\url{https://www.Legacy Survey.org/dr9/bitmasks/}} applied to their targeting. For dark time tracers (QSO, LRG, ELG), these are the bright star mask (bit 1) set for Tycho MagVT $< 13$ or Gaia $G<13$, the bright galaxy mask (bit 12), and the globular cluster mask (bit 13). For bright time (BGS), we do not apply the bright galaxy mask, as this removes many real low redshift galaxy targets. The redshifts of the galaxies that define the bright galaxy mask are predominately $z<0.15$ and thus do not share large-scale structure with any of the dark time tracers, which have all have $z>0.4$ or higher. The presence of the bright galaxies makes the photometry in their vicinity unreliable, and we thus mask the area from the dark time LSS catalogs. Further, the area must have been covered by at least one exposure in all of the $g,r,z$ bands. These masks are applied to the corresponding randoms (and any DR1 simulations), prior to determining their potential assignments.\footnote{They were already applied to the DESI target samples and thus do not need to be reapplied to the data.} 

\begin{figure*}
    \centering 
    \includegraphics[width=0.49\columnwidth]{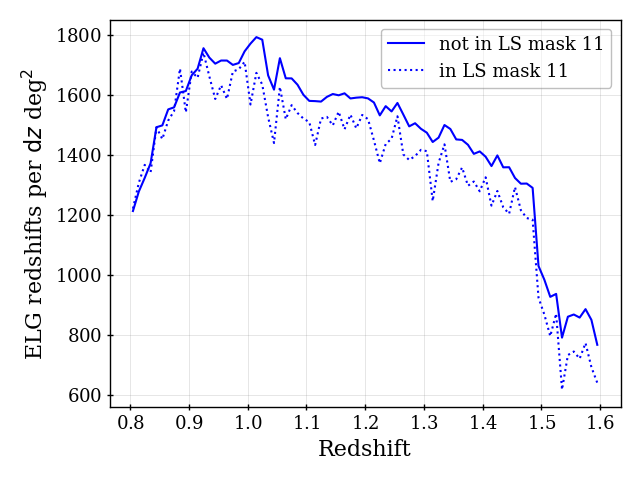}
    \includegraphics[width=0.49\columnwidth]{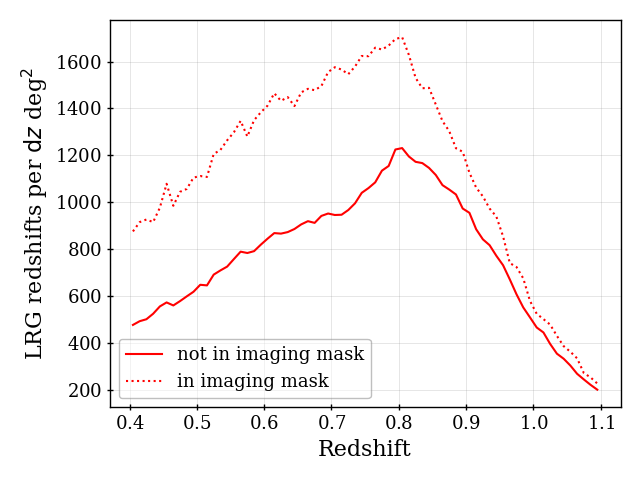}
    \includegraphics[width=0.49\columnwidth]{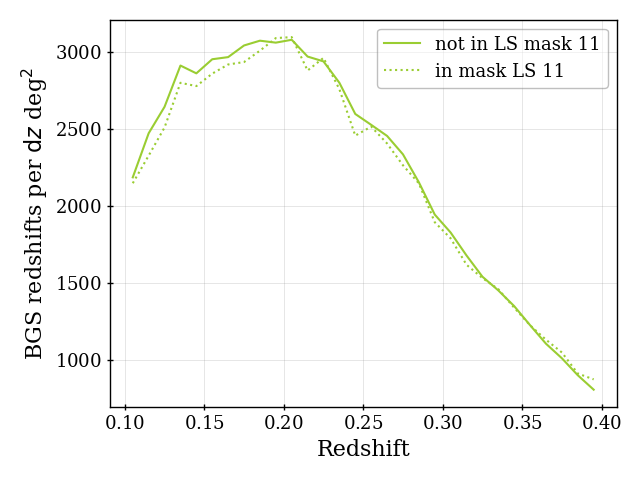}
    \includegraphics[width=0.49\columnwidth]{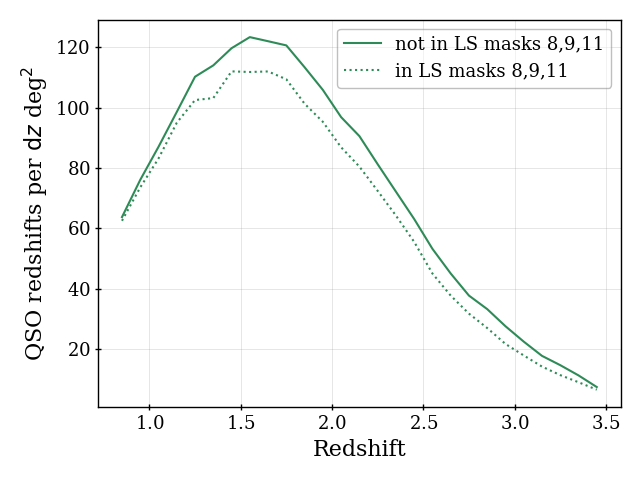}
   \caption{The projected, completeness-corrected, density of each DESI LSS tracer in DR1 inside and outside of the imaging veto masks we apply. The plotted values are determined by counting the redshifts in each area and redshift bin and dividing by the fiber assignment completeness, the area in square degrees, and the width of the redshift bin. `LS' denotes Legacy Survey, which defined a series of maskbit values. The difference between the dotted and solid curves illustrates the need to apply these masks, with the LRG case being the most extreme.}
    \label{fig:allnzveto}
\end{figure*}

We apply additional bright-object veto masks to the DR1 LSS catalogs masks that were not applied to the DESI target samples, with details that depend on the tracer, as follows:
\begin{itemize}
    \item For all dark-time tracers (QSO, ELG, LRG), we apply the custom mask described at the end of Appendix D in \cite{LRGtarget} that removes less than 0.01\% of the footprint. 

    \item For ELG, we apply an additional custom mask, which eliminates areas at the location of, e.g., Milky Way dwarf satellites and imaging ghosts that result in significant excesses of ELG targets. This custom mask is defined in \cite{TanveersELGpaper} and removes 0.1\% of the ELG footprint.

    \item For BGS and ELG, we apply the Legacy Survey {\texttt{MEDIUM}} star mask (bit 11) that masks area around bright stars based on the Gaia magnitude, up to $G<16$. The comparison of the $n(z)$ for ELG data inside and outside of this masked region is shown in the top left panel of \cref{fig:allnzveto}. The density of ELG data inside of the mask is 10\% to 20\% lower depending on the redshift. We make the simple choice to discard the 4.5\% of the footprint within the mask, though given the moderate effect, one could imagine modeling the ELG selection function within this region in the future. The bottom left panel of \cref{fig:allnzveto} compares the \texttt{BGS\_BRIGHT} data inside and outside of the imaging veto mask we apply, where we find only a small ($<10\%$) effect; while the effect is small, we still opt for the conservative choice to remove the 5.2\% of footprint data within the mask for the Y1 analyses.

    \item For QSO, we apply three additional maskbits from Legacy Survey, on top of the three applied to targeting. These are the bright star masks for WISE W1 and W2 (bits 8 and 9) and the same {\texttt{MEDIUM}} star mask applied to the BGS sample. The $n(z)$ for the QSO data outside of and inside of this masked region is shown in the bottom right panel of \cref{fig:allnzveto}; the results are corrected for completeness. One can see that the density is significantly lower, by 10\% to 20\% depending on redshift, within the masked region. (Splitting the data into the BASS/MzLS and DECam photometric regions does not affect the comparison.) Further, we find that the spectroscopic success is significantly worse for the data within the masked region: 53\% compared to 67\%. Thus, we apply these masks to the Y1 QSO data, which removes 6.3\% of the footprint. However, given the size of the differences, one can imagine that future work properly determines the selection function for these masked data and includes them in future DESI analyses.

    \item For LRG, we also apply masks for WISE and Gaia bright stars, but they are constructed as described in \cite{LRGtarget} and remove more area than the Legacy Survey bits 8, 9, and 11. The mask is particularly important for LRG, as considerably more LRG are targeted within these masks, with almost all having good redshifts; i.e., the photometry is affected within these regions in a way that produces a much denser sample than outside of them. This increased sample size can be observed in the upper right panel of \cref{fig:allnzveto}. The size of the masks determined by \cite{LRGtarget} keep the area where the LRG selection function can be determined with the methods described throughout the rest of this paper. Given that the data within the LRG mask yields such a surplus of galaxies with good redshifts, one could imagine that in future releases we could determine methods to down-select to the sample of galaxies statistically matching the intended DESI LRG sample.

\end{itemize}

Similar to LRG, ELG targets have an excess density around bright stars, beyond what would be removed by the {\texttt{MEDIUM}} mask. An extended mask was thus defined and applied in \cite{TanveersELGpaper}. However, unlike LRGs, the excess targets are not found to have redshifts within the $0.8<z<1.6$ redshift range used for DR1 ELG clustering analyses. That is, after applying completeness weights, the $n(z)$ of DESI ELGs within the extended mask region is consistent with the $n(z)$ of ELGs in the rest of the footprint. We therefore do not apply the extended mask to the DR1 ELG sample. 

\subsubsection{Masks for Imaging Properties}
\label{sec:improp_veto}
We remove portions of the footprint in the tails of the distribution of imaging conditions containing the worst data, as traced by the \textsc{Healpix} maps we use for regressions to correct for imaging systematics (described in \cref{sec:imsys}). The full details of the cuts applied to all samples and how much of the QSO footprint they remove are given in \cref{tab:propveto}.\footnote{The fractions of the footprint removed are very similar for all tracers, with slight differences due to each tracer having its own set of veto masks.} The main purpose of these cuts is to remove the small amount of data that exists as outliers in the image property space that would have the potential to affect the performance of the regressions unduly. The choice of $E(B-V)_{\rm SFD}<0.15$ was motivated in part because it matches previous SDSS analysis choices \citep{bosslsscat,ebosslss}. In total, the \textsc{Healpix} based mask removes 3.4\% of both of the QSO and BGS footprints and 3.2\% of both of the ELG and LRG footprints. 

\begin{table*}
\centering
\begin{tabular}{|l|cccc|}
\hline%\hline
Map & cut  & frac. removed N & frac. removed S & frac. removed total  \\\hline
E(B-V)$_{\rm SFD}$ & $<0.15$ mag & 0.003 & 0.019 & 0.016\\
Gaia star density & $<10^{4.4}$ & $<0.001$ & 0.006 & 0.005\\
PSF$_G$ & $<2^{\prime\prime}.4$ & 0.018 & $<0.001$ & 0.004\\
PSF$_R$ & $<2^{\prime\prime}.3$ & 0.009 & $<0.001$ & 0.002\\
PSF$_Z$ & $<2^{\prime\prime}$ & $<0.001$ & $<0.001$ & $<0.001$\\
GALDEPTH$_G$ & $>250$~nmag & 0.006 & $<0.001$ & 0.001\\
GALDEPTH$_R$ & $>80$~nmag & 0.006 & $<0.001$ & 0.001\\
GALDEPTH$_Z$ & $>30$~nmag & $<0.001$ & 0.009 & 0.007\\
PSFDEPTH$_{W1}$ & $>2$~nmag & $<0.001$ & $<0.001$ & $<0.001$\\
all & - & 0.041 & 0.032 & 0.034\\
\hline
\end{tabular}
\caption{\label{tab:propveto} The amount of area removed from the QSO footprint based on the denoted cuts on imaging property maps, as described in \cref{sec:improp_veto}. Columns show the fractions removed in the N and S imaging areas separately as well as the total. The fractional area removed from the footprints for other tracers is similar.}
\end{table*}

\section{Fiber Assignment Completeness}
\label{sec:fiberassigncomp}

The fiber assignment completeness for any arbitrary selection of targets is simply the number of those targets assigned to a fiber, divided by the total number of those targets. We denote this as $C_{\rm assign}$.
This, determined over the full DR1 footprint for each tracer, is listed in \cref{tab:Y1lss}. For the dark time tracers, the relative completeness is determined by the relative assignment priorities. Maps of the completeness for each of our DR1 DESI samples are shown in \cref{fig:allcomp}. One can see that for the three dark time tracers, the pattern is the same, but is most pronounced for the ELG sample. The SGC region has less of its footprint covered to high completeness than the NGC, which explains the difference in $n(z)$ determined without completeness corrections, shown in the right-hand panel of \cref{fig:nzall}.  

The completeness is almost entirely a function of the number of overlapping tiles, $n_{\rm tile}$, at any given location. This can be observed by comparing the completeness patterns in \cref{fig:allcomp} to the maps of $n_{\rm tile}$ for dark and bright time in the same figure.  We determine $n_{\rm tile}$ for all data and random targets within the DR1 area by counting the number of times the target was reachable, after applying the hardware veto described in \cref{sec:hardware_veto}. The priority veto is not considered and thus $n_{\rm tile}$ at any given celestial coordinate is the same for all dark time tracers. Full details of the $n_{\rm tile}$ calculation are provided in \cite{KP3s15-Ross}.

\cref{tab:statsvntile} provides completeness, area, and number of observed redshifts as a function of $n_{\rm tile}$ for each of our four tracers. From the numbers in the table, one can determine that for BGS, over half of the DR1 area is in regions with single tile coverage ($(7472.7-3382.5)=4090.2$ deg$^2$ compared to 3382.5 deg$^2$),
and for QSO, it is almost exactly half (3634.0/7249.1 = 0.501). By comparing the QSO and ELG areas, one can see that the QSO priority veto removes more than 1/3 of the footprint in areas covered by only 1 tile but only removes a few percent of the footprint in areas covered by 2 or more tiles. While the completeness statistics improve dramatically as the coverage increases, any minimum cut on $n_{\rm tile}$ removes a large fraction of the data. The effect is smallest for ELGs, but would still remove at least 1/6th of the redshifts from the sample. Thus, our fiducial choice for the DR1 LSS catalogs is not to apply any $n_{\rm tile}$ (or any other completeness) threshold.

\begin{table}
\resizebox{\columnwidth}{!}{
    \begin{tabular}{|l|r|r|r|r|r|r|r|}
    \hline
     & $n_{\rm tile}\geq 1$ & $n_{\rm tile}\geq 2$  & $n_{\rm tile}\geq 3$ & $n_{\rm tile}\geq 4$ & $n_{\rm tile}\geq 5$ & $n_{\rm tile}\geq 6$ & $n_{\rm tile}\geq 7$\\
    \hline
    \rule{0pt}{3ex} 
    {\bf BGS} & & & & & & &\\
    \hline
    \% $C_{\rm assign}$ & 63.6 & 81.5 & 90.5 & 95.7 & - & - &- \\
     Area [deg$^2$] & 7472.7 & 3382.5 & 1268.5 &  233.9 & - & - &- \\ 
    $N$ $0.1<z<0.4$  & 300,043 & 177,399 & 75,427 & 14,608 & - & - &- \\
    \hline
    \rule{0pt}{3ex} 
    {\bf LRG} & & & & & & &\\
    \hline
    \% $C_{\rm assign}$ & 69.3 & 81.8 & 90.5 & 94.3 & 96.6 & 98.2 & 99.2 \\
     Area [deg$^2$] & 5739.7 & 3390.6 & 1908.7 &  1087.7 & 506.3 & 146.0 & 16.7 \\ 
    $N$  $0.4<z<1.1$  & 2,138,627 & 1,502,311 & 938,237 & 556,506 & 264,730 & 77,483 & 9,040 \\
    \hline
    \rule{0pt}{3ex} 
    {\bf ELG} & & & & & & &\\
    \hline
    \% $C_{\rm assign}$ & 35.2 & 48.0 & 61.6 & 71.5 & 79.5 & 86.6 & 92.2 \\
     Area [deg$^2$] & 5924.0 & 3500.3 & 1969.1 &  1120.6 & 520.9 & 150.0 & 17.1 \\ 
    $N$  $0.8<z<1.6$  & 2,432,072 & 1,985,319 & 1,460,897 & 977,272 & 508,848 & 160,646 & 19,545 \\
    \hline
    \rule{0pt}{3ex} 
    {\bf QSO} & & & & & & &\\
    \hline
    \% $C_{\rm assign}$ & 87.4 & 97.5 & 98.9 & 99.2 & 99.5 & 99.7 & 99.8 \\
     Area [deg$^2$] & 7249.1 & 3634.0 & 1980.6 &  1117.7 & 516.7 & 148.5 & 17.0 \\ 
    $N$ $0.8<z<3.5$  & 1,223,391 & 682,903 & 377,235 & 214,073 & 98,901 & 28,489 & 3,219 \\
    \hline
    \end{tabular}
}
    \caption{The assignment completeness ($C_{\rm assign}$) percentage, area in square degrees, and number ($N$) of observed redshifts as a function of the number of overlapping tiles ($n_{\rm tile}$) for each of our 4 DESI DR1 tracers.  \label{tab:statsvntile}}
\end{table}

In the following three subsections, we first discuss completeness calculations determined for different resolutions, we then compare to assignment probabilities determined from repeated realizations of the DR1 fiber assignment, and we finally describe how we use the calculations to determine completeness corrections in the DR1 LSS catalogs.

\subsection{Completeness Definitions}
\label{sec:compdef}
For the DESI DR1 cosmological analyses, our fiducial approach to fiber assignment incompleteness is to divide it into two components, in a way that mimics the SDSS approach \cite{bosslsscat,ebosslss}. The full details of the calculations are provided in \cite{KP3s15-Ross} and we repeat the basic definitions and concepts here.

The first component is analogous to the SDSS `close-pair' weights. Recall that the full catalogs are split by target type and cut from the potential assignments catalog to unique \texttt{TARGETID}, after a careful sorting (described in \cite{KP3s15-Ross}) that puts each object at the most relevant combination of tile and fiber. Thus, every target (observed or not) in the full LSS catalog is associated with a single combination of tile and fiber. By definition, the unobserved targets are at combinations of tile and fiber assigned to a different observed target. For unobserved targets, reducing the potential assignments catalogs\footnote{Again, this is fully detailed in \cite{KP3s15-Ross}.} to unique \texttt{TARGETID} prioritizes the instances of tile and fiber assigned to the given type. The result is that most unobserved targets in the full LSS catalogs are at combinations of tile and fiber that were used to observe the given target type.
Every observed DESI target in each respective full catalog is given a completeness, $f_{\rm TLID}$, that is simply the inverse of the total number of unique DESI targets within the catalog at the given tile and fiber. This number is essentially the number of targets that were competing for the fiber.

The calculation of $f_{\rm TLID}$ does not account for all fiber assignment incompleteness. Some fraction of (unobserved) targets in the full catalogs will be at a tile and fiber that did not observe any target of the given type. These data thus do not influence any $f_{\rm TLID}$ calculations. 
These cases occur due to, e.g., the fiber needing to be assigned to a standard star or sky fiber to meet the minimum threshold. In the case of ELGs, it will also be due to an LRG being assigned to the combination of tile and fiber associated with the given ELG target. In such cases, the target that ultimately received the observation should only depend on randomized processes within the DESI targeting, such as the subpriority value of the target, or---for ELGs competing with LRGs---whether or not it was one of the 10\% that were boosted to the LRG priority. We therefore expect such completeness effects to be distributed equally within any given set of overlapping tiles, which we denote $t_{\rm group}$. The $t_{\rm group}$ associated with a given point on the sky is the set of tiles that had a DESI fiber postioner included in the good hardware definition that could have reached the point. It can thus be determined for the data and random catalogs based on set of tiles for each \texttt{TARGETID} that are in the respective potential assignments catalog, after applying the bad hardware veto. We thus determine a completeness, $f_{\rm tile}$, per $t_{\rm group}$ that treats the targets that influenced the $f_{\rm TLID}$ calculation as if they were observed. The groups of overlapping tiles are analogous to SDSS `sectors' and $f_{\rm tile}$ is thus analogous to $C_{\rm BOSS}$. Please see section 4.3 of \cite{KP3s15-Ross} for the full details.

\subsection{Realizations of Fiber Assignment}
\label{sec:altmtl}
A separate way to evaluate DESI assignment completeness, and to enable more than point estimates of it, is through the production of alternative realizations of assignment histories. Such realizations are produced by changing the random seed that determines quantities such as the subpriority or whether an ELG is given the same priority as an LRG. Such changing of the random seed happens when creating the initial MTL and the impact on the assignment history can be simulated by running the \textsc{fiberassign} software using the same settings and updating the `alternative' MTL in the same order as for DR1 observations. This process is described and validated in \cite{KP3s7-Lasker}; we denote it as the `altmtl' process. A total of 128 altmtl realizations were produced for DR1 LSS catalogs. We apply the same hardware veto as described in \cref{sec:hardware_veto} to the assignments determined for each realization and store the binary \texttt{True}/\texttt{False} information on whether each target was assigned for each realization within a bit array,\footnote{In practice, we store the results from the 128 realizations via two 64 bit arrays.} which we store in the column \texttt{BITWEIGHTS}. Counting DR1 observations, we have 129 total realizations and the probability of assignment for any target is simply
\begin{equation}
p_{\rm obs} = \frac{N_{\rm assign}}{129}\,,
\end{equation}
where $N_{\rm assign}$ is the number of realisations in which the target is assigned.

For observed targets, we expect $p_{\rm obs}$ to be a close match to $f_{\rm TLID}f_{\rm tile}$. Detailed comparisons of the assignment completeness determined from the altmtl realizations and $f_{\rm TLID}f_{\rm tile}$ are presented in \cite{KP3s6-Bianchi}. We describe the DESI DR1 LSS catalogs that use the altmtl data for completeness corrections in \cref{sec:pipcat}. We recommend using the altmtl version of the LSS catalogs for small-scale clustering measurements.

\subsection{Weights for Completeness }
\label{sec:comp}

To correct for the variations in completeness in the DR1 2-point functions, we produce weights based on the completeness definitions described in the previous subsection. For our fiducial LSS catalogs, we simply use 
\begin{equation}
w_{\rm comp} = 1/f_{\rm TLID}.
\label{eq:wcomp}
\end{equation}
This is analogous to the SDSS close pair weight, as described in \cref{sec:compdef}. 

The $w_{\rm comp}$ defined in \cref{eq:wcomp} does not account for the incompleteness that we have tracked via the $f_{\rm tile}$ determination. Given that we identify the same tile groupings in data and random samples, we choose to apply $f_{\rm tile}$ as a weight to the randoms. A small number of tile groupings are present only in randoms, i.e., there were no reachable targets within some tile groupings for a particular tracer. Such tile groupings are typically very small regions with many overlapping tiles. We assigned these areas $f_{\rm tile}=1$. The precise implementation is detailed further in \cref{sec:weights}.

Our fiducial method for including completeness weights in the DR1 LSS catalogs does not produce unbiased 2-point clustering statistics, because it does not account for the fact that the number of close pairs observed at small angular scales is highly incomplete, because of the physical limit on the minimum separation of neighbouring fibers in a single tile. To account for this, \cite{KP3s5-Pinon} develop a ``$\theta$-cut" method to remove small angular separations from 2-point measurements in both configuration- and Fourier-space and include the impact of removing such information into a window matrix, to be convolved with the theoretical model. We remove pairs at angular separation scales less than 0.05 degrees, as spectra of targets at separations greater than this scale can be measured simultaneously, given the distance between fiber positioners. The $\theta$-cut method is the default choice in the DESI DR1 analyses to remove any biases in derived parameters due to fiber assignment incompleteness when using large-scale 2-point clustering measurements. Residual uncertainties related to the process are studied in \cite{DESI2024.V.KP5}, by comparing results obtained from realistic DR1 simulations (the `altmtl' mocks described in \cref{sec:mockfa}) to the results of simulations without any fiber assignment incompleteness.

Alternatively, one can use the bit arrays determined from the alternative fiber assignment realizations described in the previous subsection to compute `pairwise inverse probability' (PIP) weights which can be used to obtain unbiased 2-point clustering measurements, as described in \cite{2017Bianchi}. This is not our fiducial method in the DR1 analysis for two main reasons. One is simply that the altmtl process required to obtain the bit arrays takes significant computing time\footnote{Currently, it takes several days on a NERSC Perlmutter CPU node with 128 physical cores to obtain 128 realizations, with the time dominated by the I/O feedback loop that must occur in the proper order.} and it is not feasible to run this number of realizations on each of a large number of separate simulations. For instance, in the DESI 2024 cosmological analyses, 25 simulations of DESI DR1 were used for validation and 1000 were used to help determine covariance matrices (these are described in \cref{sec:mocks}).
 The other is that PIP weights alone cannot correct for cases where there are 0 probability pairs. There are many such pairs in regions that have been covered by only one tile, which is a significant fraction of the DR1 footprint. For any group of targets that is reachable only to one single combination of tile and fiber, the probability of observing any pairs within the group is 0. The effect of the 0 probability pairs can be corrected via angular upweighting, but the results are no longer strictly unbiased. Any angular upweighting must rely on the angular clustering of the full target sample. The relative amount of area that is covered by only 1 pass will decrease as the DESI survey is completed, and the impact of 0 probability pairs will thus decrease. The fiducial methods to correct for fiber assignment incompleteness will be re-evaluated with each data release. Characterization of fiber assignment incompleteness issues in DR1, for all DESI tracers, is detailed further in \cite{KP3s6-Bianchi}. 

While it is not our fiducial choice for DESI DR1 analysis, the use of PIP weights and angular up-weighting is our only option for accurately measuring small-scale clustering. The PIP weights will account for both the $f_{\rm TLID}$ and $f_{\rm tile}$ factors. Thus, we must not weight the randoms by $f_{\rm tile}$ when obtaining PIP-weighted clustering measurements. We discuss this further in \cref{sec:weights} and \cref{sec:pipcat}.

\section{Treatment for Imaging Systematics}
\label{sec:imsys}

Our fiducial approach for mitigating the effect of imaging systematics in the clustering of DESI DR1 is to seek the minimal set of image property maps to use, paired with the simplest regression method, that allows us to reduce trends between the density of our LSS catalogs projected onto the celestial sphere (`projected density') and the full set of image property maps used for validation to a level consistent with those observed in simulations.
The complexity of how mode-removal effects are introduced by this procedure (alternatively understood as over-fitting or noise biases) bias our estimated 2-point functions increases with both the number of maps used and the complexity
of the regression method applied. 

In early versions of the catalogs, we took a different approach, where we applied the non-linear \textsc{SYSNet} neural net (NN) \cite{Rezaie23} and \textsc{Regressis} random forest (RF) \cite{Chaussidon21QSOsys} regressions to all of the maps potentially relevant for a given tracer. When doing so and applying the process to simulations with no systematic contamination, we found that the mode removal effects on the LRG and BGS tracers were greater than the estimated removal of systematic contamination; i.e., the mitigation method imparted more bias in the 2-point functions than it removed. After learning this, we adopted a procedure where we first tested the performance of the linear regression method used for the final eBOSS catalogs \cite{BautistaDR14LRG,ebosslss} (with the code integrated into the DESI framework) and only used one of the non-linear methods if it was determined to be necessary based on the null tests described below.

We perform the regressions for each tracer using the following redshift bins and regression techniques:
\begin{itemize}
    \item BGS $0.1<z<0.4$, linear
    \item LRG $0.4<z<0.6$, $0.6<z<0.8$, $0.8<z<1.1$, linear
    \item ELG $0.8<z<1.1$, $1.1<z<1.6$, \textsc{SYSNet}
    \item QSO $0.8<z<1.3$, $1.3<z<2.1$, $2.1<z<3.5$ \textsc{Regressis}
    
\end{itemize}
For all samples but QSO these redshift bins are the same as those used for clustering measurements.
In all cases, the regressions determine a model for how the observed density varies with the imaging properties included in the regression model. The maps of imaging properties that we use have \textsc{Healpix} resolution Nside=256. The inverse of the model determined in each \textsc{Healpix} pixel and for the particular redshift bin, is added as a weight $w_{\rm imsys}$ to the LSS catalogs and recommended for use in all subsequent calculations.

For all tracers, we performed null tests where we determined the normalized number density versus the value of the imaging property (with the potential to cause systematic variation), in 10 evenly spaced bins of imaging property, both with and without using the determined imaging systematic weight $w_{\rm imsys}$ in the calculation. The counts always include completeness weights. To enable comparisons while the `clustering' catalogs (see \cref{sec:weights}) remained blinded, the full catalogs (with all vetos applied) were used. A weight, $w_{\rm FKP, 2D}$, similar to the FKP weight added to the clustering catalogs as described in \cref{sec:FKP} was calculated and applied to the counts. Instead of allowing the number density to evolve with redshift, for simplicity, we chose a constant number density, $n_0$ for each tracer, with values $5\times10^{-4}$($h$/Mpc)$^3$ for BGS and ELG, $4\times 10^{-4}$($h$/Mpc)$^3$ for LRG, and $2\times 10^{-5}$($h$/Mpc)$^3$ for QSO. We then used
\begin{equation}
    w_{\rm FKP, 2D} = 1/[1+n_0 \langle C_{\rm assign}\rangle(n_{\rm tile})P_0],
\end{equation}
where $\langle C_{\rm assign}\rangle(n_{\rm tile})$ is the mean completeness at a given (discrete) number of overlapping tiles. 
Similarly, we refactored the completeness weights to depend on $n_{\rm tile}$ (as done in \cref{sec:FKP}).
 In each bin, we then estimated the uncertainty based on the Poisson error determined from the weighted counts\footnote{When doing so, we erroneously divided by the mean completeness weight for the full sample, as determined before the refactoring. This will make the uncertainties under-estimated, but was consistently applied to the data and mocks and thus does not impact the null tests that depended on the comparison of data and mock results.}. The approach of using FKP-weighted counts in the uncertainty calculation has been shown to provide approximately correct uncertainties \citep{RossDR12}. Since these are only approximately correct, we applied the same methodology to mocks and compared the recovered $\chi^2_{\rm null}$ statistics---where the null expectation is a constant---from the data to the distribution of values from 25 mocks from the DESI DR1 AbacusSummit suite with realistic fiber assignment applied (these are the `altmtl' mocks described in \cref{sec:mocks}). In all cases, the regressions have been run on the mocks using the same maps and settings as for the DESI data and $w_{\rm imsys}$ obtained and applied to the calculations. However, unlike the data, the mocks have no systematic contamination. 

The validation null tests were always performed against the full set of maps that were considered potentially relevant (even those not included in the regression). When blinded, we classified the null tests as passed if the sum of the $\chi^2_{\rm null}$ for the data across all tested maps was less than that from at least one of the 25 mocks and each of the individual maps tested had a $\chi^2_{\rm null}$ less than at least two of the 25 mocks. \cref{fig:bgs_imsys_mocks} shows examples of this test for the BGS sample. Cases where these two criteria were not met were investigated further. Statistically, given the number of maps tested, one would expect a small number of cases that do not pass. We investigated the severity of each failure and, e.g., in cases where the map was not originally included to be regressed against, we tested whether adding it significantly improved the results. 
This determined the final choices of the maps and the regression methods to be applied to the catalogs, which were fixed based on tests on the blinded data and never changed on the unblinded data.  
The tests on the blinded data were applied to an earlier iteration of the DR1 mocks and the results of the null tests change slightly when the catalog versions are updated. In the subsections that follow, we present the results obtained with the final data and mock versions and pay particular attention to results that are outside of the expectations provided by the mock results and in some cases fail the above criteria.

The following maps are always included in the null tests presented for each tracer in the subsections that follow: 
\begin{itemize}
    \item The stellar density (deg$^{-2}$) as determined from Gaia stars \cite{GaiaDR2} with $12<G<17$;  we label it as \texttt{STARDENS}.
    \item Galactic extinction in $E(B-V)$ magnitudes, as estimated in \cite{chiang_csfd} by removing cosmic infrared background contamination from the widely applied `SFD' dust map of \cite{schlegel1998maps}; we label it as \texttt{EBVnoCIB}.
    \item The HI column density from \cite{HI.Bekhti}, which studies \cite{lenz2017new} have shown correlates with Galactic extinction but not with extra-Galactic sources; we label it as \texttt{HI}.
    \item The imaging depths and PSF sizes in the $g$, $r$, $z$ bands (as determined by the DR9 Legacy Survey). We label these as \texttt{DEPTH\_<band>} and \texttt{PSF\_<band>}.\footnote{See the definitions at \url{https://www.Legacy Survey.org/dr9/files/\#randoms-1-fits}. We use the galaxy depths for all tracers except the QSO, for which we use the PSF depths.}
    \item The difference between the SFD $E(B-V)$ (applied to DESI targeting) and the $E(B-V)$  determined by \cite{KP3s14-Zhou} using spectra of DESI stars Legacy Survey photometry\footnote{Note that for the derivation of DESI $E(B-V)$, instead of using different extinction coefficients for the BASS/MzLS and DECam regions as in \cite{KP3s14-Zhou}, here we use the DECam coefficients for BASS/MzLS to be consistent with the extinction correction in DESI target selection.}. Assuming the map determined from DESI stars is truth, any trend is the result of applying an incorrect Galactic extinction correction to DESI targeting. This produced two separate maps, one based on $g-r$ and the other $r-z$. Here, we report the results obtained from $g-r$, which produces the less noisy map, and label it $\Delta$\texttt{EBV GR}. 
\end{itemize}
We describe how these maps are created in \cref{sec:improp_maps}. For LRGs, we also add the depth in $W1$, while for QSO we add the depth in $W1$ and $W2$. When we regress against the imaging depth, we always apply the nominal Galactic extinction corrections to the depth maps, but in the null tests presented throughout this section, we do not. This was an arbitrary choice as we expect to pass the null test either way, assuming systematic trends have been removed. In the following subsections, we describe the regressions performed and the results per tracer.

\subsection{BGS}
\begin{figure}
    \centering 
    \includegraphics[width=\columnwidth]{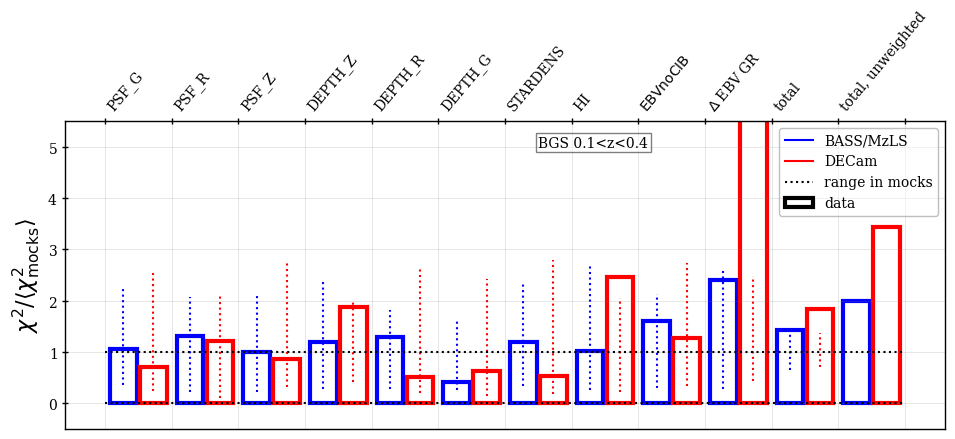}
    \includegraphics[width=\columnwidth]{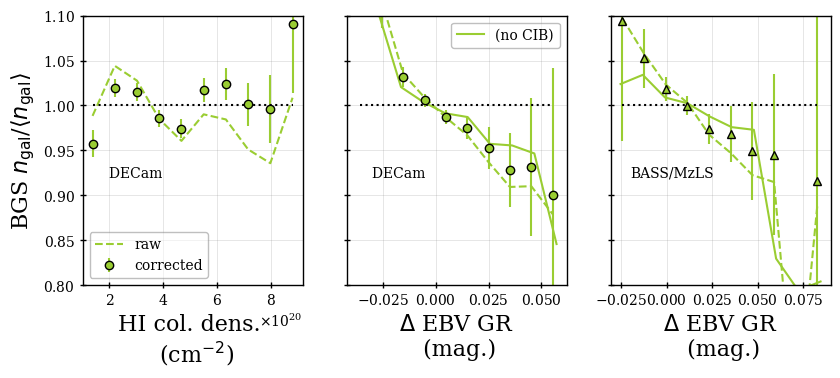}
    \caption{In the top panel boxes show $\chi^2_{\rm null}$ values for the null test for BGS data, determined for each of the image property maps individually and the total across maps, in units of the mean $\chi^2_{\rm null}$ values for the same tests evaluated on the 25 null, i.e., uncontaminated mocks. Also shown are the totals obtained from the unweighted data. Results for the BASS/MzLS and DECam regions are shown separately in blue and red respectively. Dashed lines indicate the range of values obtained from the 25 mocks. The bottom panel shows the three cases where the null tests against imaging systematics are worst, in comparison to the expectation from mock analysis. The dashed curves display the results when no weights for imaging systematics were applied to the data, while the points with error bars display the cases where they were. The points thus display the residual trends in the final catalogs. The results plotted with a solid curve replace the SFD $E(B-V)$ map with the $E(B-V)_{\rm no CIB}$ when determining $\Delta$\texttt{EBV GR}.
    }
    \label{fig:bgs_imsys_mocks}
\end{figure}

For our BGS sample, we use only three maps and the linear method for the regression to determine the imaging systematic weights. The three maps used are the $r$-band depth, the stellar density, and the \texttt{HI} column density.

The top panel of \cref{fig:bgs_imsys_mocks} summarizes the imaging systematic validation tests for the BGS sample. As for \cref{fig:lrg_imsys_mocks}, we show the $\chi^2_{\rm null}$ obtained from the data relative to the mean of that from the mocks, displaying the results for the data with bars and the range of values obtained from the individual mocks with dotted lines. The results are shown for the individual maps used for the BGS null test, all with the fiducial imaging systematic weights applied. If the three above-listed maps are sufficient, we expect satisfactory results when testing all of the maps.
We also show the results summed over all maps without (last column) and with (second to last column) the imaging systematic weights. One can observe that the improvement from using the weights is close to a factor of 2 for DECam and more moderate for BASS/MzLS.

For particular maps, three cases are highlighted. One is the HI column density in the DECam region, which is included as one of the three in the regressions. The trend against HI before and after applying weights can be observed in the bottom left panel of \cref{fig:bgs_imsys_mocks}: the $\chi^2_{\rm null}$ improves from 51.3 to 26.8. 
One can observe the result is just outside of the range found in the 25 mocks. The other two cases are similarly strong trends in the BGS density against the $\Delta$ \texttt{EBV GR} in both the DECam and BASS/MzLS regions. The result is more significant in the DECam region, as it includes more data. This map was not used in the regressions, as we expect the difference in $E(B-V)$ to include CIB contamination and the BGS density to correlate with CIB contamination. We use a solid curve to display the results for $\Delta$ \texttt{EBV GR} when substituting \texttt{EBVnoCIB} for the fiducial SFD $E(B-V)$. 
The trend becomes more moderate but remains significant. In terms of the $\chi^2_{\rm null}$ we determine, the fiducial value is 43.0 for the DECam region and reduces to 24.8 when using \texttt{EBVnoCIB}. 
For the data in the BASS/MzLS region, the corresponding $\chi^2_{\rm null}$ are 17.9 (which is within the range found in the mocks) and 9.7, respectively. This improvement in the $\chi^2_{\rm null}$ in both regions suggests the trends with $\Delta$ \texttt{EBV GR} are primarily driven by CIB contamination in SFD $E(B-V)$ and any regression applied that would remove the trends is likely to remove real clustering modes (traced by CIB).

 DESI BAO \cite{KP3s2-Rosado} and Full-Shape analysis ~\cite{KP5s6-Zhao} studies demonstrate that the results using DR1 BGS samples are robust even when not using the imaging systematic weights. For any analysis that instead proves sensitive to the imaging systematic weights, we suggest robustness tests against including the $E(B-V)$ difference when determining the imaging systematic weights.

\subsection{LRG}
\begin{figure}
    \centering 
    \includegraphics[width=0.9\columnwidth]{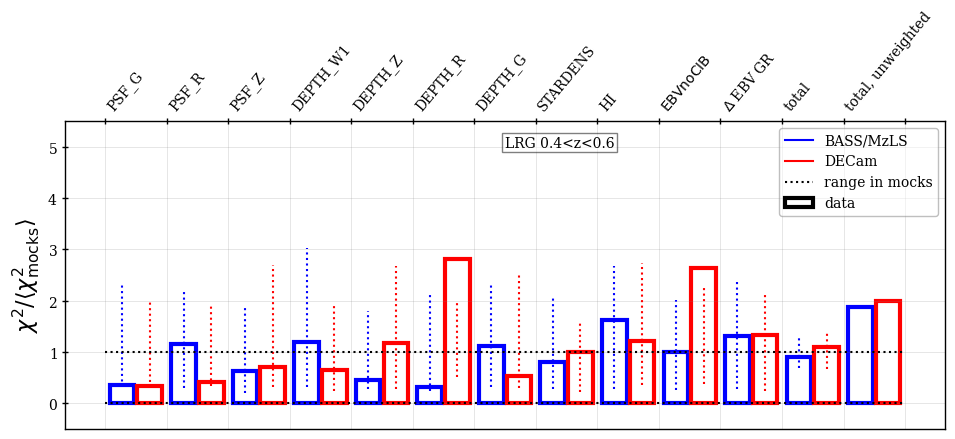}
    \includegraphics[width=0.9\columnwidth]{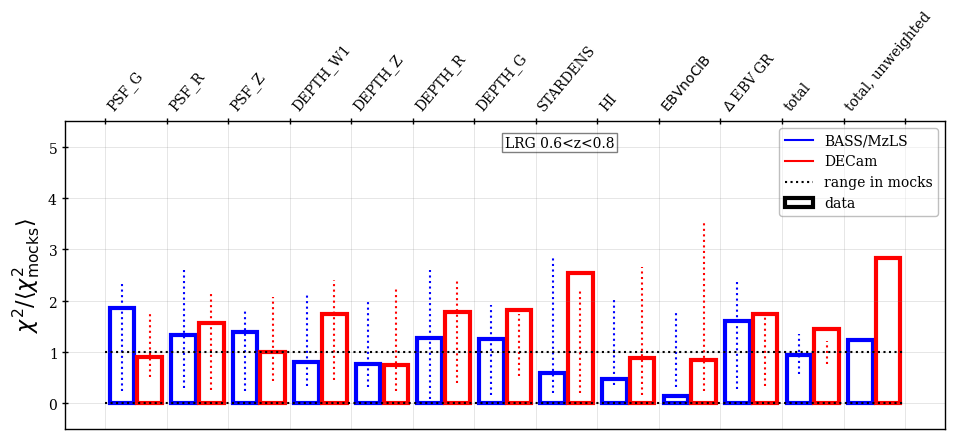}
   \includegraphics[width=0.9\columnwidth]{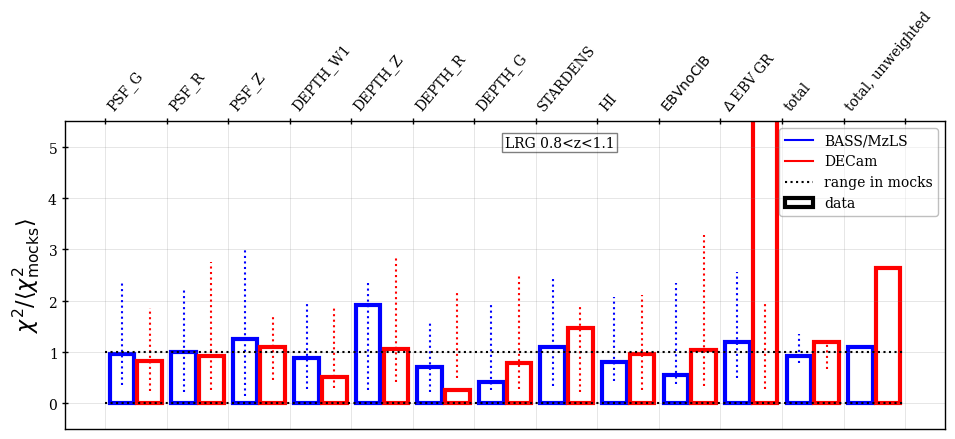}
    \caption{ Same as the top panel of \cref{fig:bgs_imsys_mocks}, but for the three LRG redshift bins. From top to bottom, the panels show results for each of the redshift bins $0.4<z<0.6$, $0.6<z<0.8$, and $0.8<z<1.1$ used in the analysis.
    }
    \label{fig:lrg_imsys_mocks}
\end{figure} 

\begin{figure}
\centering
\includegraphics[width=\columnwidth]{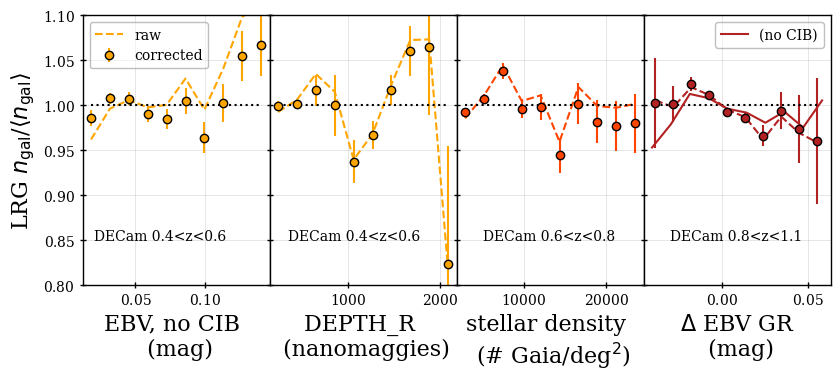}
    \caption{The same as the bottom panel of \cref{fig:bgs_imsys_mocks}, but for the four worst LRGs cases. }
    \label{fig:lrg_imsys_resid}
\end{figure} 

For LRGs, we apply the linear regression method. The regressions are performed separately in each of the three LRG redshift bins. In all three, we regress against maps of stellar density, HI column density, $z$-band galaxy depth, $r$-band PSF size, and $W1$ PSF depth. These five maps are sufficient to pass all null tests in the BASS/MzLS imaging area, and are motivated by the results of studies by \cite{Rezaie23} and \cite{KP3s13-Kong} on the LRG target sample. Based on the null tests in the DECam imaging area, one additional map was needed in each redshift bin, applied only to the regression in the DECam area: $r$-band galaxy depth for $0.4<z<0.6$, $g$-band galaxy depth for $0.6<z<0.8$, and the $z$-band PSF size for $0.8<z<1.1$.

\cref{fig:lrg_imsys_mocks} shows the $\chi^2_{\rm null}$ values obtained for each tested map, relative to the mean $\chi^2_{\rm null}$ obtained across the 25 mocks. The results for the data are shown in bars, while the range of values obtained from each of the individual mocks is displayed with dotted lines. In the final two columns, we show the results summing across all of the maps, one when applying the weights and the other without applying weights. One can observe that the use of imaging weights decreases the total $\chi^2$ by more than a factor of two for all redshift ranges for the data in the DECam region, and for the $0.4<z<0.6$ data in the BASS/MzLS region. However, 
the weights make little difference for LRG data at $z>0.6$ in the BASS/MzLS region.

There are four combinations of imaging property map and redshift range where the $\chi^2_{\rm null}$ 
from the DR1 LRG data is significantly higher than obtained for any mock realization. These are all in the DECam region, which has the majority of the DR1 data and thus the least statistical uncertainty on these tests.  We plot the trends for each in \cref{fig:lrg_imsys_resid}. Two cases are for the $0.4<z<0.6$ redshift bin. One is the SFD $E(B-V)$ map with CIB removed (\texttt{EBVnoCIB}), which was used for validation but not in the regression. The trend is improved considerably by the imaging weights compared to  
the case without weights applied (`raw', using the dashed curve): the $\chi^2_{\rm null}$ improves from 50.9 to 21.1 when using the weights obtained from the linear regression (`corrected', points with error bars). The other case is for a trend with the depth in $r$-band, which we do include in the regression for this redshift bin. One can observe that a downward fluctuation at a depth of $\sim$ 1000 nanomaggies drives the discrepancy with the null expectation. The $\chi^2_{\rm null}$  
improves from 30.4 to 20.6 when comparing the unweighted to weighted cases.

For the $0.6<z<0.8$ redshift bin, none of the 25 mocks have a greater $\chi^2_{\rm null}$ than found when testing the DR1 data against Gaia stellar density. The stellar density is included in the regression. One can observe in \cref{fig:lrg_imsys_resid} that the size of the $\chi^2_{\rm null}$ is driven by large fluctuations at particular stellar density values. The $\chi^2_{\rm null}$  
improves from 52.1 to 35.8 when comparing the unweighted to weighted cases. Notably, the total sum of the $\chi^2_{\rm null}$ for the DR1 LRG data with $0.6<z<0.8$ is greater than found in any mock. We discuss this at the end of the subsection.

Finally, for the $0.8<z<1.1$ bin, there is a significant trend in the DECam data with the difference in $E(B-V)$ determined using the spectra of DESI stars and the SFD map used in targeting. It can be seen in the right panel of \cref{fig:lrg_imsys_resid}. One can observe that there is a clear trend showing an approximately 5\% decrease in the DR1 LRG density. One can further observe that the weights make little difference; the $\chi^2$ is indeed slightly worse with weights, 33.4 compared to 31.6. This map was not considered for use for regressions against the LRG data, as one component of the difference in $E(B-V)$ should be CIB contamination, which we expect to be correlated with the large-scale structure traced by LRGs. We can test this by determining $\Delta$\texttt{EBV GR} using \texttt{EBVnoCIB} %$E(B-V)_{\rm no CIB}$ 
instead of the fiducial SFD $E(B-V)$. When we do so, the result is plotted using a solid curve in \cref{fig:lrg_imsys_resid} and we find the $\chi^2$ is reduced to 13.2. The dramatic reduction in $\chi^2$ suggests that the trend is indeed primarily driven by the CIB contamination component of $\Delta$\texttt{EBV GR} and is thus not a trend we wish to regress out of the data, as it would remove real large-scale structure.

The residual trends for LRGs that have been identified as potentially significant are small compared total improvement provided by the systematic weights. DESI studies of the BAO~\cite{KP3s2-Rosado} and full-shape~\cite{KP5s6-Zhao} fits demonstrate that the cosmology results using our DR1 LRG samples are robust even when not using the imaging systematic weights. However, further study of the residual trends may be necessary to obtain robust $f_{\rm NL}$ results from the DESI DR1 LRG samples, as studied in \cite{ChaussidonY1fnl}. For any analysis that proves sensitive to the inclusion of imaging systematic weights, we suggest robustness tests against splitting out the DES region, including the $E(B-V)$ difference when determining the weights, and any residual trends as a function of the $z$-band flux threshold, as motivated by the results of \cite{KP3s13-Kong}. Many such tests are performed in \cite{ChaussidonY1fnl}.

\subsection{QSO}
\begin{figure}
    \centering 
    \includegraphics[width=\columnwidth]{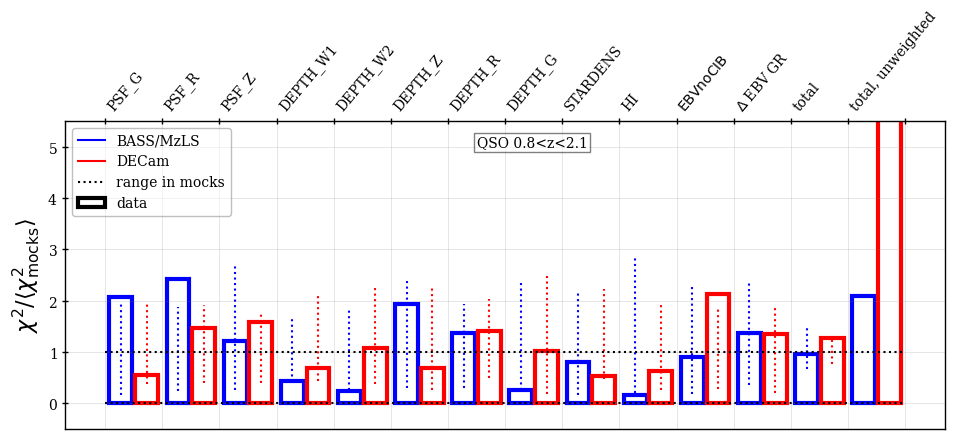}
    \includegraphics[width=\columnwidth]{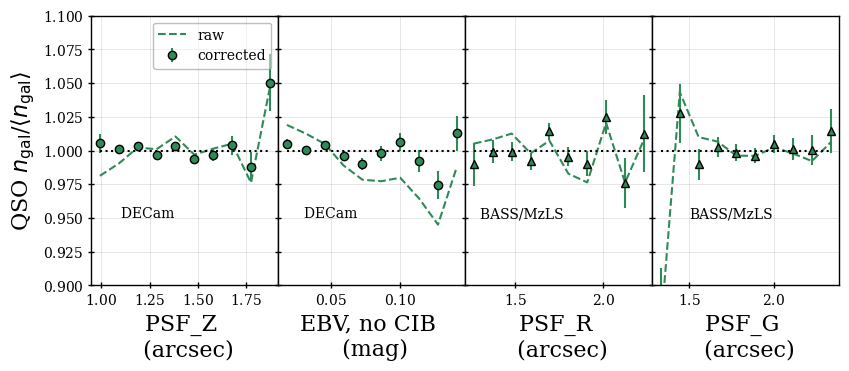}
    \caption{Same as for \cref{fig:bgs_imsys_mocks}, but for QSO. }
    \label{fig:qso_imsys_mocks}
\end{figure}

For QSO, we apply the RF regression, using all the maps that we include for validation, except for the \texttt{EBVnoCIB} map. The RF regression method was developed with DESI QSO in mind \cite{Chaussidon21QSOsys}. Another unique aspect of the QSO regressions is that the DES and DECaLS regions are fit independently. In the validation plots, we still combine the two regions into the full DECam region.

The top panel of \cref{fig:qso_imsys_mocks} summarizes the imaging systematic validation tests for the QSO sample, in the same manner as already shown for LRG and BGS in the previous two subsections. One can observe that the overall improvement of the $\chi^2_{\rm null}$ across all of the maps is greater than a factor of five for the DECam region and a factor of two for the BASS/MzLS region. In the DECam region, the total $\chi^2_{\rm null}$ after applying weights is still right at the maximum edge of the mock distribution, which motivates careful study of potential systematic effects on the clustering of the sample. The importance of any residual uncertainty on 2-point measurements is studied in the context of $f_{\rm NL}$ in \cite{ChaussidonY1fnl}, structure growth measurements in \cite{KP5s6-Zhao}, and BAO measurements in \cite{KP3s2-Rosado}.

In terms of specific maps, the bottom panel of \cref{fig:qso_imsys_mocks} presents the four cases where the data result is most extreme compared to the mock distribution. In all cases, one can observe that the trends are significantly improved compared to the unweighted (`raw') case. In the DECam region, for the $z$-band PSF size the $\chi^2_{\rm null}$ improves from 38.5 to 15.3, and for \texttt{EBVnoCIB} it goes from 166 to 19.8. In the BASS/MzLS region, for the $r$-band PSF size the $\chi^2_{\rm null}$ improves from 22.2 to 14.5, and for the $g$-band PSF size, it goes from from 15.9 to 13.1. Given the improvement in the trends and the fact that all of these high $\chi^2_{\rm null}$ values are not far from the maximum found in the mocks, we do not find any maps to point to particular concerns with the DESI DR1 QSO sample.

\subsection{ELG}
\begin{figure}
    \centering 
    \includegraphics[width=\columnwidth]{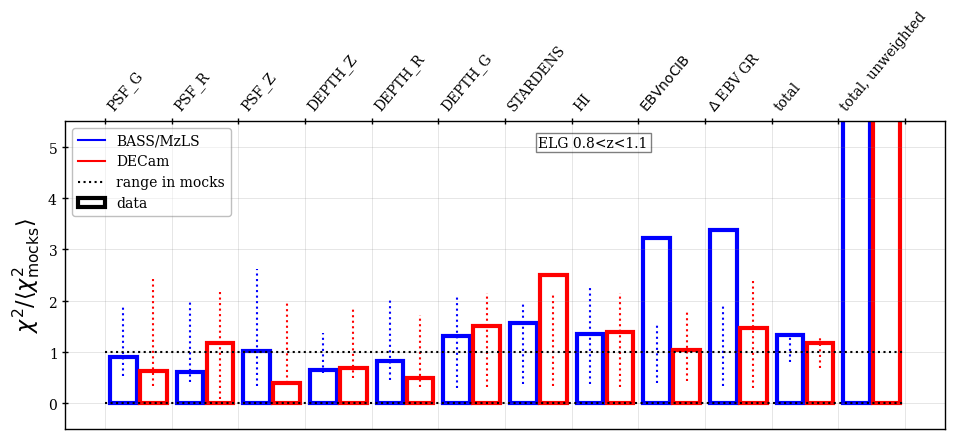}
    \includegraphics[width=\columnwidth]{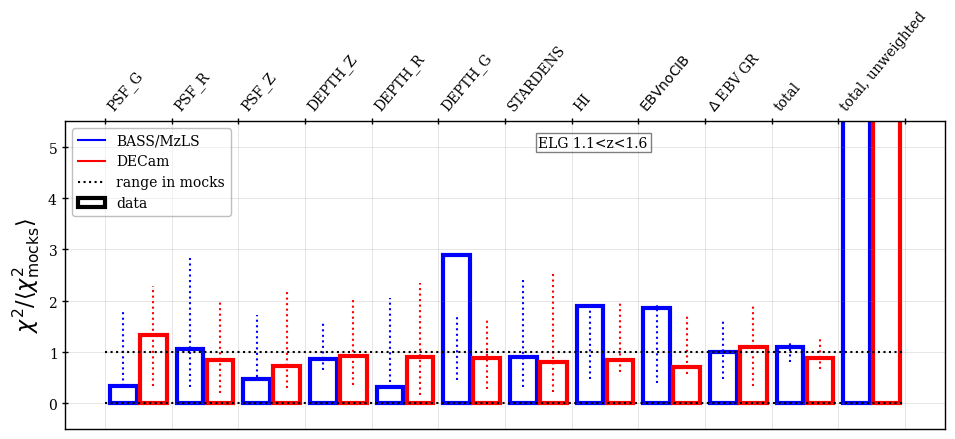}
    \includegraphics[width=0.9\columnwidth]{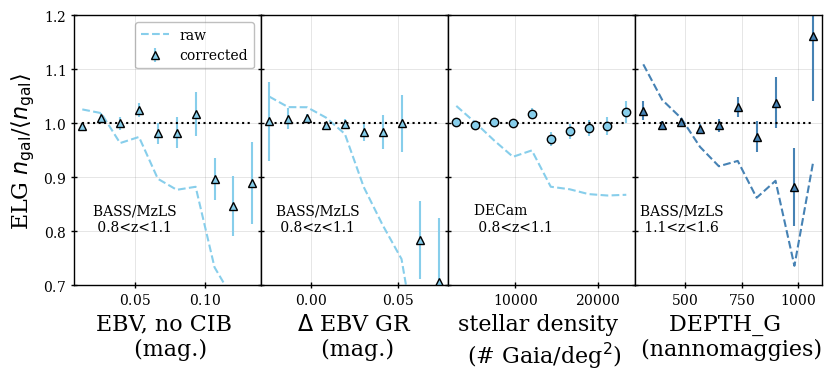}
    \caption{Same as for \cref{fig:bgs_imsys_mocks,fig:qso_imsys_mocks}, but for ELG.}
    \label{fig:elg_imsys_mocks}
\end{figure}

For DESI ELGs, we use the \textsc{SYSNet} NN regression to obtain weights to correct for the trends with imaging properties. All of the maps used for validation are used for training, due to severe trends with imaging properties for this tracer, except for \texttt{EBVnoCIB} (which is expected to be redundant with \texttt{HI}). The full details on the settings used are provided in \cite{KP3s2-Rosado}

\cref{fig:elg_imsys_mocks} shows the results of imaging validation tests performed on the DR1 ELG sample, in a similar manner as presented in the previous three subsections. In the top panel, the data results are compared to the distribution of results recovered from our 25 mock samples. In the top two panels, one can see that for both the $0.8<z<1.1$ and $1.1<z<1.6$ redshift ranges, the total $\chi^2_{\rm null}$ results for the data are consistent with those from the mocks, and that there is more than a factor of five improvement in the total $\chi^2_{\rm null}$ comparing the weighted and unweighted cases.

The results when testing against four individual maps show $\chi^2_{\rm null}$ values that are outside of the bounds recovered from the 25 mock realizations. The trends in these cases are displayed in the bottom panel of \cref{fig:elg_imsys_mocks}. In all cases, there is a large improvement in the trends when comparing before (`raw') and after (`corrected') applying the weights for imaging systematics.  For the case of the \texttt{EBVnoCIB} map tested against the data with $0.8<z<1.1$ in the BASS/MzLS region, the $\chi^2_{\rm null}$ is 23.8 with weights and 189.3 without. Similarly, for $\Delta$ \texttt{EBV GR} in the same redshift range and region, the $\chi^2_{\rm null}$ is 18.2 with weights and 182.6 without. For both, one can observe deviations at the high end of the map quantity; only 2.7\% of the BASS/MzLS region has a \texttt{EBVnoCIB} value greater than 0.1 and only 0.5\% of the region has a $\Delta$\texttt{EBV GR} greater than 0.05. For the stellar density, $\chi^2_{\rm null}$ in the DECam region and $0.8<z<1.1$ is 13.4 after weighting and 654.2 before. Finally for the DR1 data with $1.1<z<1.6$ in the BASS/MzLS region, the weighted $\chi^2_{\rm null}$ is 12.5 and the unweighted is 204.6.

While the imaging systematic mitigation can produce results that are consistent with our mock tests, the impact of imaging systematics on the DESI ELG sample is by far the most severe of any of the DESI tracers. Further details on the ELG density variations, including how well they are predicted by image simulations, changes in the expected redshift distribution, and the relative impact on clustering measurements can be found in \cite{KP3s2-Rosado}. There, it is also demonstrated that despite the severity of the trends, there is negligible impact on BAO measurements. The potential impact on studies of the full shape of ELG clustering statistics %Full-Shape analysis 
is investigated in \cite{KP5s6-Zhao}, where a method to account for systematic uncertainty related to imaging systematic is validated and applied to obtain DR1 results. We recommend similar testing and rigor be adopted for any LSS studies of the DR1 ELG sample.

\section{Treatment for Spectroscopic Systematics}
\label{sec:specsys}

When observed, some DESI targets spuriously receive the wrong redshift estimate, or are unable
to receive a reliable redshift estimate because they are too faint or were not observed with sufficient effective exposure time. The latter effect can cause spatially dependent variations in the density of confirmed DESI targets, tracing structure in the focal plane or variations related to spectroscopic observing conditions on larger scales. We therefore correct for these variations with a set of weights, described in Section~\ref{subsec:zfail_wts}.
We do not correct for the former effect of incorrect redshift estimates, but in Sec.~\ref{subsec:z_estimation} we characterize
the precision and accuracy of redshifts by tracer, and also estimate the rates of catastrophic redshift failures, where the measured redshift has a large offset from the truth.

\subsection{Redshift Failure Weights}
\label{subsec:zfail_wts}

The elimination of spurious fluctuations in the observed tracer density correlated with properties relating to how DESI spectra were observed, and the impact of this elimination, are detailed in \cite{KP3s3-Krolewski,KP3s4-Yu}. Significant trends were found and investigated for numerous properties, including the effective observing time, the survey speed, number of exposures or nights required to fully observe a target, the date of observation, and the focal plane position. One outcome of the investigation was the identification of issues with the DR1 processing of data taken on the night of Dec.\, 12th, 2021, and the ultimate replacement of that data in our DR1 LSS catalogs, as described in \cref{sec:dataspec}. Despite finding many significant trends, both studies show that the impacts of these trends on the two-point clustering statistics, even when uncorrected, are negligible.

For the corrections we ultimately apply to the DR1 LSS catalogs, for all tracers we model the success rate, $G_z$, as a function of the template signal-to-noise ratio, \texttt{TSNR2}. These \texttt{TSNR2} values are determined for each tracer type and per coadded spectrum, but are independent of the specific properties of the target (e.g., its brightness) observed to produce the spectrum; they use a fixed template (different for each tracer type) for the signal and the estimated noise and the result is thus proportional to an estimated effective observing time. Each is defined in \cite{DESIpipe} and we use the \texttt{TSNR2} associated with the particular tracer (i.e., \texttt{TSNR2\_ELG} for ELGs). 
For the BGS, LRG, and QSO samples, we use the fiberflux of the observed target as a secondary variable in the modeling \cite{KP3s3-Krolewski}. For ELGs, $G_z$ depends strongly on the precise redshift of the galaxy, due to noise from sky lines interfering with the ability to detect the [OII] doublet. The success rate is thus modeled as a function of \texttt{TSNR2\_ELG} and the galaxy redshift, as described in \cite{KP3s4-Yu}.

In all cases, to apply weights we find the relative success at the given redshift or \texttt{FIBERFLUX}, $\phi_{\rm fib}$. For BGS, LRG, and QSO, the success model is asymptotic and the weight can thus be defined as
\begin{equation}
    w_{\rm zfail}(S_{\rm ratio},\phi_{\rm fib}) = G_z (\infty,\phi_{\rm fib})/G_z( S_{\rm ratio},\phi_{\rm fib}),
\label{eqn:wzfail}
\end{equation}
where $S_{\rm ratio}$ is the \texttt{TSNR2} value and $G_z$ is a function modelling the success rate as a function of \texttt{FIBERFLUX} and \texttt{TSNR2}. As \texttt{TSNR2} goes to infinity, $G_z$ must asymptote to a constant value at fixed \texttt{FIBERFLUX}; we use an error function  to achieve this behavior, fit it to the observed redshift failure rates, and evaluate it in the limit of $\texttt{TSNR2} \rightarrow \infty$ to determine the numberator of Eq.~\ref{eqn:wzfail}. For the ELGs, a linear relationship is determined for the expected success rate as a function of \texttt{TSNR2\_ELG} at each redshift. These linear relationships are normalized to have a mean of one, weighting by the effective observing time. 

For all tracers, the modeling approach described above means that we do not over-correct for samples that have an overall poor spectroscopic success rate (i.e., when $G_z( (\infty,\phi_{\rm fib})) $ is much less than 1); e.g., in the most extreme case, QSO at low fiberflux have a $\sim30\%$ success rate and rather than upweight them all by a factor of $\sim$ 3, we only correct for the $\sim$ 10\% variation in the success rate of the subsample as a function of effective observing time. The weighted observed density for any selection in fiberflux will have a null trend with TSNR2, and thus any impact of the DESI observing pattern is expected to be nulled, under the assumption that the fiberflux dependence captures any trends with redshift\footnote{This not being the case for ELGs necessitated the specific redshift dependence of their weights.} and TSNR2 captures the DESI observing pattern.

\subsection{Accuracy and Precision of Redshift Estimation}
\label{subsec:z_estimation}

Uncertainties in the estimation of the DESI redshift impact the measured clustering, but are not accounted for in the LSS catalogs. These uncertainties include the random scatter from standard measurement noise, systematic biases in the redshifts, and also catastrophic errors that are due to, e.g., mis-identification of emission lines or confusion with sky lines.

For LRG, redshift accuracy can be assessed
using repeated observations and is estimated at a standard deviation of 40--60 km s$^{-1}$ 
\cite{Yu23,VIGalaxies.Lan.2023} (note that different statistics are used, with \cite{Yu23} quoting the standard deviation and \cite{VIGalaxies.Lan.2023} quoting the normalized median absolute deviation). Its dependence on magnitude is measured in \cite{VIGalaxies.Lan.2023} and on redshift in \cite{Yu23}, who show that the distribution of redshift errors
is best fit by a Lorentzian distribution.
Comparison to BOSS redshifts in \cref{sec:compare_to_sdss}
shows that only 0.3\% of the LRG targets in common differ by more than 0.002 in redshift ($\sim$250 km s$^{-1}$ at typical LRG redshifts), with a mean difference of 4 km s$^{-1}$ (very similar to the mean difference of 3.3 km s$^{-1}$ found in \cite{VIGalaxies.Lan.2023} from comparison to DEEP2).

Since BGS is brighter than LRG, its redshifts are more accurate:
the standard deviation is 10 km s$^{-1}$ with little dependence on $r_{\textrm{fiber}}$ \cite{VIGalaxies.Lan.2023}, with a mean difference of 6.5 km s$^{-1}$ compared to DEEP2 or 2 km s$^{-1}$ compared
to BOSS, and a 0.2\% outlier fraction.

While ELGs are faint, their redshifts are measured
more accurately than LRGs due to their strong emission lines. Their redshift errors are 8--10 km s$^{-1}$ and best fit with a Lorentzian \cite{VIGalaxies.Lan.2023,Yu23,ELG.TS.Raichoor.2023}, again from repeat observations. The impact of ELG catastrophic
errors are extensively studied in \cite{KP3s4-Yu}, who find
0.27\% of ELG redshifts have catastrophic redshift errors
$|\Delta v|>1000$ km s$^{-1}$. \cite{KP3s4-Yu} characterize
the origins of these catastrophic errors, and their impact on the galaxy clustering two-point statistics. 

Quasar redshifts are more challenging, both due to the difficulty of centroiding the broad emission lines and due to systemic shifts in the high-ionization CIV line, which is observable by DESI at $z > 1.6$. The more reliable MgII line redshifts out of the DESI wavelength range at $z = 2.5$, making $z > 2.5$ particularly problematic.
The \redrock\ quasar templates were updated
from the Early Data Release \cite{RedrockQSO.Brodzeller.2023}, improving the redshift accuracy and bias compared to early studies in \cite{Yu23,QSO.TS.Chaussidon.2023,VIQSO.Alexander.2023}, which used the old BOSS quasar templates.
The LSS catalogs use
the updated templates presented and validated in \cite{RedrockQSO.Brodzeller.2023}, whereas the Ly$\alpha$ analysis additionally applied a correction
for the evolution of the Ly$\alpha$ mean flux, which further improves the quasar redshifts at $z > 1.8$ \cite{KP6s4-Bault}.
Redshift precision is assessed by comparing redshifts in the SV Repeat Exposures, with an overall normalized median absolute deviation (NMAD) of 57 km s$^{-1}$, which rises with redshift from a low of $\sim$ 20 km s$^{-1}$ at $z \sim 0.8$ to a peak at $\sim 125$ km s$^{-1}$ at $z \sim 1.8$, and declines thereafter \cite{RedrockQSO.Brodzeller.2023}.
Redshift precision as a function of redshift is shown in Fig.~7 of \cite{RedrockQSO.Brodzeller.2023}.
The distribution of redshift errors has very heavy,
non-Gaussian tails, and are better fit by a Lorentzian \cite{Yu23} or a combination of three Gaussians \cite{QSO.TS.Chaussidon.2023}.
Quasar redshift accuracy is assessed via small-scale cross-correlations
with LRG, ELG, and the Ly$\alpha$ forest \cite{RedrockQSO.Brodzeller.2023}: biases in quasar redshift measurements will lead to a radial shift in the peak of the cross-correlation from zero.
These biases are shown in Table 6 of \cite{RedrockQSO.Brodzeller.2023} and are 30---60 km s$^{-1}$ at $0.8 < z < 2.1$.
Finally, catastrophic redshift errors are assessed from the comparison
of DESI automated and VI redshifts in \cite{VIQSO.Alexander.2023} (using the old quasar templates), with 0.7\% (4.5\%) of quasars having offsets $> 3000$ (1000) km s$^{-1}$ at $z < 2.1$ and 1.8\% (12.2\%) at $z > 2.1$. This paper uses the old BOSS quasar templates, but \cite{RedrockQSO.Brodzeller.2023} finds that the catastrophic rate is very similar between the old and the new templates.

Random redshift errors have a similar impact on the two-point functions as the small-scale Finger of God velocity dispersion, though smaller because the redshift errors
are much smaller than the virial velocities. 
For the BAO analysis of the clustering of BGS, LRG, ELG and QSO tracers, the small-scale velocity dispersion
is modelled by a Lorentzian multiplying the broad-band 
power spectrum with free parameter $\Sigma_s$ which is taken to have a Gaussian prior \cite{DESI2024.III.KP4}.
The redshift error can be absorbed by $\Sigma_s$ and the BAO damping parameter $\Sigma_\parallel$ at the level of 1-2 $h^{-1}$ Mpc for quasars (and smaller for other tracers); as shown in \cite{KP4s2-Chen}, the BAO results are insensitive to this level of change in $\Sigma_s$ and $\Sigma_\parallel$.
Likewise, the impact of redshift errors on the Ly$\alpha$ BAO analysis is presented in \cite{Youles22} and the impact on the BAO peak position is found to be negligible. For the analysis of the full shape of clustering statistics for discrete tracers,
modifications to the simple Lorentzian Finger of God form (e.g.\ due to the heavy tails in the redshift error distribution) are effectively encapsulated by additional EFT counterterms
\cite{KP5s1-Maus}. The impact of ELG catastrophic errors
on the power spectrum and correlation function is extensively studied in \cite{KP3s4-Yu}, who find shifts of $< 0.2\sigma$ in the two-point functions, at a level that does not affect the derived parameters.

\section{Summary of Weights and Normalization of Randoms}
\label{sec:weights}
The completeness weights $w_{\rm comp}$ (defined in \cref{sec:comp}),  redshift failure weights $w_{\rm zfail}$ (defined in \cref{sec:specsys}, and imaging systematic weights $w_{\rm imsys}$ (defined in \cref{sec:imsys}) are combined to a total weight $w^{\prime}_{\rm tot}$ to correct for how the DESI selection function changes as a function of position,
\begin{equation}
    w^{\prime}_{\rm tot} = w_{\rm comp}w_{\rm zfail}w_{\rm imsys}.
\end{equation}

In this section, we first describe how redshift and weight information are added to the DESI DR1 LSS random catalogs. We then describe how the weights that optimally balance number density variations are determined for both data and randoms, and how this requires a refactoring of $w^{\prime}_{\rm tot}$.

\subsection{Randoms}
\label{sec:ranz}

To assign redshifts to the random catalogs introduced in \cref{sec:target} while matching their redshift distribution to that of the data, we assign them redshifts and all associated weights from randomly chosen galaxies in the data catalogs. 
The random catalogs will thus have all of the same weights as the galaxies, but $w_{\rm comp}$, $w_{\rm zfail}$, and $w_{\rm imsys}$ are present purely to make sure that their weighted (normalized) redshift distribution matches the data. For ELG, LRG, and BGS, the assignment of these quantities to the randoms is done separately in the BASS/MzLS and DECam photometric regions. For QSO, the DECam region is further split into the DES region and DECaLS regions as the photometric selection of QSO is different in each region. We then multiply the sampled $w^{\prime}_{\rm tot}$ value by $f_{\rm tile}(t_{\rm group})$ (\cref{sec:compdef}). Finally, we normalize the weights\footnote{The weights described in the following subsection do not affect the normalization, as they are applied equally to data and randoms.} on the randoms such that the ratio of weighted data to weighted random counts is the same in each photometric region\footnote{The regions are the North and South for all tracers except QSO, which further divide the South region into DES and not DES.}. Further details on the process are given in Section 7.2 of \cite{KP3s15-Ross}.

Assigning redshifts to the random points as described above is often described as `shuffled' randoms and matches the approach applied to SDSS \cite{bosslsscat,ebosslss}
Out of the available options, it was determined to be the least biased method for BOSS CMASS galaxies \cite{RossDR9}. However, given the radial distribution of the randoms exactly matches the data by construction, this method nulls purely radial clustering modes and introduces a radial integral constraint bias \cite{demattiaIC}, whose effects must be modeled. 
We summarize the method and present details on all window effects accounted for in the DESI DR1 analyses in \cref{sec:pk}.

\subsection{Weights to Optimally Balance Sampling Rate}
\label{sec:FKP}

Based on the principles first presented in \cite{FKP}, to increase the expected signal-to-noise of our clustering measurements, we apply weights to data and randoms based on the observed number density as a function of redshift and $n_{\rm tile}$, which we denote $w_{\rm FKP}$. The completeness variations in DESI DR1 are large (see \cref{fig:allcomp}). We thus incorporate the mean completeness (see \cref{sec:comp}) as a function of $n_{\rm tile}$ into our $w_{\rm FKP}$ calculation.
The process is fully detailed in section 7.3 of \cite{KP3s15-Ross} and we repeat some details below. In order to do so while accounting for the completeness weights\footnote{They are all greater than 1 and thus otherwise up-weight incomplete regions.}, we first divide the total data weight column $w^{\prime}_{\rm tot}$, for both data and random by the mean completeness weight as a function of $n_{\rm tile}$, $\langle w_{\rm comp}\rangle (n_{\rm tile})$. Thus, the final $w_{\rm tot}$, included for both data and randoms and meant to account for selection function variations, becomes
\begin{equation}
    w_{\rm tot} = w^{\prime}_{\rm tot}/\langle w_{\rm comp}\rangle (n_{\rm tile})\,.
\end{equation}
This is the column \texttt{WEIGHT} in the DR1 `clustering' catalogs. In this way, the total data counts are normalized such that they sum to approximately the number of observed objects, but still fully account for the impact of variations due to fiber assignment incompleteness, imaging systematics, and redshift systematics.

The factoring of the weights applied above allows us to determine $w_{\rm FKP}$, taking the variation in observed number density as a function of redshift and survey coverage into account. First, we define 
\begin{equation}
    n_x(z,n_{\rm tile}) = n(z)\langle C_{\rm assign}\rangle(n_{\rm tile}),
\end{equation}
where $C_{\rm assign}(n_{\rm tile})$ assignment completeness values are determined in the same way as in \cref{tab:statsvntile} (selecting the particular discrete $n_{\rm tile}$ rather than applying a threshold)\footnote{The values of $C_{\rm assign}$ and $w_{\rm comp}$ vary with $n_{\rm tile}$ for the same reason, the fiber}.
In this manner, $n_x(z,n_{\rm tile})$ is the expected number density at the location of any galaxy or random point. We then define $w_{\rm FKP}$ following the standard convention
\begin{equation}
    w_{\rm FKP}(z,n_{\rm tile}) = 1/[1+n_x(z,n_{\rm tile})P_0].
\end{equation}
The value of $P_0$ is chosen separately for each tracer, given an approximate nominal value of the power spectrum monopole $P_0(k=0.15 h{\rm Mpc}^{-1})$. The values used in the DR1 analysis are $P_{0,\rm BGS} = 7000\;({\rm Mpc}/h)^3$, $P_{0,\rm LRG} = 10,000 \;({\rm Mpc}/h)^3$, $P_{0, \rm ELG} = 4000 \;({\rm Mpc}/h)^3$ and $P_{0, \rm QSO} = 6000 \;({\rm Mpc}/h)^3$. These values are only roughly consistent with the actual clustering amplitude of the respective DESI samples and it is likely that slightly more optimal choices can be adopted in future DESI analyses. These $w_{\rm FKP}$ are meant to improve the expected signal-to-noise of 2-point function measurements at the scales used for BAO analyses. For such calculations, one can simply multiply $w_{\rm tot}$ by $w_{\rm FKP}$ to obtain the total weight to apply to each data/random point. Analyses that use different scales or clustering estimates might wish to derive their own, more optimal, weighting to apply. Indeed, the primordial non-Gaussianity analysis presented in \cite{ChaussidonY1fnl} derives alternative weights.

\section{Comparison with SDSS}
\label{sec:compare_to_sdss}

\begin{figure*}
    \centering 
    \includegraphics[width=0.49\columnwidth]{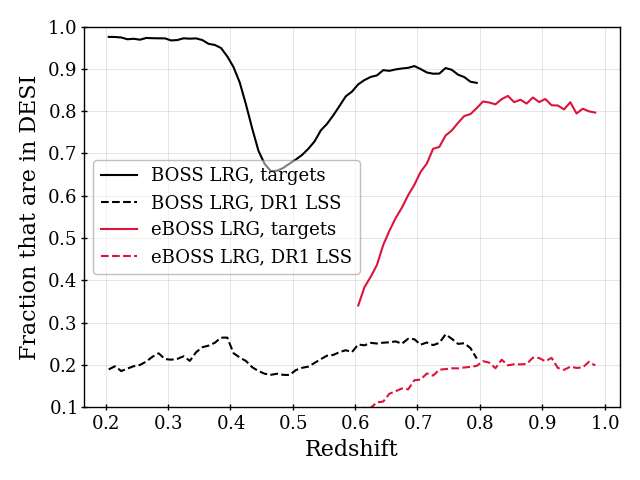}
    \includegraphics[width=0.49\columnwidth]{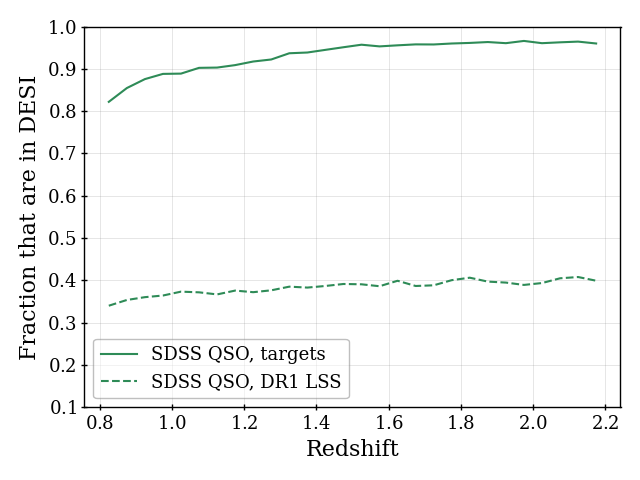} 
    \caption{The fraction of SDSS galaxies and quasars in (e)BOSS LSS catalogs that are in DESI samples. Solid curves show the fraction of the SDSS data that are DESI targets. The dashed curves show the fraction that are in DR1 DESI LSS catalogs. }
    \label{fig:sdssretar}
\end{figure*} 

\begin{figure*}
    \centering 
    \includegraphics[width=0.49\columnwidth]{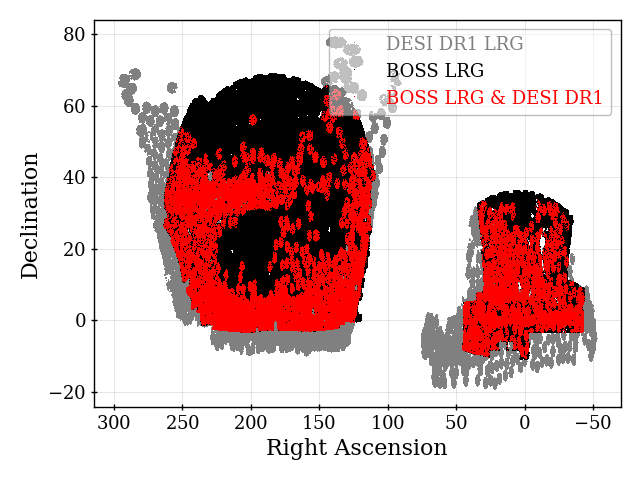}
    \includegraphics[width=0.49\columnwidth]{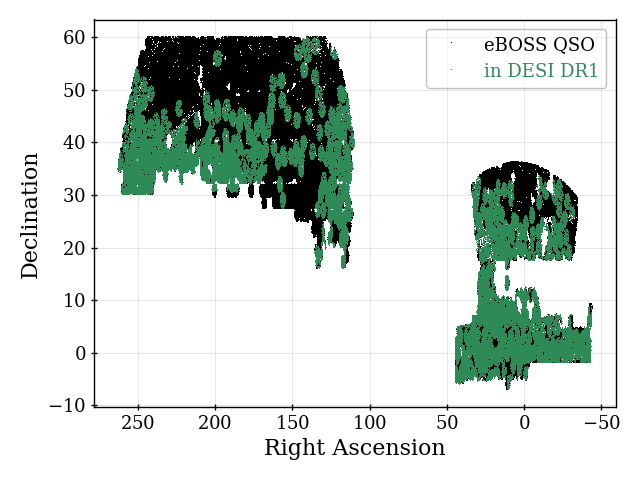} 
    \caption{The celestial coordinates of LRGs (left) and quasars (right). For LRGs, we show the coordinates of all DESI DR1 galaxies in the LSS catalogs in gray. The coordinates of all BOSS galaxies in the LSS catalogs are in black and all galaxies that are in both BOSS and DR1 are shown in red. The right-hand plot shows similar information for QSO, but we do not plot the whole DR1 footprint (which covers the same area as the LRGs do.)}
    \label{fig:sdssoverlap}
\end{figure*} 

The area on the sky covered by the SDSS (e)BOSS LSS catalogs \cite{bosslsscat,ebosslss} is a subset of that covered by DESI. DESI targeting does not take into account whether or not SDSS has already observed any potential target. Thus, a large fraction of the (e)BOSS galaxies and quasars are DESI targets. This is quantified in \cref{fig:sdssretar}, where we match between (e)BOSS LSS catalogs and DESI data by treating any targets within 1 arcsecond of each other as a match. The fractions that are matched to DESI targets are shown with solid curves. For BOSS, we match to the combination of BGS and LRG targets. The matched fraction is greater than 0.95 at low redshift, as almost all BOSS data at such redshifts have $r$-band magnitudes brighter than 19.5. For redshifts $z>0.4$, the DESI LRG sample dominates the matches. For both BOSS and eBOSS, the matched fraction increases with redshift. Despite the DESI number density being greater by $\sim$ 25\%, only 65\% of BOSS galaxies are targeted as DESI LRGs at $z=0.5$. Within the eBOSS QSO LSS catalog redshift range of $0.8<z<2.2$, the fraction that are DESI targets is always greater than 0.8 and is greater than 0.95 for $z>1.6$.

The dashed curves in \cref{fig:sdssretar} show the fractions of the galaxies and quasars in the (e)BOSS LSS catalogs that are matched to galaxies and quasars in the DESI DR1 LSS catalogs. Due to the redshift bounds of the DESI LSS catalogs, all SDSS LRGs with redshifts $z<0.4$ are matched to BGS galaxies, and those with $z>0.4$ are matched to DESI LRGs. The matched fraction is consistently just greater than 0.2 for LRGs and close to 0.4 for quasars. Given the overall fraction of SDSS QSOs that are DESI targets is nearly 1, and that the completeness of the DESI DR1 QSO sample is 0.87, one can infer that DESI DR1 overlaps with just less than half of the eBOSS quasar LSS sample. A similar calculation for BOSS yields a result closer to 2/5th. The overlap can be seen in \cref{fig:sdssoverlap}, by comparing the colored points to the black ones. Based on the gray points, one can also see that a considerable amount of the DR1 area is outside the SDSS footprint.

\begin{table}
\centering
    \begin{tabular}{|l|l|l|}
    \hline
     Tracer and Redshift bin & $N$ DESI  & $N$ SDSS \\
    \hline
    BGS $0.1<z<0.4$ & 300,043 & 82,398  \\
    LRG $0.4<z<0.6$  & 506,911 & 138,405  \\
    LRG $0.6<z<0.8$  & 771,894 & 70,979  \\
    LRG $0.8<z<1.1$  & 859,822 & 10,474  \\
    QSO $0.8<z<2.1$  & 856,831 & 124,726  \\
    \hline
    \end{tabular}
    \caption{The number of DESI galaxies and QSO and the number of them that are repeated SDSS observations, in each of the redshift bins used for the 2024 cosmological analyses, \label{tab:SDSSrepeats}}
\end{table}

We next compare the fraction of DESI DR1 galaxies that are repeats of SDSS LSS data. The numbers are found in \cref{tab:SDSSrepeats}. For BGS, 27\% of the sample was already observed by BOSS (. For the LRGs, in the $0.4<z<0.6$ redshift bin we again find that 27\% of the sample was already observed by BOSS. For $0.6<z<0.8$, 9\% of the sample was already observed by BOSS and eBOSS (51,175 in BOSS and 19,804 in eBOSS), while for $0.8<z<1.1$, just 1\% of the sample was already observed by BOSS and eBOSS (550 in BOSS and 9924 in eBOSS). Finally, the fraction for DESI QSO in the range $0.8<z<2.1$ the fraction is 15\%.

In total, 190,277 BOSS galaxies are matched to DR1 DESI LRGs. Of these, only 602 (0.3\%) have a redshift that differs by more than 0.001$(1+z_{\rm DESI})$. For those that differ by less than 0.001$(1+z_{\rm DESI})$, the mean $(z_{\rm SDSS}-z_{\rm DESI})/(1+z_{\rm DESI})$ is 1.4$\times 10^{-5}$ and the standard deviation is 1.7$\times 10^{-4}$. Comparing the DESI DR1 BGS and BOSS redshifts, the differences are all smaller: the outlier rate is 0.2\%, and after rejecting outliers the mean difference is 4$\times 10^{-6}$, and standard deviation 1.3$\times 10^{-4}$. We expect that this level of difference will be negligible for most uses of the DESI LSS catalogs.

The measurement of quasar redshifts is much more uncertain than for galaxies, as discussed in \cref{sec:specsys}. Thus, for comparing the DESI and SDSS quasar redshifts, we increase the outlier cut to $|z_{\rm SDSS}-z_{\rm DESI}|/(1+z_{\rm DESI})>0.01$ and find that 0.5\% of the DESI sample is an outlier. After removing outliers, the standard deviation of $(z_{\rm SDSS}-z_{\rm DESI})/(1+z_{\rm DESI})$ is $1.2\times10^{-3}$ and the mean difference, $(z_{\rm SDSS}-z_{\rm DESI})$ (with no scaling factor) is $-8\times 10^{-4}$. These differences are consistent with the DESI quasar redshift results discussed in \cref{sec:specsys}.

\section{2-point Functions}
\label{sec:2pt}
For the DR1 clustering analysis, the information from the DESI galaxy and quasar samples is condensed into redshift-space 2-point measurements. To this end, angular positions and redshifts are converted into Cartesian coordinates using the fiducial cosmology model (see \cref{sec:intro}). We apply estimators in both configuration- and Fourier-space, to measure the correlation function and power spectrum respectively. These binned measurements of the 2-point functions are compared to physical models in the cosmology analyses; doing so requires both a covariance matrix of the binned measurements and knowledge of the window function that converts a smooth analytic model to the expectation of the binned measurement and accounts for the effects of the survey window and estimator bias. Each of these pieces is detailed in the following subsections.

\subsection{Clustering Estimators}

We describe separately below the methods for estimating the correlation function and the power spectrum.

\subsubsection{Correlation Function}
\label{sec:xicalc}
The anisotropic 2-point correlation function $\xi(s, \mu)$ is a measure of the excess probability of finding two galaxies (at positions $\Vec{x}_{1}$, $\Vec{x}_{2}$) at a separation distance $s = \| \Vec{s} \|$ and with cosine angle $\mu = \hat{s} \cdot \hat{\eta}$ between their separation vector $\Vec{s} = \Vec{x}_{2} - \Vec{x}_{1}$ and the line-of-sight from the observer, $\Vec{\eta}$. We use the so-called Landy-Szalay estimator~\cite{LandySzalay}, with the mid-point line-of-sight $\hat{\eta} = (\Vec{x}_{1} + \Vec{x}_{2}) / 2$ convention:
\begin{equation}
\widehat{\xi}(s, \mu) = \frac{DD(s, \mu) - DR(s, \mu) - RD(s, \mu) + RR(s, \mu)}{RR(s, \mu)}
\label{eq:ls_estimator}
\end{equation}
where the notation $XY(s, \mu)$ corresponds to the weighted number of pairs of objects $X$ and $Y$ in a $s, \mu$ bin, divided by the total weighted number of pairs. Specifically, $DD(s, \mu)$ is the weighted number of data (galaxy and quasar) pairs and $RR(s, \mu)$ is the weighted number of pairs of objects in the random catalog which samples the selection function. For the autocorrelation (single-tracer) measurements performed in our analysis, by symmetry $DR(s, \mu) = RD(s, \mu)$. For standard estimates, pair weights are the product of the total individual weights $w_{1}$ and $w_{2}$ of the two objects in the pair, themselves obtained as the product of systematic correction weights and FKP weights. When BAO reconstruction is applied, both data $D$ and randoms $R$ are shifted to (partially) undo non-linear structure formation and redshift-space distortions. In notation introduced in \cite{Pad12}, the shifted data is denoted as $D$ and the shifted randoms are denoted as $S$, and the Landy-Szalay estimator is modified to be
\begin{equation}
\widehat{\xi}(s, \mu)_{\rm recon} = \frac{DD(s, \mu) - DS(s, \mu) - SD(s, \mu) + SS(s, \mu)}{RR(s, \mu)},
\label{eq:ls_estimator_recon}
\end{equation}
that is, the convention is to use the counts from the unshifted randoms as the normalization.

We use $200$ $s$-bins up to $200\; \mathrm{Mpc}/h$ (subsequently regrouped in $4 \: \mathrm{Mpc} / h$ bins) and $200$ $\mu$ bins in $[-1, 1]$.
The anisotropic correlation function $\widehat{\xi}(s, \mu)$ is further projected onto the basis of Legendre polynomials to estimate the monopole ($\ell = 0$), quadrupole ($\ell = 2$) and hexadecapole ($\ell = 4$) moments of the correlation function:
\begin{equation}
\widehat{\xi}_{\ell}(s) = \frac{2 \ell + 1}{2} \int {\rm d}\mu\,\widehat{\xi}(s, \mu) \mathcal{L}_{\ell}(\mu)\;,
\end{equation}
where $\mathcal{L}_\ell(\mu)$ is the Legendre polynomial of order $\ell$. In practice, the above integral is turned into a finite sum over $\mu$-bins, weighted by the integral $(2\ell + 1) / 2 \int {\rm d}\mu\,\mathcal{L}_{\ell}(\mu) $ of $\mathcal{L}_{\ell}(\mu)$ over the bin  (ensuring that $\ell > 0$ multipoles of a purely isotropic signal are $0$).

To optimize computing time we use the technique of~\cite{SplitRandom}, which consists of summing $DR$, $RD$, and $RR$ pair counts obtained using several random catalogs each of a size approximately matching that of the data catalog. In practice, for each tracer, there are up to 18 random catalogs available, each with a density of 2500 deg$^{-2}$. We thus vary the number used for pair-counts for each tracer, choosing the number that provides at least 50x the number of data points for each case. This comes to  1, 8, 10, and 4 of the LSS catalog random files for BGS, LRG, ELG, and QSO respectively.
 Correlation function estimates are obtained for each galactic cap and redshift range within $\sim 10$ minutes on a NERSC Perlmutter 4 A100 GPU node. NGC and SGC measurements are then combined by summing the pair counts computed within each region.

As explained in~\cite{KP3s5-Pinon}, to mitigate the effect of fiber collisions, specifically the loss of galaxy pairs at small separations, so-called $\theta$-cut 2-point correlation function multipoles are estimated by removing pairs at small angular separation $\theta < 0.05 \; \mathrm{deg}$ from all terms of~\cref{eq:ls_estimator}. This $\theta$-cut, and to a lesser extent pair count binning, make the relation between the average of the measured correlation function $\langle \hat{\xi}_{\ell}(s) \rangle$ and the theory $\xi_{\ell^\prime}(s^\prime)$ non-trivial, such that $\langle \hat{\xi}_{\ell}(s) \rangle = W_{s, s^\prime}^{\ell \ell^\prime} \xi_{\ell^\prime}(s^\prime)$, with a window matrix $W_{s, s^\prime}^{\ell \ell^\prime}$ that is computed as detailed in~\cite{KP3s5-Pinon}.

Correlation function measurements are performed with \textsc{pycorr}\footnote{\url{https://github.com/cosmodesi/pycorr}} which wraps a version of the \textsc{Corrfunc} package~\cite{Corrfunc} modified to also run on GPU and support alternative lines-of-sight (first-point, end-point) and weight definitions (PIP and angular up-weights, see~\cref{sec:pipcat}) and the $\theta$-cut, along with jackknife utilities.

We use the methods described above to produce the following three types of measurements of the $\ell=0,2,4$ multipoles of the 2-point correlation function from the DESI DR1 LSS catalogs, split into the redshift bins listed in \cref{tab:zbins} and weighting by $w_{\rm tot}*w_{\rm FKP}$:
\begin{itemize}
    \item {\bf `raw'}: These are obtained directly from the catalogs, using all weighted pair-counts. 
    \item {\bf `$\theta$-cut'}: These are obtained when removing all pair-counts with an angular separation $\theta<0.05^\circ$.
    \item {\bf `reconstructed'}: These are obtained using the catalogs produced after applying BAO reconstruction. The $\theta$-cut is not applied.    
\end{itemize}

In \cref{sec:clus}, we compare the raw clustering measurements from the data to those obtained from simulations of DESI DR1. The reconstructed measurements are presented in \cite{DESI2024.III.KP4}, where they are used to measure the BAO scale and compared to the raw pre-reconstruction results. All of these measurements will be released publicly with DR1.

\subsubsection{Power Spectrum}
\label{sec:pk}

The power spectrum estimator~\cite{yamamoto2006} is based on the FKP field~\cite{FKP}:
\begin{equation}
F(\vec{r}) = n_{D}(\vec{r}) - \alpha_{R} n_{R}(\vec{r}).
\label{eq:fkp}
\end{equation}
where $n_{D}(\vec{r})$ and $n_{R}(\vec{r})$ are the weighted number of galaxies and randoms painted on a grid, and $\alpha_R = \sum_{i=1}^{N_D} w_{d, i} / \sum_{i=1}^{N_R} w_{r, i}$ is the ratio of the total weight of galaxies to that of randoms. The power spectrum multipoles can then be estimated as:
\begin{equation}
\hat{P}_{\ell}(k) = \frac{2 \ell + 1}{A N_{k}} \sum_{\vec{k}} \sum_{\vec{r}_{1}} \sum_{\vec{r}_{2}} F(\vec{r}_{1}) F(\vec{r}_{2}) \mathcal{L}_{\ell}(\hat{k} \cdot \hat{\eta}) e^{i \vec{k} \cdot (\vec{r}_{2} - \vec{r}_{1})} - \mathcal{N}_{\ell}
\label{eq:fkp_estimator}
\end{equation}
where the sum $1 / N_{k} \sum_{\vec{k}}$ should be interpreted as an average over the Fourier-space $\vec{k}$ grid within a given bin of $k \equiv \| \vec{k} \|$, and $\sum_{\vec{r}_{1}}$, $\sum_{\vec{r}_{2}}$ as sums over the configuration space grid.  
The normalization term is given by the sum over grid cells $A = \alpha_R / dV \sum_i n_{D, i} n_{R, i}$ with fixed cell size of $dV^{1/3} = 10 \; \mathrm{Mpc}/h$ (which in practice approximates the norm of the window matrix in the middle of the $k$-range of interest). The shot noise term $\mathcal{N}_{\ell}$ is non-zero only for the monopole:
\begin{equation}
\mathcal{N}_{0} = \frac{1}{A} \left[ \sum_{i=1}^{N_D} w_{D, i}^2 + \alpha_R^2 \sum_{i=1}^{N_R} w_{R, i}^2 \right]\,.
\end{equation}

We choose the direction to the first galaxy $\hat{r}_{1}$ as the `first-point' line-of-sight $\eta$, such that the above double integral can be split into:
\begin{equation}
\hat{P}_{\ell}(k) = \frac{2 \ell + 1}{A N_{k}} \sum_{\vec{k}} F_{0}(\vec{k}) F_{\ell}(-\vec{k}) - \mathcal{N}_{\ell}\,,
\label{eq:power_spectrum_multipoles}
\end{equation}
with:
\begin{equation}
F_{\ell}(\vec{k}) = \sum_{\vec{r}} F(\vec{r}) \mathcal{L}_{\ell}(\hat{k} \cdot \hat{r}) e^{i \vec{k} \cdot \vec{r}}.
\label{eq:fkp_multipoles}
\end{equation}
$F_{\ell}(\vec{k})$ is then recast as a sum of Fast Fourier Transforms (FFT) following~\citep{Hand2017:1704.02357v1}. 
As a default, we use the TSC (triangular shaped cloud) scheme to paint galaxies onto the grid and obtain $n_{D}(\vec{r})$, $n_{R}(\vec{r})$. The Fourier-space grid is therefore compensated for by the kernel $\prod_{i=x,y,z} \text{sinc}(2\pi k_{i}/k_{N})^{p}$ with $p = 3$, see~\cite{Jing2005:astro-ph/0409240}, with $k_{N}$ the Nyquist frequency of the grid. We also implement the interlacing method to mitigate aliasing. The interlacing at order $n$ consists in painting the density field shifted by $[0, 1,..., n-1]$ mesh cell size in $x$, $y$ and $z$ directions, average the FFT multiplied by the appropriate phase terms, and (if required) FFT back to configuration-space, see~\cite{Sefusatti2015:1512.07295}. Given the TSC assignment kernel, we found interlacing of order $3$ to be sufficient to achieve percent precision up to the Nyquist frequency. 

When BAO reconstruction is applied, we use the shifted data and randoms to compute the FKP field in~\cref{eq:fkp}, while, similarly to the case of the correlation function, we estimate the normalization factor $A$ and the window matrix with the standard (unshifted) random catalogs.

In computing the gridded densities, we use physical box sizes of $4000 \; \mathrm{Mpc}/h$, $7000 \; \mathrm{Mpc}/h$, $9000 \; \mathrm{Mpc}/h$, $10000 \; \mathrm{Mpc}/h$ for BGS, LRG, ELG, QSO respectively, and a grid cell size of $6 \; \mathrm{Mpc}/h$, resulting in a Nyquist frequency of $k_{N} \simeq 0.5 \; h/\mathrm{Mpc}$. Since the computing cost is to first order independent of the number density of points, for these measurements we use the maximum number of random catalogs available (18), which corresponds to more than 100$\times$ the data density for all tracer types.

Power spectrum estimates are obtained for each galactic cap and redshift bin with MPI within $\sim 5$ minutes (including $\theta$-cut, as described below) on a Perlmutter CPU node. NGC and SGC measurements are then combined by averaging the two power spectra (with weights corresponding to the normalization amplitude $A$ in each cap). In the following, the power spectrum estimates thus obtained are called \emph{raw} power spectrum measurements.

As for the correlation function, the ensemble average of power spectrum multipoles $\hat{P}_{\ell}(k)$ are related to the theory $P_{\ell^\prime}(k^\prime)$ as $\langle \hat{P}_{\ell}(k) \rangle = W_{k, k^\prime}^{\ell \ell^\prime} P_{\ell^\prime}(k^\prime)$. We estimate the window matrix $W_{k, k^\prime}^{\ell \ell^\prime}$ from the 2-point selection function which we compute by concatenating the power spectra of the random catalogs obtained with box sizes $20 \times$, $5 \times$ and $1 \times$ the nominal box size used for the data power spectrum measurements. Following~\cite{BeutlerWindow}, and as detailed in~\cite{KP3s5-Pinon}, this window matrix also includes first-order wide-angle effects. As for the correlation function and as explained in~\cite{KP3s5-Pinon}, we also provide so-called $\theta$-cut power spectrum estimates by removing from the standard estimator above the power spectrum of galaxy pairs at angular separation $\theta < 0.05 \; \mathrm{deg}$. The window matrix calculation is modified accordingly by removing such pairs from the calculation of the 2-point selection function.

In practice, to estimate power spectra we use \textsc{pypower},\footnote{\url{https://github.com/cosmodesi/pypower}} which is a modified version of the \textsc{nbodykit} implementation~\cite{Hand2017:1712.05834v1} to account for our new definition of the normalization $A$, and implements cross-power spectra estimation, $\theta$-cut, and utilities to estimate the window matrix.

The veto masks described in \cref{sec:footprint}---and more importantly, the implementation of the $\theta$-cut---mix small and large scale modes, such that the window matrix $W_{k, k^\prime}^{\ell \ell^\prime}$ is very non-diagonal. Convolution of this non-diagonal window with the theory power $P_{\ell^\prime}(k^\prime)$ would therefore naively require specifying $P_{\ell^\prime}(k^\prime)$ to scales $k^\prime > 0.5 \; h/\mathrm{Mpc}$ that cannot be described by perturbation theories. Although \cite{KP3s5-Pinon} show that in practice this is not a problem for DESI, in principle it is preferable not to have a window matrix which produces sensitivity to details of the theory model above the maximum $k$ where the theory is reliable. Therefore, following \cite{KP3s5-Pinon}, we \emph{rotate} the data vector $\vec{P}$ (composed of the binned estimated power spectrum multipoles with $N_{\rm bins}$ bins), the window matrix $\mathrm{W}$ and the covariance matrix $\mathrm{C}$ to the corresponding transformed quantities
\begin{align}
\vec{P}^\prime& = \tens{M}\vec{P} - s_\ell \tens{m}_{{\rm o}, \ell}\,,\\
\tens{W}^\prime& = \tens{MW} - \tens{m}_{{\rm o}, \ell}\tens{m}_{{\rm t}, \ell}^{T}\,,\\
\tens{C}^\prime& = \tens{MCM}^T\,,
\end{align} 
where $\tens{M}$ (of shape $(N_{\rm bins}, N_{\rm bins})$)
, $\tens{m}_{\rm o, \ell}$ (each of shape $N_{\rm bins}$, with $\ell$ running over $3$ multipoles) and $\tens{m}_{\rm t, \ell}$ (each also of shape $N_{\rm bins}$) are chosen as explained in \cite{KP3s5-Pinon} by an optimisation procedure in order to make the rotated window $\tens{W}^\prime$ as compact as possible and the rotated covariance $\tens{C}^\prime$ as close to diagonal as possible. 

\begin{figure}
\centering
\includegraphics[width=0.9\columnwidth]{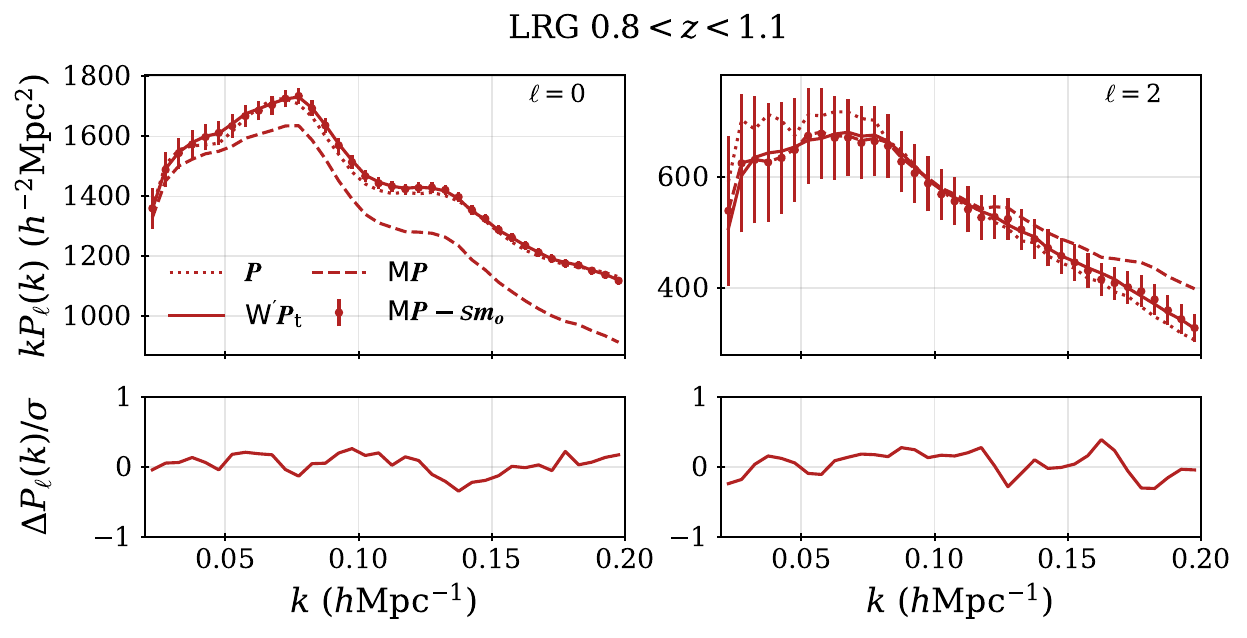}
.\caption{Validation of the rotated power spectrum window matrix for $0.8<z<1.1$ LRGs. Top row: power spectrum monopoles (left) and quadrupoles (right). Dotted curves show the mean measured $\vec{P}_{t}$ from the 25 \texttt{AbacusSummit} LRG cubic mocks, solid lines show $\vec{P}_{t}$ multiplied by the rotated window matrix $\tens{W}^\prime\vec{P}_{t}$. The data points and dashed lines respectively show the mean of the corresponding rotated 25 `complete' power spectrum measurements with ($\vec{P}^\prime = \tens{M} \vec{P} - s\vec{m}_{\rm o}$) and without ($\vec{P}^\prime = \tens{M} \vec{P}$) the best fit $s\vec{m}_{\rm o}$ correction determined from the corresponding 25 cutsky mocks. Error bars on the data points indicate the uncertainty in the power spectrum measurements, estimated from the \texttt{EZmocks} (see~\cref{sec:cov}). Bottom row: the difference $\tens{W}^\prime\vec{P}_{t} - \vec{P}^\prime$ between the rotated windowed expectation and measurements with the $s\vec{m}_{\rm o}$ correction, in units of the measurement uncertainty. }
\label{fig:WindowMatrixLRG}
\end{figure}

The amplitude vector $\vec{s}$ (of size $3$) is a free parameter, to be marginalized in the fit assuming a given prior.
To specify this prior, we fit $\vec{s}$ to minimize the difference (within scales $k < 0.2\;h/\mathrm{Mpc}$) between the rotated power spectrum multipoles from the \texttt{AbacusSummit} `complete' cutsky mocks (described in \cref{sec:mockfa} below) measured with the $\theta$-cut, contained in $\vec{P}^\prime$, and the power spectrum multipoles $\vec{P}_{t}$ of the corresponding cubic box mocks multiplied by the rotated window $\tens{W}^\prime$. The agreement between $\tens{W}^\prime \vec{P}_{t}$ and $\vec{P}^\prime = \tens{M}\vec{P} - s_\ell \vec{m}_{{\rm o}, \ell}$ is illustrated for the high LRG redshift bin in \cref{fig:WindowMatrixLRG}, and is smaller than $1/5$th of the measurement uncertainties for DR1 data: i.e., given the number of mocks used in this test we are unable to detect any systematic bias produced by this method. We then assume a Gaussian prior on $\vec{s}$ with both mean and standard deviation given by the previously obtained best-fit value. In practice, $\vec{s}$ is directly marginalized over at the power spectrum level, by providing $\tens{M}\vec{P} - \vec{s} \tens{m}_{\rm o}$ as rotated power spectrum measurement and adding to the rotated covariance matrix $C^\prime$ the contribution $\vec{m}_{{\rm o}, \ell} s_\ell^2 \vec{m}_{{\rm o}, \ell}^T$. 

\begin{figure}
\centering
\includegraphics[width=0.49\columnwidth]{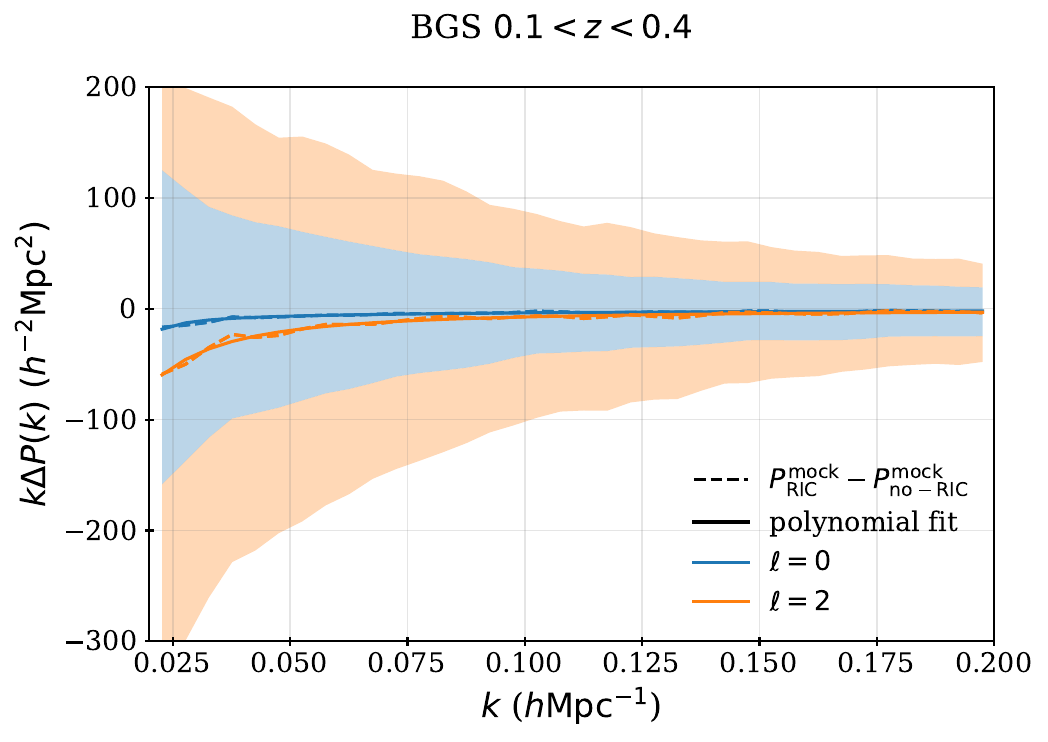}
\includegraphics[width=0.49\columnwidth]{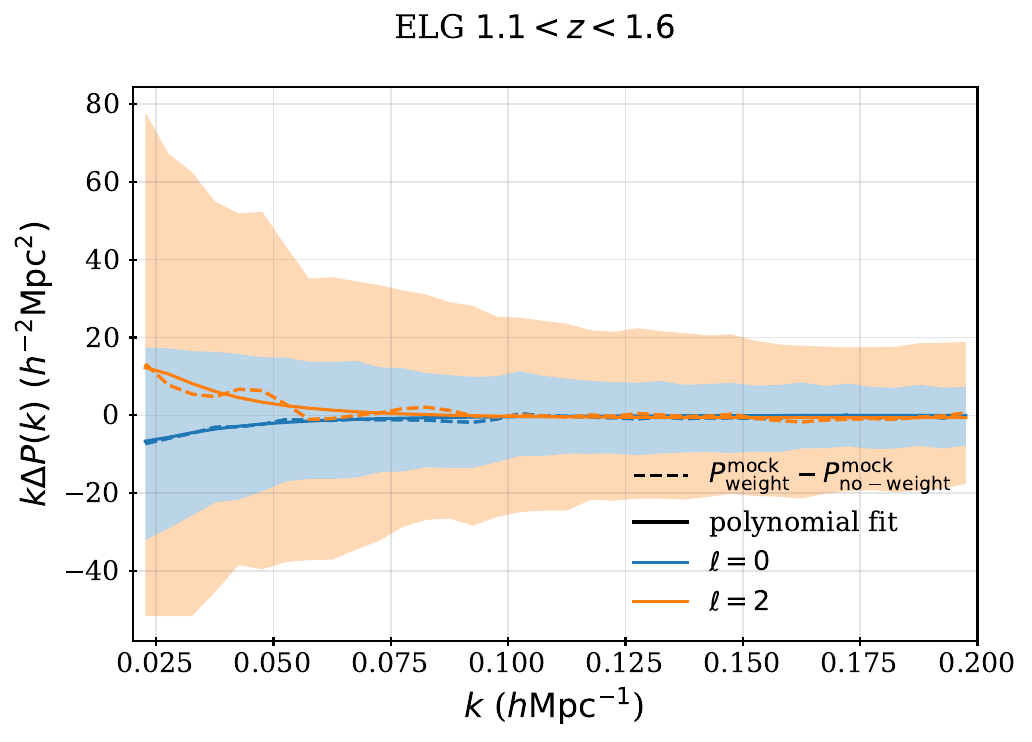}
\caption{Left: dashed curves show the difference of power spectrum multipoles with and without radial integral constraint effect (monopole in blue, quadrupole in orange), measured from the mean of 50 \texttt{EZmocks} for the BGS sample as described in the text; solid curves show polynomial fits to these differences. The blue and orange shaded regions indicate the measurement uncertainties, for monopole and quadrupole respectively, for the DR1 data, estimated from the \texttt{EZmocks} as described in \cref{sec:cov}. Right: as for left panel, but dashed curves show the difference of the power spectrum multipoles from \texttt{AbacusSummit} `altmtl' cutsky mocks for the ELG sample with and without  imaging systematic weights, illustrating the angular mode removal induced by the imaging systematic correction. }
\label{fig:TemplateCorrection}
\end{figure}

Finally, we implement two further systematic corrections to the measured power spectrum. First, as discussed in~\cref{sec:ranz}, since the redshift distribution of randoms is constructed to exactly to match that of data, radial modes in the measured power spectrum or correlation function are nulled, known as the radial integral constraint (RIC)~\cite{demattiaIC}. To model this, we produce \texttt{EZmock} realisations of the measurements with and without the RIC effect, by creating random catalogs in which redshifts are taken from the corresponding mock data realization using the shuffle method in the former case and by sampling random redshifts from the smooth redshift selection applied in creating the mocks in the latter case. We then measure the difference $\Delta P_{\mathrm{RIC}} \equiv P^{\mathrm{mock}}_{\mathrm{RIC}} - P^{\mathrm{mock}}_{\mathrm{no-RIC}}$ in the mean of the power spectra over 50 \texttt{EZmocks} with and without the RIC, fit this difference with a polynomial of the form $c_{-5}k^{-5} + c_{-3}k^{-3} + c_{-2}k^{-2}$ to obtain a template, and subtract this fitted template from the rotated power spectrum measured from the data. The results of this procedure are illustrated in the left panel of \cref{fig:TemplateCorrection} for the BGS sample. The dashed lines in the plot show the measured power difference $\Delta P_{\mathrm{RIC}}$ and the solid lines show the polynomial fits. The RIC removes power for radial modes, i.e. at $\mu = 1$, where the Legendre polynomials of order $\ell = 0, 2$ are both positive, resulting in damped monopole and quadrupole power on large scales. We note that the polynomial model is a good fit to the observed RIC effect relative to the measurement uncertainties in the DR1 data indicated by the shaded bands.

The second correction is for the effect of an angular integral constraint (AIC) introduced by the use of the imaging systematic weights discussed in \cref{sec:imsys}. These are determined by regressing the observed galaxy density with a set of maps encoding imaging properties. Depending on the exact regression method (linear, \textsc{SYSNet}, or \textsc{Regressis}), this may result in added noise and the removal of large scale angular modes.
To estimate this, we use 25 \texttt{AbacusSummit} `altmtl' cutsky mocks without imaging contamination (see \cref{sec:mockfa}) and as for the RIC correction, we use the difference in the mean power spectrum measurements obtained with and without the imaging weights to create a template model that is subtracted from the rotated power spectrum measured from the DR1 data. An example of this correction is shown in the right panel of \cref{fig:TemplateCorrection} for the ELG sample (which applied \textsc{SYSNet}). The AIC removes power for angular modes, i.e. at $\mu = 0$, where the Legendre polynomials of order $\ell = 0$ and $\ell = 2$ are positive and negative respectively. Thus on large scales it damps power in the monopole and enhances it in the quadrupole. Again the polynomial model is seen to be a good description of the observed mode-removal effect from the AIC given the measurement precision of DESI DR1.

This correction for the AIC is applied only for imaging weights derived from the \textsc{SYSNet} and \textsc{Regressis} methods for the ELG and QSO samples, as for linear regression the mode removal effect is much smaller and is barely detectable even in the mean power spectrum over 25 mocks.

For both the AIC and RIC, the correction to be applied depends on the geometry and the large-scale clustering of the given sample. This implies that the accuracy of the correction depends on how closely the mocks match the data. We expect any residual uncertainty from this dependence to be negligible for most analyses, given that the estimated size of the effect is already sub-dominant compared to the uncertainty of the DR1 $P(k)$ measurements (as can be seen in \cref{fig:TemplateCorrection}).

To summarise the contents of this section: we use the methods described above to produce the following four types of measurements of the $\ell=0,2,4$ multipoles of the power spectrum from the DESI DR1 LSS catalogs, split into the redshift bins listed in \cref{tab:zbins} and weighting by $w_{\rm tot}*w_{\rm FKP}$:
\begin{itemize}
    \item {\bf `raw'}: These are obtained from the direct application of the standard estimator \cref{eq:fkp_estimator} to the data, without further corrections. 
    \item {\bf `$\theta$-cut'}: These are obtained after removing the impact of all pair-counts with an angular separation $\theta<0.05^\circ$.
    \item {\bf `$\theta$-cut+rotation+RIC+AIC'}: These results are obtained from applying the window rotation and corrections for the radial integral constraint and angular integral constraint to the `$\theta$-cut' measurements.
    \item {\bf `reconstructed'}: These are obtained from the DR1 catalogs after applying BAO reconstruction shifts, but without using the $\theta$-cut, rotation, or RIC and AIC corrections.    
\end{itemize}

In \cref{sec:clus}, we compare the raw clustering measurements obtained from the data and from simulations of DESI DR1. The reconstructed measurements are presented in \cite{DESI2024.III.KP4}, where they are used to measure the BAO scale and compared to the raw pre-reconstruction results. The $\theta$-cut+rotation+RIC+AIC measurements are presented in \cite{DESI2024.V.KP5} and used for the cosmological constraints obtained from the full-shape of the power spectrum \cite{DESI2024.VII.KP7B}. All flavours of the measurements will be released publicly with DR1.

\subsection{Covariances}
\label{sec:cov}

Analytic or semi-empirical models of the covariance of DESI DR1 2-point function measurements are produced in configuration-space \cite{KP4s7-Rashkovetskyi} and Fourier-space \cite{KP4s8-Alves}. These models are validated in \cite{KP4s6-Forero-Sanchez} via comparison to the clustering statistics measured from 1000 mock realizations (the \texttt{EZmocks}) of the DR1 data described in \cref{sec:mocks}. There, it is found that configuration-space \textsc{RascalC} semi-empirical covariance matrices constructed to fit to the \texttt{EZmocks} match the covariance matrices constructed from the \texttt{EZmocks} themselves to within the expected statistical scatter. However, the \textsc{RascalC} covariance matrices tuned to fit the DESI DR1 data predict higher variance than is seen in the DR1 \texttt{EZmocks}. We interpret this as the result of the failure of the \texttt{EZmocks} to precisely match all aspects of the DR1 data---in particular, as described in \cref{sec:mockfa} below, they only approximately reproduce the true effects of fiber assignment in the data. Since the \textsc{RascalC} covariance matrices are flexible enough to allow recalibration from the DR1 data, we use them for all configuration-space analysis and recommend all reanalyses of the DR1 2-point clustering results do the same.

The efforts to model the covariance in Fourier-space were less successful in reproducing the numerical covariance obtained from the \texttt{EZmocks}, even when tuned to these mocks. For the Fourier-space covariance matrices, we therefore use the mock-based numerical covariance, but rescale it in order to account for the mismatch between the \texttt{EZmocks} and the DR1 covariances seen in configuration space. The rescaling applied is determined through comparison of the configuration-space mock-based covariance $\tens{C}_{S,\xi}$ and the \textsc{RascalC} version $\tens{C}^{-1}_{R}$ determined for the DR1 data: we use what \cite{KP4s7-Rashkovetskyi} refer to as the `reduced $\chi^2$' for the comparison of these covariance matrices:
\begin{equation}
    \chi^2_{\rm red}(\tens{C}_{R},\tens{C}_{S,\xi}) = \frac{1}{N_{\rm bins}}{\rm tr}(\tens{C}^{-1}_{R}\tens{C}_{S,\xi})\,,
\end{equation}
and multiply the mock-based Fourier-space \texttt{EZmock} covariance by $1/\chi^2_{\rm red}$ to account for the enhanced total variance observed in the DR1 data in configuration-space. To determine $1/\chi^2_{\rm red}$ for all tracers, we use the covariance elements corresponding to the monopole and quadrupole moments in the range $20<s<200\;h^{-1}$Mpc (without BAO reconstruction) in the expression above. We find that the obtained value of $1/\chi^2_{\rm red}$ varies only at the percent level when restricting to only the monopole elements or when reducing the range of scales. The enhancement factor applied to each tracer and redshift bin is given in Table \ref{tab:covfac}.

\begin{table*}
\centering
\begin{tabular}{|l|ccccccc|}
\hline%\hline
Tracer &  BGS & LRG1 & LRG2 & LRG3 & ELG1 & ELG2 & QSO \\\hline
$1/\chi^2_{\rm red}$ & 1.39 & 1.15 & 1.15 & 1.22 & 1.25 & 1.29 & 1.11\\
\hline
\end{tabular}
\caption{The factor by which we multiply the mock-based Fourier-space covariance matrix when fitting DR1 data.
\label{tab:covfac}}
\end{table*}

The covariance matrices used to test cosmological models against DESI DR1 power spectra measurements in \cite{DESI2024.V.KP5,DESI2024.VII.KP7B} include additional terms that account for observational and theoretical systematic uncertainties. The full details are provided in \cite{DESI2024.V.KP5}. Notably, the method to account for residual imaging systematic uncertainty, validated in \cite{KP5s6-Zhao}, is included as a component of the covariance matrix. 

\section{DESI DR1 Simulated Data}
\label{sec:mocks}

A full description of the simulations used and the process applied to them to obtain mock DESI DR1 LSS catalogs (mocks) is provided in \cite{KP3s8-Zhao}. The mocks are
produced in two sets: one based on 25 realisations of the (2$h^{-1}$Gpc)$^3$ \texttt{AbacusSummit} N-body simulations \cite{abacus1, abacus2} tiled to cover the DESI DR1 volume, and the other based on (6$h^{-1}$Gpc)$^3$ \texttt{EZmocks} simulations \cite{EZmocks} that provide 1000  realizations of DESI DR1 without requiring any replication. Each type of mock is used for different applications, depending on the level of survey realism we need and the scales of analysis. \texttt{Abacus} mocks, coming from an N-body simulation, will reproduce the small-scale clustering with much more accuracy than \texttt{EZmocks}, but the computing cost to generate them limits us to 25 realizations that require some replication to simulate DESI DR1.
 On the other hand, the \texttt{EZmocks} were fast to produce, such that the computing cost for each input realization was sub-dominant compared to the post-processing steps of LSS catalog generation and analysis, but they are less precise at small scales. 
 
In what follows, we detail the general steps that are applied to all mocks to turn them into mock DESI DR1 LSS catalogs. We divide the discussion into two subsections. The first describes how the target samples are simulated. The second describes the different ways in which the fiber assignment process is applied to the target samples, and how the LSS catalog pipeline is applied.

\subsection{Simulating Target Samples}

 To properly simulate the DR1 data sample, we must simulate the dark- and bright-time target samples. First, we calibrate the halo occupation distributions separately for each target type (BGS, LRG, ELG, and QSOs) based on the EDR clustering signal for dark time tracers \cite{abacushod2, abacusHODELG, bgs_hod} and using early versions of the full DR1 BGS sample (without any absolute magnitude cut) for BGS, using the methods described in \cite{Smith23bgsmocks}. We use simulation boxes at different redshift snapshots. These include $z=0.2$ for BGS, $z=0.5,0.8$ for LRGs, $z=0.95,1.325$ for ELGs, and $z=1.4$ for QSO. We convert the box coordinates into angular sky coordinates and `real-space' redshifts---converted directly from radial comoving distances---with respect to a chosen observer position using the fiducial DESI DR1 cosmology.
 %(With the application of replication to the \texttt{AbacusSummit} boxes to cover the full DR1 area.)
To this end, for each Galactic hemisphere, we place an observer at one corner of the (6$h^{-1}$Gpc)$^3$ box (in the case of \texttt{AbacusSummit}, obtained by tiling multiple copies of the smaller simulated box) and specify a line-of-sight such that the boxes cover the full DR1 comoving volume (see \cite{KP5s6-Zhao}). 

To simulate the effect of redshift space distortions, we further adjust the redshift coordinate of each tracer to obtain the `apparent' redshift as follows,
 \begin{equation}
 z = z_{\rm r} + (\boldsymbol{u} \cdot \hat{\boldsymbol{r}}_{\rm c}) (1 + z_{\rm r} ) / c ,
 \end{equation}
where $z_{\rm r}$ denotes the `real-space' redshift, $\boldsymbol{u}$ indicates the peculiar velocity of the tracer, and $\hat{\boldsymbol{r}}_{\rm c}$ is the unit line-of-sight vector in comoving space.
 
For LRGs, the $z=0.5$ box is used for simulation data with $z<0.6$ and the $z=0.8$ box is used for $z>0.6$. For ELGs, we use the $z=0.95$ snapshot for $z<1.1$ and $z=1.325$ for $z>1.1$. These redshift splits allow some of the evolution in the samples determined by \cite{abacushod2, abacusHODELG} to be included in our DR1 mocks.\footnote{Including evolution at a better resolution in redshift is a goal for future DESI analyses.}

We then subsample these catalogs to match the $n(z)$ distribution estimated for a complete sample divided by the spectroscopic success rate; i.e., for dark time, we take the curves in the left-hand panel of \cref{fig:nzall} and divide them by the spectroscopic success fraction reported in \cref{tab:Y1lss}. For the ELGs, we use the $n(z)$ estimated separately in the DECam and BASS/MzLS regions, but for the rest of the tracers we use the same $n(z)$ over the entire footprint. All dark-time tracers are combined into a single target file and formatted to match the data model required to properly run DESI fiber assignment. For bright-time, the \texttt{AbacusSummit} mocks simulate all BGS targets and assign them absolute magnitudes. However, for the \texttt{EZmocks}, only a sample matching the $M_r < -21.5$ sample we analyze is simulated. This has some consequences for the fiber assignment on the mocks, which we describe in the following subsection.

\subsection{Fiber Assignment and LSS Pipeline} 
\label{sec:mockfa}
\begin{figure}
    \centering    \includegraphics[width=0.65\columnwidth]{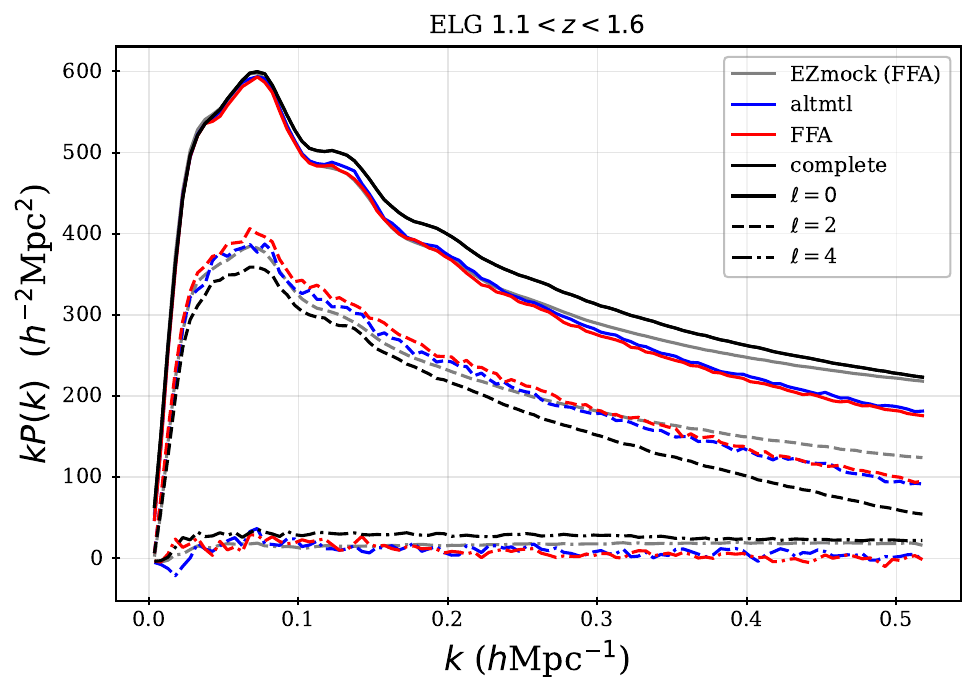}
\caption{The mean multipoles, with respect to the line of sight, of the redshift-space power spectrum of $1.1<z<1.6$ ELGs measured from four different types of mocks. The gray curves show the mean results from 1000 \texttt{EZmock} with the FFA. The other curves are all the mean of the 25 \texttt{AbacusSummit} mocks, with the three different labeled versions of DESI fiber assignment applied, as described in the text.}
    \label{fig:PkmockELG}
\end{figure} 
  
For DESI DR1, we have developed and applied three separate treatments to account for DESI fiber assignment in the mock catalogs, which we label as mock ``flavors''. In all cases, we obtain the potential assignments using the same methods as applied to obtain the DR1 random samples. We are thus able to utilize the same random angular coordinates as used for the real data \texttt{full\_noveto} samples (see \cref{sec:LSScat}), and with the only difference in the veto masks being the realization specific priority veto (see \cref{sec:prioritymask}). The footprint of all of the mock flavors and the real data are thus identical and use the same hardware and imaging veto masks as the DR1 data. The only variation concerns the priority veto mask, which we describe below. 

The three flavors are:
\begin{itemize}
    \item {\bf complete:} For the `complete' mocks, all potential assignments are treated as observed. All completeness weights are thus 1 and there is no priority veto mask to apply.
    
    \item {\bf altmtl:} For the `altmtl' mocks, we follow the method \cite{KP3s7-Lasker} described in \cref{sec:altmtl} to run the mock data through the same fiber assignment loop as the real data, considering the same hardware status and tile ordering as the DR1 processing and observing. This method allows us to run the LSS catalog pipeline in the same way as applied to the DR1 data. We expect all completeness statistics to match those of the DESI data up to the level that the underlying simulations match the DR1 data target samples\footnote{One example of the simulations being imperfect is they are not full lightcones and thus, e.g., the `redshift' at which QSO and ELG targets overlap do not correspond to the redshift output of the simulations.}.
    
    \item {\bf FFA:} The large computation time needed for the altmtl method makes running it on all 1000 \texttt{EZmocks} impractical. Therefore we developed a ``fast-fiberassign'' (FFA) \cite{KP3s6-Bianchi} method to simulate DESI fiber assignment. The method uses a shallow learning algorithm to produce a fiber assignment emulator that takes the local angular density, determined via friends-of-friends with a linking length that is tuned per tracer, and $n_{\rm tile}$ as input variables. The training and assignment are applied individually to each tracer type. This means that the priority veto mask does not get applied; instead, the average effect of, e.g., QSO on LRG targets as a function of $n_{\rm tile}$ is learned, and then the FFA assignments are based on the local density of the given tracer type and $n_{\rm tile}$. Similarly, each ``assigned" target is given a probability of assignment, $p_{\rm obs}$, and $w_{\rm comp}$ for FFA is simply its inverse. There is no decomposition into $f_{\rm TLID}$ and $f_{\rm tile}$.
    
\end{itemize}

We produce \texttt{AbacusSummit} mocks in all three flavors. For the \texttt{EZmocks}, we only apply FFA. For altmtl and FFA mocks, to simulate redshift failures, we simply select a random fraction of the ``observed'' redshifts to be failures, using a rate that provides a match to the number of good redshifts observed in each DR1 sample, to better than  99\%. All flavors produce `\texttt{full\_HPmapcut}' catalogs for data and randoms that are then passed to the LSS catalog pipeline in the same way to apply redshift cuts and the steps described in \cref{sec:weights}. The results are catalogs with the same datamodel as the real DR1 LSS catalogs, that are closely matched in the number of redshifts and the footprint they occupy. This allows the same clustering analysis pipeline to be applied consistently to all of them. For altmtl and FFA, we expect a close match between the signal-to-noise of the resulting mock clustering measurements and real DR1 data. 

\cref{fig:PkmockELG} shows the mean multipole moments of the power spectrum obtained for ELG mocks with $1.1<z<1.6$. The results averaged across 1000 \texttt{EZmocks} are shown in gray; they were only generated using FFA. The complete mocks, with curves shown in black, have all targets given redshifts. One can see clear offsets between the complete mocks and all other versions, for all multipoles. This shows how much fiber assignment biases the power spectrum estimation. It is shown in \cite{KP3s5-Pinon} that when removing small-scale angular pairs via the $\theta$-cut method the power spectra obtained for complete and altmtl mocks are in good agreement. One can observe a close match between the results that apply FFA (red) and those that apply the full realism of the altmtl (blue). Their agreement is further studied, including for all tracers, in \cite{KP3s6-Bianchi}. The \texttt{EZmocks} do not approximate small-scale clustering accurately, but one can see that they produce power spectra that are a good match to those from \texttt{AbacusSummit} out to $k\sim 0.25\;h\,$Mpc$^{-1}$. However, despite the match in clustering amplitude, we find that the FFA mocks return clustering measurements with less variance than the altmtl mocks, see \cref{sec:cov}. Further details on the \texttt{EZmocks}, for all tracers, can be found in \cite{KP3s8-Zhao} and \cite{KP4s6-Forero-Sanchez}. In the following section, we will compare the clustering measurement of the DR1 data to the mean obtained from altmtl mocks, for all tracers.

\section{DESI DR1 raw 2-point Clustering Measurements}
\label{sec:clus}
In this section, we present the `raw' multipoles of 2-point clustering of DESI tracers in Fourier- and configuration-space and compare to the same measurements obtained from DESI DR1 mock LSS catalogs. As discussed in \cref{sec:2pt}, these `raw' measurements refer to the 
results obtained without applying any $\theta$-cut or rotation and without BAO reconstruction. We always apply the weighting $w_{\rm tot}w_{\rm FKP}$ to data and random points. The 2-point functions displayed throughout this section are not those used for the BAO measurements of \cite{DESI2024.III.KP4} (they used BAO reconstruction) and are not those used by \cite{DESI2024.V.KP5} to constrain cosmological models (they applied  the $\theta$-cut, RIC and AIC corrections, and rotations). However, \cite{DESI2024.V.KP5} obtain their 2-point measurements from the same LSS catalogs and redshift bins as we use in this section and we thus expect any comparison with similarly treated simulated data would yield consistent results.\footnote{Ref. \cite{DESI2024.III.KP4} used an earlier version of the catalogs, v1.2 compared to v1.5, and recovered nearly identical BAO scale measurements from the two versions.}

We compare the results of the DESI DR1 2-point clustering to the mean measured on the 25 `altmtl' mocks; these have had fiber assignment and an  LSS pipeline applied that is fully consistent with the DR1 LSS pipeline. We thus expect the results to match to the DESI data, statistically, if the underlying input mocks are properly matched, as all window functions and integral constraints that could bias the clustering measurements are the same. We expect that any modeling framework that can obtain unbiased results on these mocks would also obtain unbiased results on the data.\footnote{Here, we are assuming the modeling framework has been demonstrated to already perform well on mocks without such observational complexity.} Specifically, we obtain $\chi^2$ values via
\begin{equation}
    \chi^2 = (\vec{X}_{\rm data} -A\bar{\vec{X}}_{\rm mock}) \tens{C}^{-1} (\vec{X}_{\rm data} -A\bar{\vec{X}}_{\rm mock})^{T},
\end{equation}
where $\vec{X}$ represents a given multipole of a 2-point measurement and $A$ is a scaling parameter that we use to test for the level of consistency in the clustering amplitude of the mocks and DESI data. The covariance matrix is obtained as described in \cref{sec:cov} and we do not apply any corrections to the Fourier-space results based on biases due to the use of 1000 mock realizations (given at most 80 measurement bins, any corrections are less than 10\%).

We present the results of each tracer within the subsections that follow, going in order of redshift. We make comparisons over a range of scales---$20 < s < 200\Mpch$ and $0<k<0.4\hMpc$ in configuration and Fourier space respectively---that is is larger than the ranges used in \cite{DESI2024.III.KP4} and \cite{DESI2024.V.KP5} for cosmological fits, and we also present comparisons for the hexadecapole, which is not used in those papers. In cases where the clustering results from the mocks disagree with the data, we test whether a rescaling factor, $b_r^2$ for the monopole and $b_r$ for the quadrupole, significantly improves the agreement. When significant disagreement remains after the rescaling test, we test restricting the range of scales used for comparison. The results are summarized in Table \ref{tab:2ptcomp}. In all cases, we use the covariance matrices described in \cref{sec:cov}. In general, if we are able to find good agreement between the simulations and data with only moderate bias factors, this serves as a validation of the approach to produce the simulations, which relied on fits to the EDR clustering measurements \cite{bgs_hod,abacushod2,abacusHODELG}. Alternatively, assuming the inputs to the simulations were determined properly, finding good agreement implies that the fiber assignment and LSS catalog pipeline applied to the simulations indeed matches the process applied to the data and we are thus able to simulate these important effects.

\subsection{BGS}

\begin{figure*}
    \centering    \includegraphics[width=0.49\columnwidth]{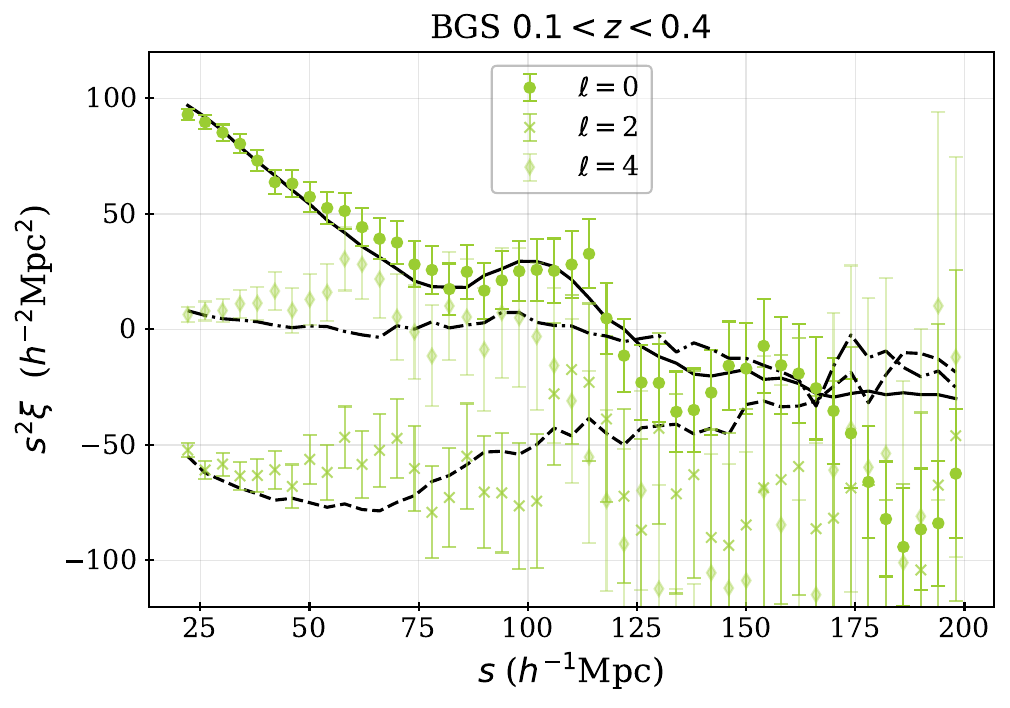}
    \includegraphics[width=0.49\columnwidth]{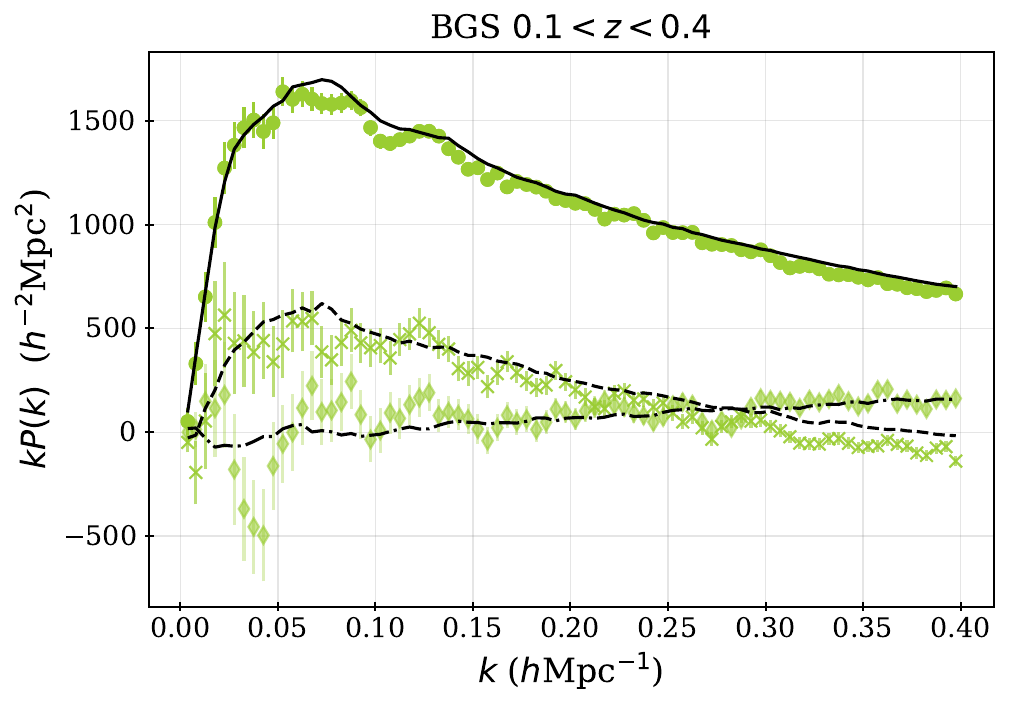}
\caption{`Raw' multipoles (with no $\theta$-cut or rotation applied), with respect to the line of sight, of the redshift-space 2-point correlation function and power spectrum. The curves display the mean on the same quantity measured on 25 `altmtl' mocks (see text for details). \cref{tab:2ptcomp} presents $\chi^2$ values for the comparison displayed here (over the full range of scales) and also the values obtained when reducing scale ranges and/or applying scaling factors to the mock results.
}
    \label{fig:2ptBGS}
\end{figure*} 

\cref{fig:2ptBGS} displays the 2-point clustering measurements for our DR1 BGS sample (points with error-bars), compared to the mean of the 25 `altmtl' mocks (curves). In configuration space, we find excellent statistical agreement, given that in the range $20<s<200\Mpch$ no multipole has a $\chi^2$ per degrees of freedom ($\chi^2/$dof) that is greater than 1.1, as can be seen in the first row of \cref{tab:2ptcomp}. In Fourier-space, the $\chi^2/$dof is greater than 1.5 for both the monopole and quadrupole. For the monopole, it can be reduced to\footnote{There are 80 measurement bins and we remove 1 dof based on the bias parameter fit.} 102.8/79 by scaling the mean of the mocks by a factor 0.983$^2$, and improved further to\footnote{We did not re-fit, but used the bias parameter fit to the full range and thus quote the value for 39 dof from 40 measurement bins.} 37.8/39 when restricting to $k<0.2\hMpc$. This implies good agreement in the shape of the monopole of power spectrum measurements, especially at larger scales, and that the bias of the mocks is $\sim$1.5\% higher than the data.

When consistently (to leading order) scaling the quadrupole by $b_r=0.983$, the $\chi^2$ slightly decreases to\footnote{The bias parameter was fit only to the monopole, so the dof are the same as the number of measurement bins, 80.} 127.2/80. This implies that how the Fourer-space quadrupole changes as a function of $k$ is inconsistent with the data. However, we find that for $k<0.2\hMpc$, the $\chi^2/$dof is 31.46/40 for the quadrupole. Thus, despite the substantial scatter around the mean that can be observed in quadrupole for $k<0.2\hMpc$, the results are statistically consistent with our DR1 mocks. A similar scatter is observed for the hexadecapole, but results in acceptable $\chi^2$ in all cases. 

\subsection{LRG}

\begin{figure*}
    \centering    \includegraphics[width=0.49\columnwidth]{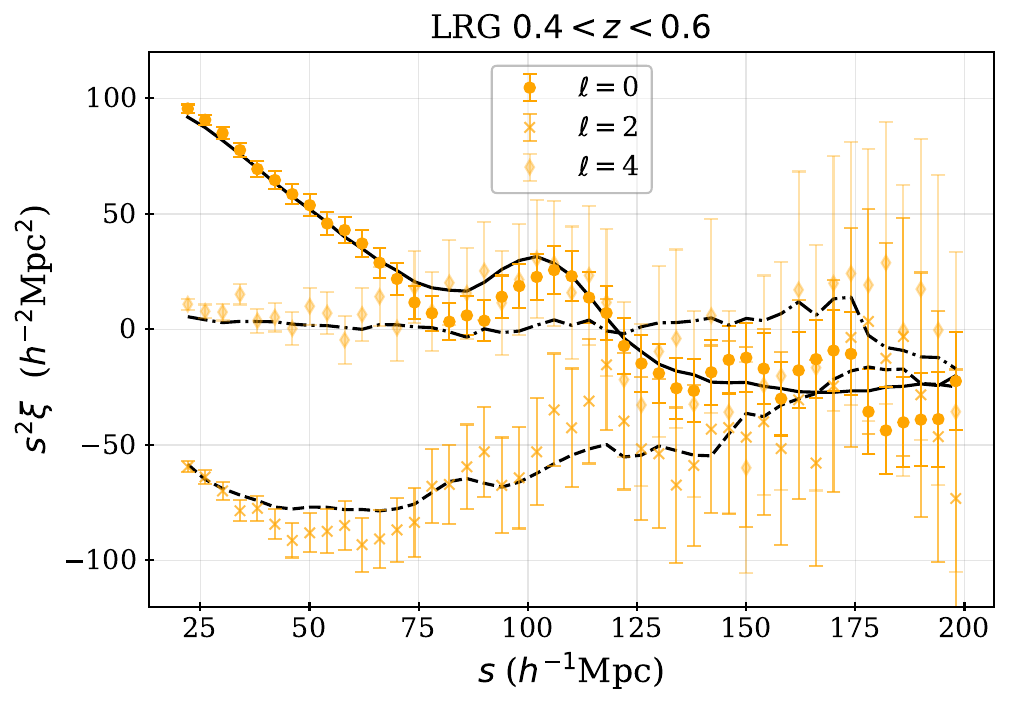}
    \includegraphics[width=0.49\columnwidth]{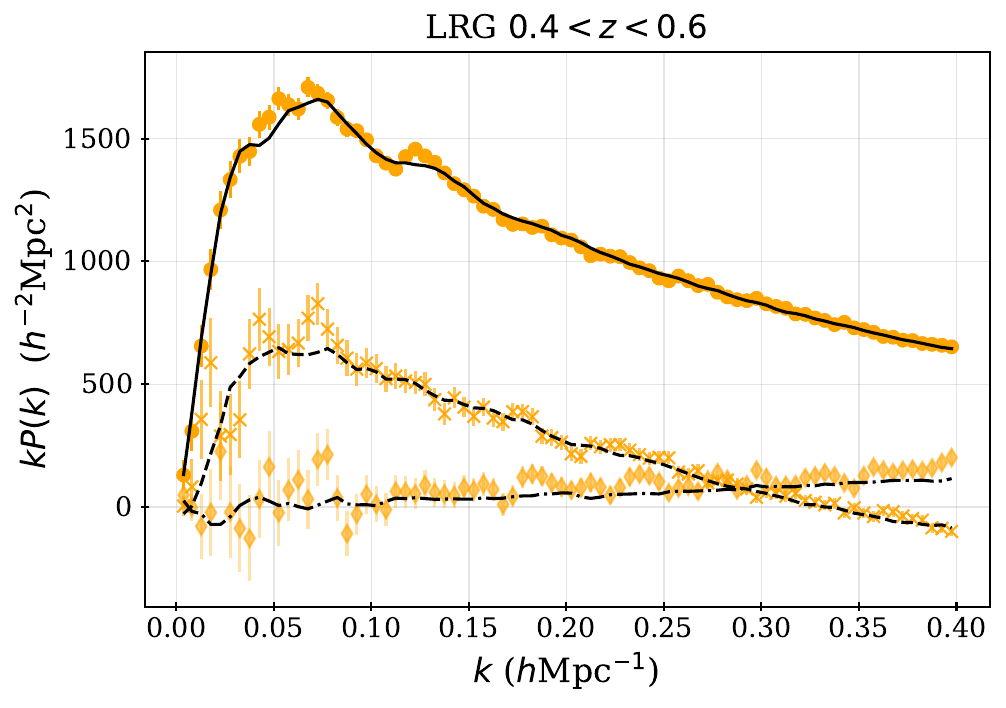}
    \centering    \includegraphics[width=0.49\columnwidth]{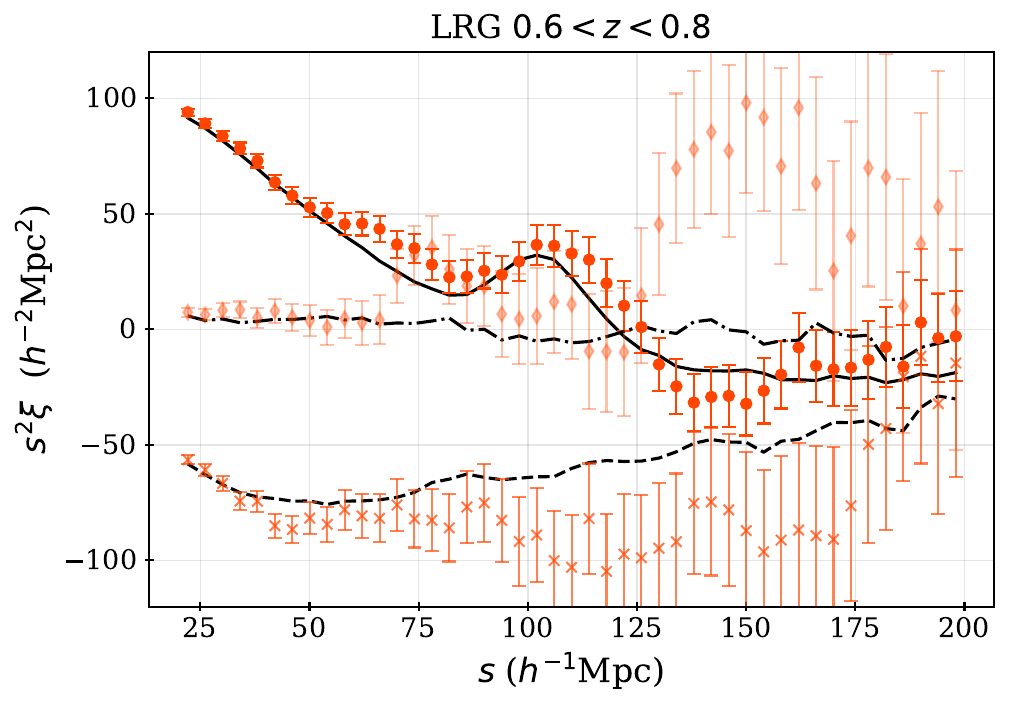}
    \includegraphics[width=0.49\columnwidth]{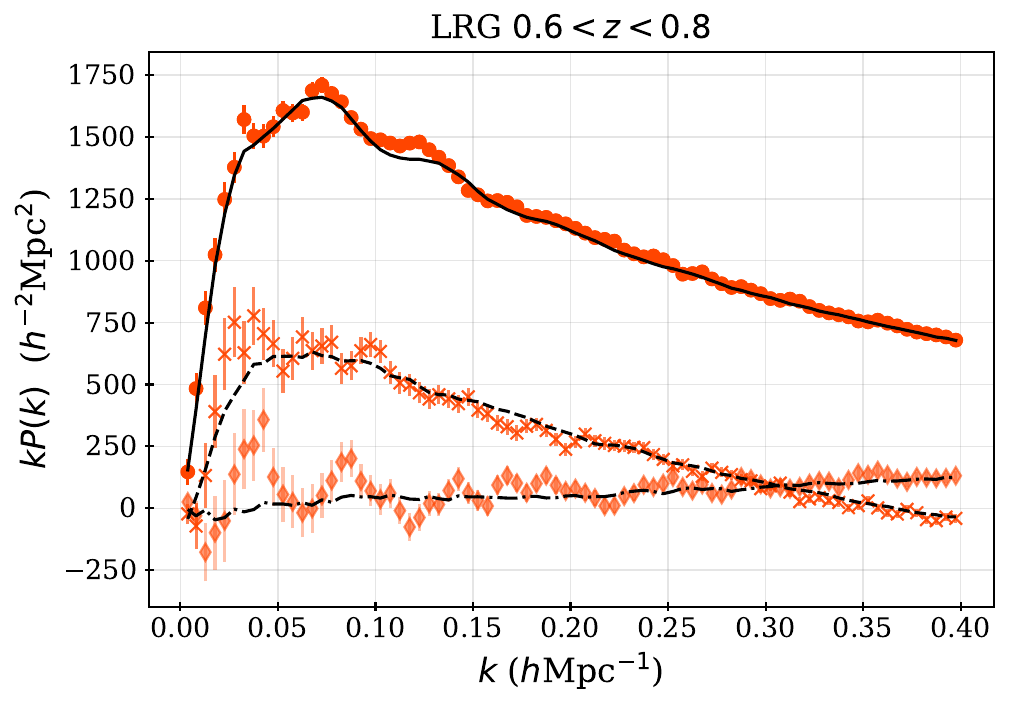}
    \centering    \includegraphics[width=0.49\columnwidth]{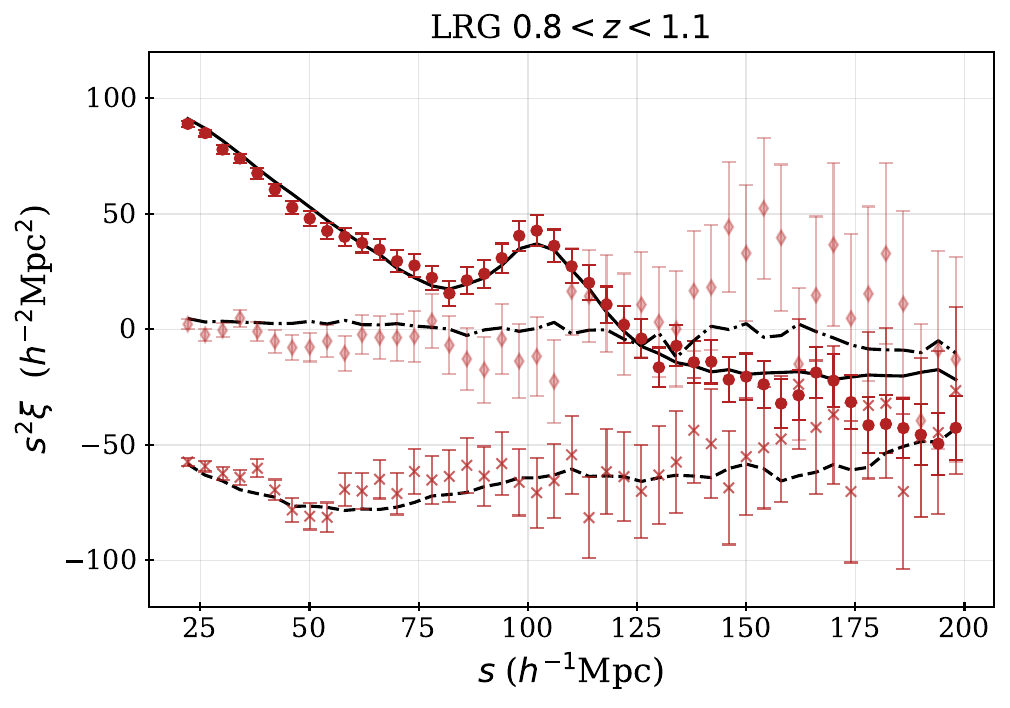}
    \includegraphics[width=0.49\columnwidth]{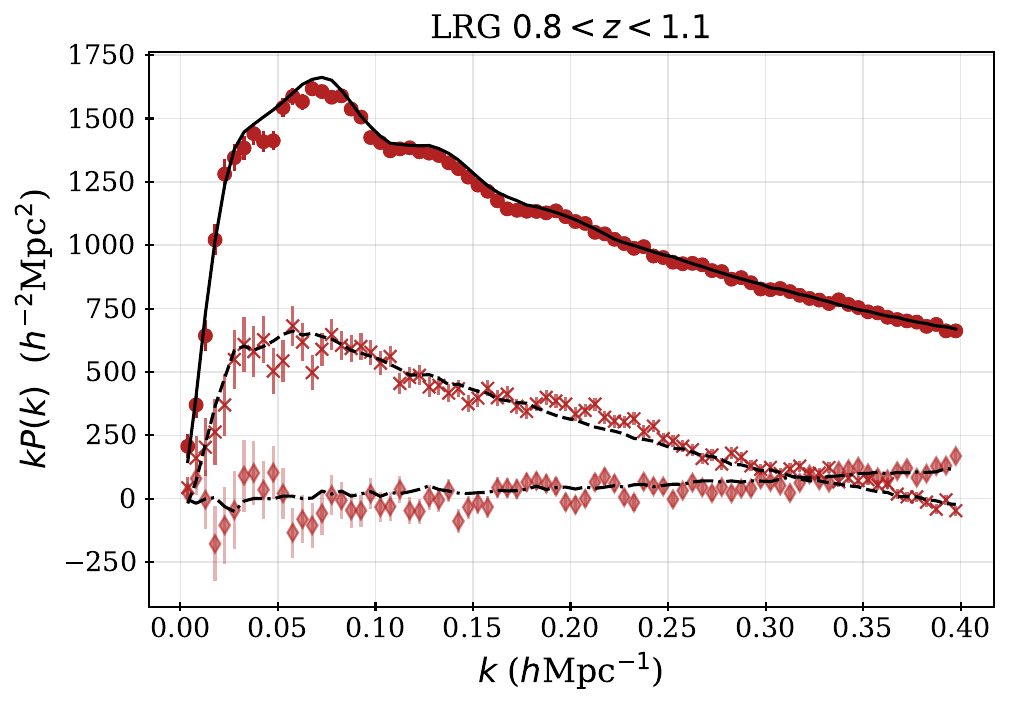}
\caption{Same as \cref{fig:2ptBGS}, but for LRGs.}
    \label{fig:2ptLRG}
\end{figure*}

\cref{fig:2ptLRG} displays the 2-point clustering measurements for our DR1 LRG sample, split into three redshift bins, $0.4<z<0.6$, $0.6<z<0.8$, and $0.8<z<1.1$. The left-hand panels display the results in configuration space, and we observe excellent statistical agreement between the mean of the mocks and the DR1 LRG data. The greatest $\chi^2/$dof found in \cref{tab:2ptcomp} is 54.6/45 for the monopole of the $0.6<z<0.8$ sample. One can observe, however, that the BAO peak appears at greater $h^{-1}$Mpc values in the data compared to the mocks. This corresponds to a smaller distance to these galaxies than expected in our fiducial DESI cosmology and is fully quantified (and enhanced using BAO reconstruction) in \cite{DESI2024.III.KP4}. 

The configuration space measurements have a strong correlation between $s$ bins. The impact of this is illustrated by the fact that all of the hexadecapole measured for the $0.6<z<0.8$ redshift bin appears $\sim2\sigma$ greater than the mock expectation for all $s>125\Mpch$, yet the $\chi^2$/dof is only 49.2/45. Similarly, the quadrupole in the same redshift bin appears coherently lower at all scales $s>40\Mpch$, yet the $\chi^2/$dof is only 40.6/45.

The statistical agreement between the DR1 LRGs and the altmtl mocks is somewhat worse for Fourier space, but the $\chi^2/$dof is less than or equal to 89.1/80 for 6 of the 9 cases considered. In the $0.6<z<0.8$ redshift bin, the monopole $\chi^2$ reduces from 103.8 to 97.0 when the mean of the mocks is multiplied by 1.006$^2$; i.e., the results are thus simply consistent with a 0.6\% difference between the bias of the data and mock LRG.

The $\chi^2/$dof values are greatest for LRGs in the $0.8<z<1.1$ redshift bin. Scaling the mean of the mocks can only improve the $\chi^2$ of the monopole to 122.1 and consistently scaling the quadrupole makes its $\chi^2$ slightly worse. The results in this bin thus are not consistent with a small difference in the linear galaxy bias. However, the $\chi^2/$dof are considerably improved when one considers only the $k<0.2\hMpc$ range. In this case, and when assuming a linear bias factor of 0.99 (fit to the monopole), the results are 47.7/39, 51.6/40, and 44.9/40 for the $\chi^2/$dof for the monopole, quadrupole, and hexadecapole, respectively.

\subsection{ELG}
\begin{figure*}
    \centering
    \includegraphics[width=0.49\columnwidth]{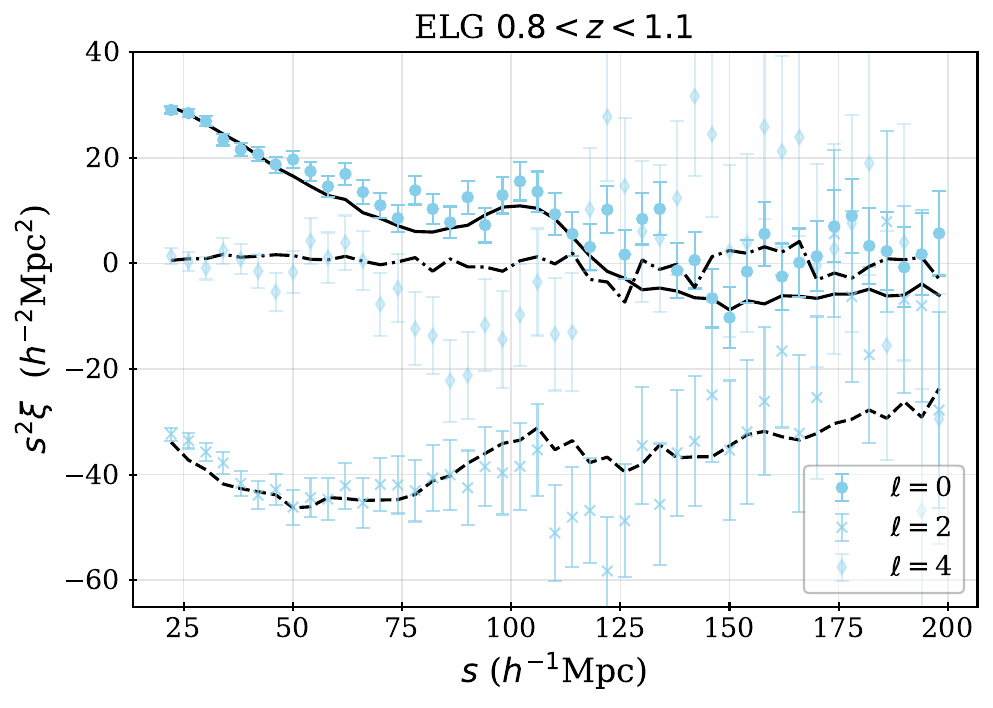}
    \includegraphics[width=0.49\columnwidth]{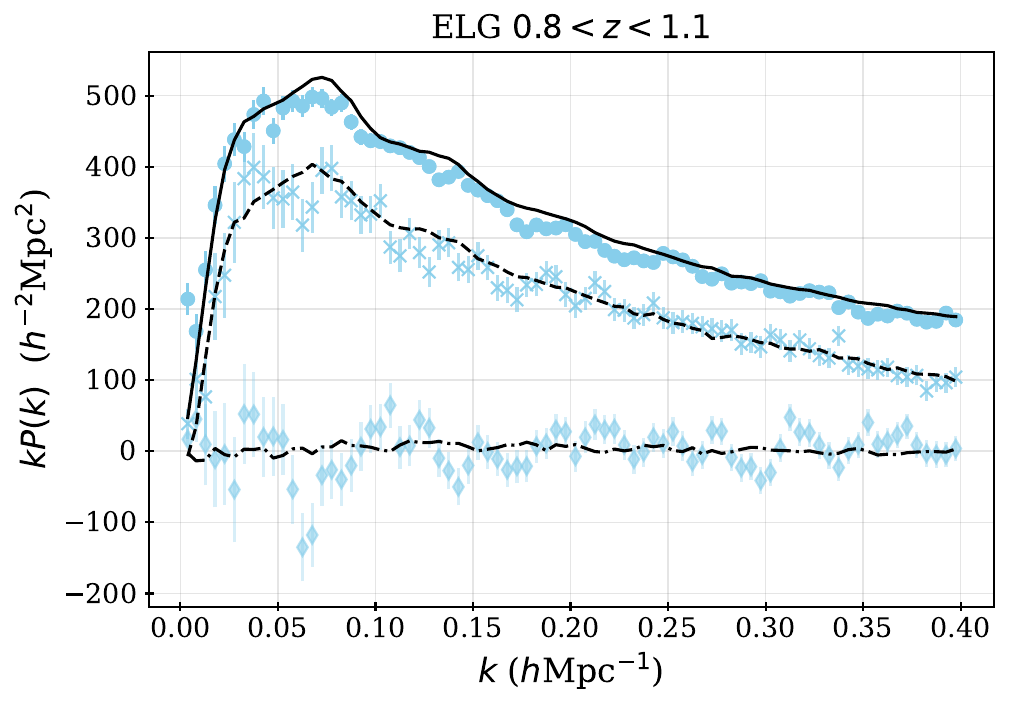}
    \includegraphics[width=0.49\columnwidth]{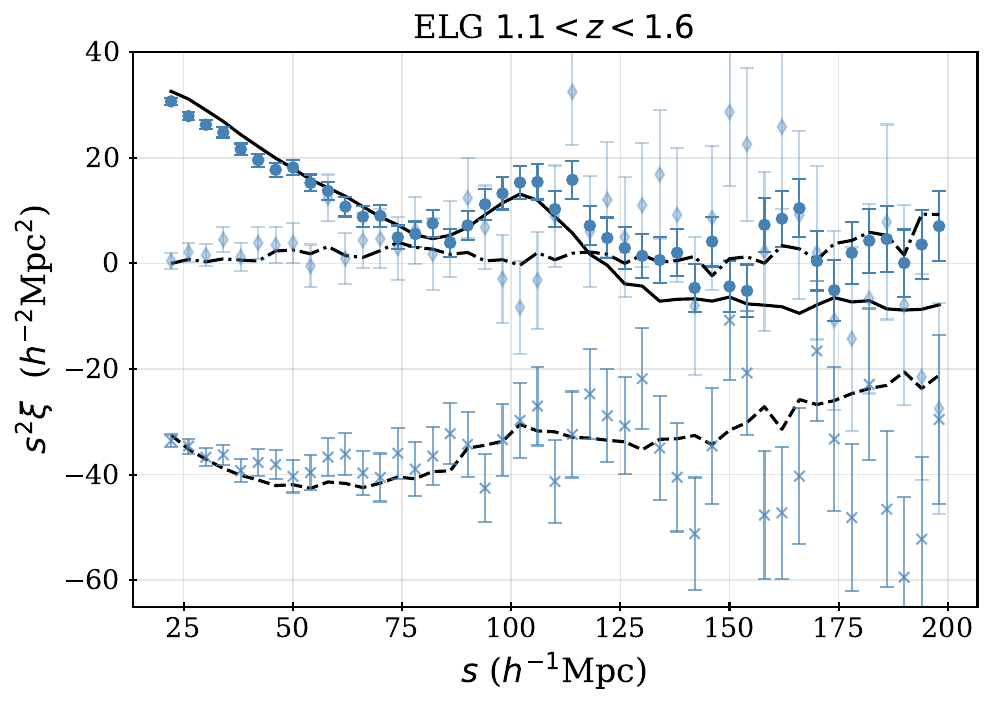}
    \includegraphics[width=0.49\columnwidth]{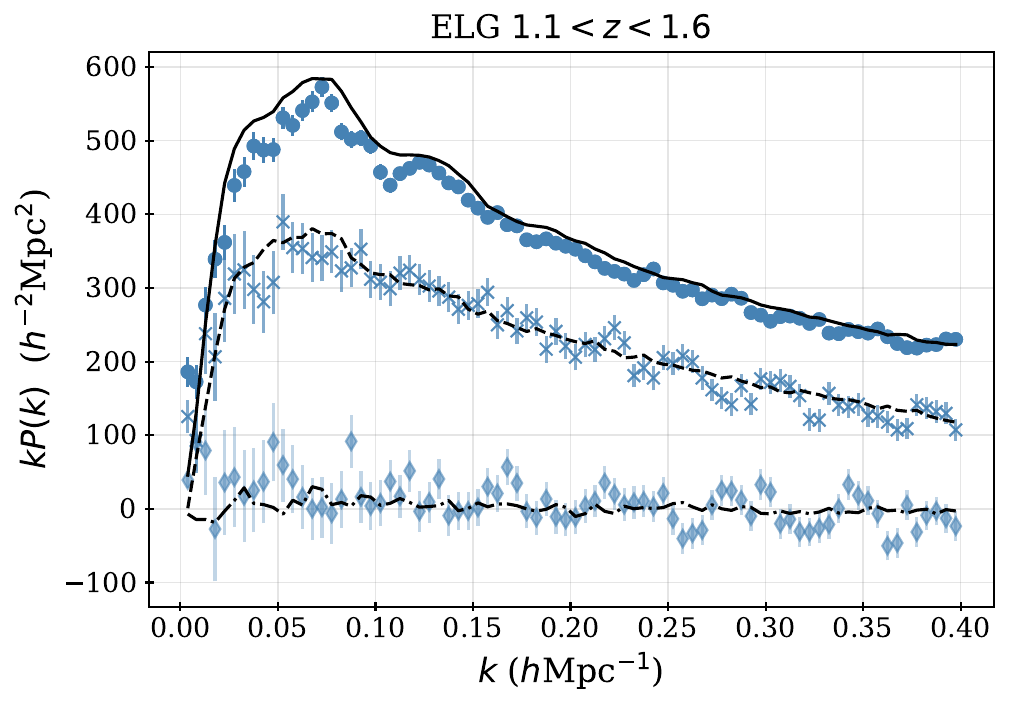}

\caption{Same as \cref{fig:2ptBGS}, but for ELGs.}
    \label{fig:2ptELG}
\end{figure*} 
\cref{fig:2ptELG} displays the 2-point clustering measurements for our DR1 ELG sample, split into two redshift bins, $0.8<z<1.1$ and $1.1<z<1.6$. In configuration space, the $\chi^2/$dof (again found in \cref{tab:2ptcomp}) is somewhat high for the monopole in both redshift bins. In the $0.8<z<1.1$ bin, the $\chi^2$/dof = 68.2/45 has a probability to exceed (PTE) value of 0.014, but we do not find significant improvement when rescaling the mean of the mocks. However, in the $1.1<z<1.6$ bin, the $\chi^2$ improves from 78.7 to 56.5, when the mocks are rescaled by 0.96$^2$. Consistently scaling the quadrupole by the 0.96 factor yields a slight increase in the $\chi^2$ from 59.1 to 59.8. The $1.1<z<1.6$ configuration space results are thus roughly statistically consistent with the mocks (the $\chi^2$ PTE values are 0.03 for $\xi_0$+$\xi_2$ and 0.05 for $\xi_0$+$\xi_2$+$\xi_4$, ignoring covariance between multiples), except for a 4\% mismatch in the linear bias.

In Fourier space, the disagreement between the monopoles of the data and the mean of the mocks is greater than in configuration space. For the monopole, the $\chi^2/$dof is nearly 3 in both redshift bins. 
In both redshift bins, we find that the combination of a small re-scaling of the mock clustering and restricting to $k>0.02\hMpc$ yields greatly improved $\chi^2$ values. In the $0.8<z<1.1$ bin, rescaling the monopole by 0.976$^2$ improves the $\chi^2$ from 220.9 to 151.6. Applying the same factor in configuration space produces a negligible change, as the $\chi^2$ decreases from 68.2 to 69.0; the Fourier-space results are clearly much more sensitive to the overall clustering amplitude. When further cutting the four measurement bins with $k<0.02\hMpc$ and applying the same 0.976 factor, the $\chi^2$ for the 76 measurement bins are 87.8, 69.7, and 93.1 for the Fourier-space monopole, quadrupole, and hexadecapole of the ELGs with $0.8<z<1.1$. The results in the $1.1<z<1.6$ bin are similar. Rescaling the mock monopole can only reduce the $\chi^2$ to 154.4 using a factor 0.979$^2$. Consistently scaling the quadrupole by 0.979 only reduces its $\chi^2$ to 133.7. However, restricting to $k>0.02\hMpc$ yields $\chi^2$ values of 92.0, 79.8, and 87.6 for the 76 measurement bins for the monopole, quadrupole, and hexadecapole. The results suggest that the effective galaxy bias used for DR1 ELG simulations is just over 2\% too high and that residual imaging systematics impact the measurements at the largest scales (lowest $k$), as we detail further below.

Excess clustering at large scales is often due to observational systematics. The ELG clustering is affected strongly by imaging systematics, which is detailed in \cite{KP3s2-Rosado}. It is likely that the excess power at $k<0.02\hMpc$ observed in comparison to the mocks is due to residual observational systematics that are present even after the imaging corrections described in previous sections. The strongest source of systematic variation is from the systematic change in Galactic extinction, which can be observed in the trends displayed in \cref{fig:elg_imsys_mocks}. The $\chi^2$ differences between the clustering measured without accounting for this systematic variation and that with it is are greater than 2500 in both redshift bins \cite{KP3s2-Rosado}. Despite this enormous difference, the BAO results are nearly identical, suggesting that the changes in the clustering are fully absorbed by the broadband terms in the BAO modeling. However, any studies that use broadband information must be more careful. In order to obtain structure growth measurements and account for residual systematic uncertainties, \cite{KP5s6-Zhao} develop a method that allows an additional systematic component to be added to the modeling. The systematic component is based on a smooth fit to the difference between the weighted and unweighted data in Fourier space. We recommend a similar treatment (and validation) for any study that uses the broadband information of the DESI DR1 ELG 2-point clustering measurements.

\subsection{QSO}

\begin{figure*}
    \centering    \includegraphics[width=0.49\columnwidth]{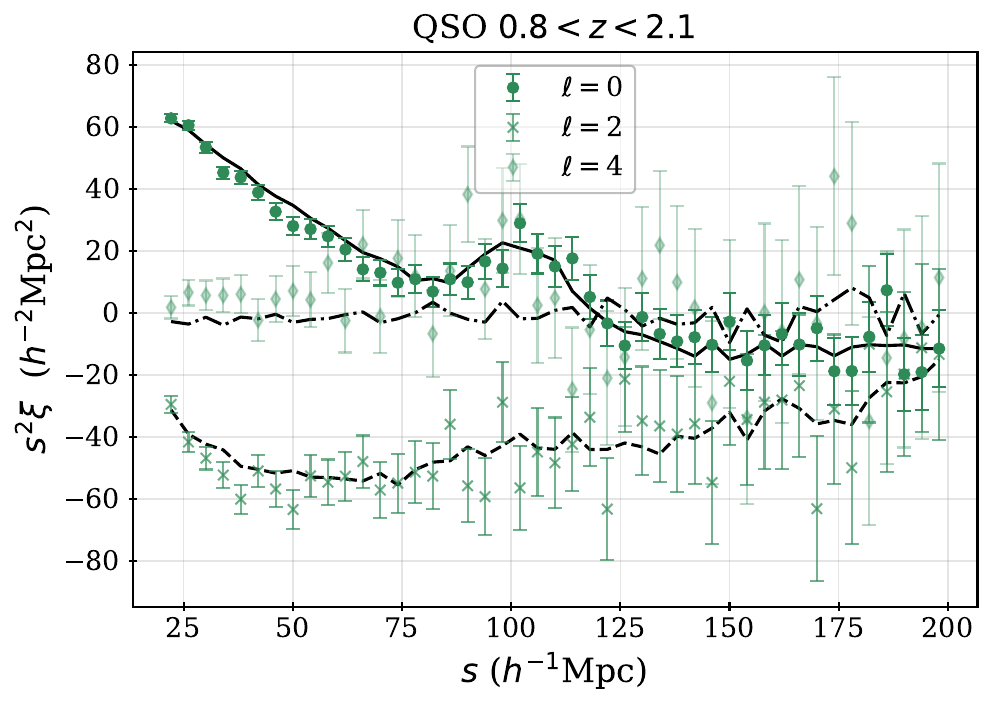}
    \includegraphics[width=0.49\columnwidth]{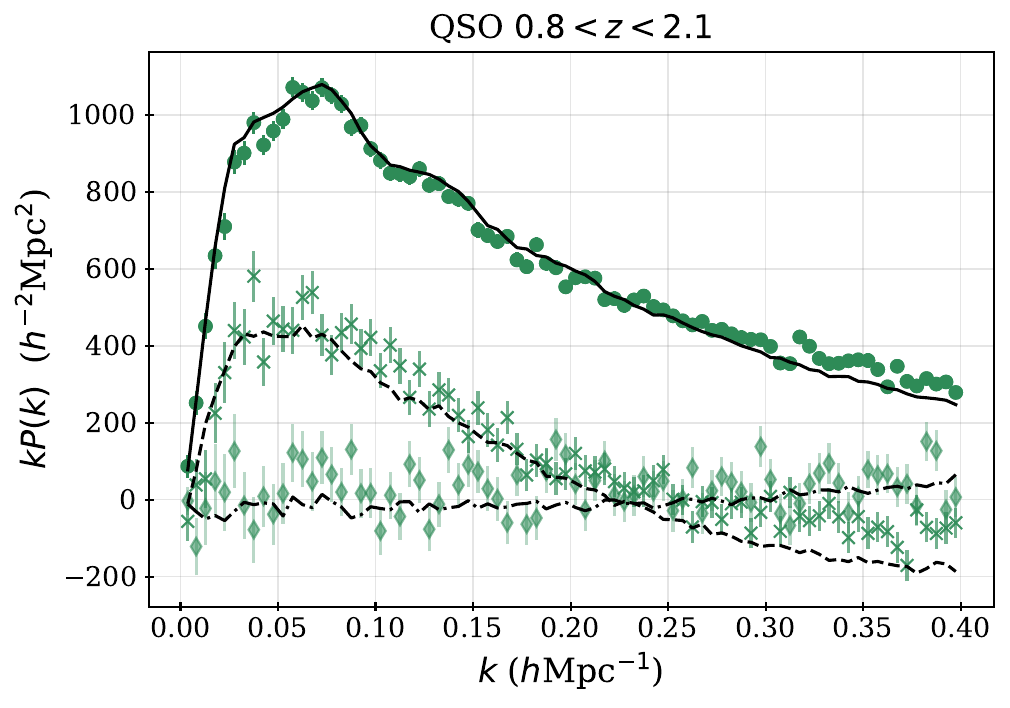}
\caption{Same as \cref{fig:2ptBGS}, but for QSO.}
    \label{fig:2ptQSO}
\end{figure*} 

\cref{fig:2ptQSO} displays the 2-point clustering measurements for our DR1 QSO sample (points with error-bars), compared to the mean of the 25 `altmtl' mocks (curves). In configuration-space, the multipoles are statistically consistent, with only the hexadecapole having a $\chi^2/$dof (once more found in \cref{tab:2ptcomp}) that is greater than 1.

In Fourier space, the quadrupole is most inconsistent. One can observe the amplitude in the mocks is lower and that it crosses 0 at a lower $k$ value. A linear rescaling thus does not provide significant improvement. Cutting to $k<0.2\hMpc$ does provide significant improvement for the monopole and quadrupole, as the $\chi^2/$dof become 65.6/40 and 47.7/40, while the hexadecapole becomes 58.1/40. The monopole $\chi^2$ can be reduced to 49.3 when applying scaling factor 0.988$^2$, which slightly increases the quadrupole $\chi^2$ to 49.6. Further restricting to $k<0.15 h$Mpc$^{-1}$ (and applying no amplitude factor) yields $\chi^2/$dof 35.0/30, 40.1/30, and 36.5/30.

The mismatch between the DESI DR1 QSO clustering in data and simulations is clearly strongly scale dependent. The amplitude of the mock $P(k)$ becomes increasingly too small going to high $k$, and the effect is most prominent in the quadrupole. A significant factor in the QSO clustering is the redshift uncertainty. The characteristics of the mismatch are consistent with the expected results if slightly too large of a redshift uncertainty was included in the mocks. Otherwise, the results are in reasonable agreement.

\begin{table*}
\centering
\begin{tabular}{|l|cccccc|}
\hline%\hline
Tracer & statistic & range & $b_f$ & $\chi^2$/dof mono. & $\chi^2$/dof quad. & $\chi^2$/dof hexadeca. \\\hline
BGS & $\xi(s)$  & 20-200 & 1 & 49.6/45 & 50.5/45 & 41.0/45\\
BGS & $P(k)$  & 0-0.4 & 1 & 126.5/80 & 127.7/80 & 69.8/80\\
BGS & $P(k)$  & 0-0.4 & 0.983 & 102.8/79 & 127.2/80 & 69.8/80\\
BGS & $P(k)$  & 0-0.2 & 0.983 & 37.8/39 & 31.6/40 & 29/40\\
LRG1 & $\xi(s)$  & 20-200 & 1 & 42.1/45 & 35.5/45 & 54.1/45\\
LRG1 & $P(k)$  & 0-0.4 & 1 & 88.7/80 & 79.7/80 & 87.9/80\\
LRG2 & $\xi(s)$  & 20-200 & 1 & 54.6/45 & 40.6/45 & 49.2/45\\
LRG2 & $P(k)$  & 0-0.4 & 1 & 103.8/80 & 81.7/80 & 79.0/80\\
LRG2 & $P(k)$  & 0-0.4 & 1.006 & 97.0/79 & 82.0/80 & 79.0/80\\
LRG3 & $\xi(s)$  & 20-200 & 1 & 42.1/45 & 51.2/45 & 47.8/45\\
LRG3 & $P(k)$  & 0-0.4 & 1 & 105.5/80 & 121.1/80 & 89.1/80\\
LRG3 & $P(k)$  & 0-0.4 & 0.993 & 95.0/79 & 121.5/80 & 89.1/80\\
LRG3 & $P(k)$  & 0-0.2 & 0.99 & 37.7/39 & 43.5/40 & 40.9/40\\
ELG1 & $\xi(s)$  & 20-200 & 1 & 68.2/45 & 41.2/45 & 58.8/45\\
ELG1 & $P(k)$  & 0-0.4 & 1 & 220.9/80 & 86.4/80 & 104.8/80\\
ELG1 & $P(k)$  & 0-0.4 & 0.976 & 151.6/79 & 83.6/80 & 104.8/80\\
ELG1 & $P(k)$  & 0.02-0.4 & 0.976 & 87.8/79 & 69.7/80 & 93.1/80\\
ELG2 & $\xi(s)$  & 20-200 & 1 & 78.7/45 & 59.1/45 & 46.5/45\\
ELG2 & $\xi(s)$  & 20-200 & 0.96 & 56.5/44 & 59.8/45 & 46.5/45\\
ELG2 & $P(k)$  & 0-0.4 & 1 & 234.3/80 & 135.7/80 & 107.5/80\\
ELG2 & $P(k)$  & 0-0.4 & 1 & 234.3/80 & 135.7/80 & 107.5/80\\
ELG2 & $P(k)$  & 0-0.4 & 0.979 & 154.4/79 & 133.7/80 & 107.5/80\\
ELG2 & $P(k)$  & 0.02-0.4 & 0.979 & 92.0/75 & 79.8/80 & 87.6/80\\
QSO & $\xi(s)$  & 20-200 & 1 & 35.7/45 & 31.5/45 & 49.7/45\\
QSO & $P(k)$  & 0-0.4 & 1 & 145.3/80 & 163.0/80 & 115.9/80\\
QSO & $P(k)$  & 0-0.2 & 1 & 65.6/40 & 47.7/40 & 58.1/40\\
QSO & $P(k)$  & 0-0.2 & 0.988 & 49.3/39 & 49.6/40 & 58.1/40\\
QSO & $P(k)$  & 0-0.15 & 1 & 35.0/30 & 40.1/30 & 36.5/40\\
QSO & $P(k)$  & 0-0.15 & 0.988 & 26.2/29 & 41.8/30 & 36.5/40\\
\hline
\end{tabular}
\caption{Assessment of the agreement between the mean of altmtl mock clustering statistics and the results obtained with the data. The value of $b_f$ represents how different the linear bias of the mocks is from the data; i.e., a value of 0.99 implies the data has a 1\% lower linear bias factor than was input into the mocks. These values were fit to the monopole and thus remove 1 dof from that comparison. The scales used for the 2-point correlation function, $\xi(s)$, is the redshift-space separation in Mpc/$h$, while for the power spectrum, $P(k)$, it is the wave number in $h{\rm Mpc^{-1}}$. The redshift ranges used for LRG1, LRG2, and LRG3 are $0.4<z<0.6$, $0.6<z<0.8$, and $0.8<z<1.1$. For ELG1 and ELG2, the redshift ranges are $0.8<z<1.1$ and $1.1<z<1.6$.
\label{tab:2ptcomp}}
\end{table*}

\section{Conclusions}
\label{sec:conclusions}

We have presented the details of LSS catalogs obtained from the DESI DR1 data, and their validation for use in DESI 2024 cosmological analyses. The catalogs contain over 5.7 million unique tracers and will be made publicly available with DESI DR1.

We have presented the methodology for the measurement of the multipoles of 2-point functions used in DESI 2024 cosmological analyses in \cref{sec:2pt}. The raw clustering measurements obtained from these LSS catalogs are biased compared to naive theoretical expectations, due to a combination of effects from fiber assignment incompleteness, survey geometry, and integral constraint effects. Where necessary, corrections for these effects, through the combination of the application of window functions to models and correction terms to the 2-point clustering are detailed in \cref{sec:2pt}. Specifically:
\begin{itemize}
    \item The raw clustering measurements of post-reconstruction configuration-space 2-point functions were demonstrated to provide unbiased BAO measurements \cite{DESI2024.III.KP4}.
    \item Fiber assignment incompleteness biases both the total and relative amplitudes of the measured 2-point multipoles. When fitting the full-shape of DESI 2-point functions in \cite{DESI2024.III.KP4}, this effect is mitigated by the removal of clustering information from angular scales less than 0.05 degrees, directly in the estimator for the measurements and in the window function for the model, with full details presented in \cite{KP3s5-Pinon}.
    \item The survey geometry is accounted for using a standard window function application in Fourier-space.
    \item The LSS catalog randoms sample the data to obtain radial information. This induces a radial integral constraint (RIC) and nulls purely radial modes in the clustering measurements. The corrections applied for imaging systematics induce stronger angular integral constraints (AIC) than would otherwise be present. We describe how an empirical correction based on the results of simulations is applied to the measured power spectra used for cosmological constraints in companion papers. Any clustering analyses that use DESI DR1 LSS catalogs may have to derive their own corrections, specific to their own clustering measurements.
\end{itemize}

We have summarized how the selection functions and corrections for systematic sources of number density variation are determined for each tracer and provided recommendations for addressing the remaining systematic uncertainties and known biases. These include:
\begin{itemize}
    \item Observational systematic uncertainties remain in the data both due to DESI spectroscopic observations and issues in the imaging data used to obtain DESI targets.
    \item Imaging systematics have a significant impact on DESI clustering measurements, especially at low-$k$. This is quantified in \cite{KP3s2-Rosado}, where the fiducial measurements are compared to those without the imaging systematic weights derived in \cref{sec:imsys} are presented. No impact was found on BAO measurements. However, \cite{KP5s6-Zhao} find a significant impact for ELG and QSO full-shape measurements and derive a method to marginalize over the residual uncertainty from imaging systematics.
    \item We recommend any clustering analysis that uses DESI DR1 LSS catalogs perform similar studies of its sensitivity to imaging systematics.
    \item As summarized in \cref{sec:specsys}, significant trends in the DESI spectroscopic success with observing conditions and with instrumental properties are found and documented in \cite{KP3s3-Krolewski, KP3s4-Yu}. However, the impact on 2-point clustering measurements was shown to be negligible, as was the effect of catastrophic redshift errors. We recommend similar tests to be performed on the impact on any higher-order or alternative kind of clustering statistic.
 \end{itemize}

In \cref{sec:mocks}, we have summarized the simulations of the DESI DR1 data that have been produced. In \cref{sec:clus}, we compared the 2-point clustering measurements of the simulations that include realistic application of DESI fiber assignment (the `altmtl' mocks) to the DESI DR1 data measurements. The 2-point clustering of DESI DR1 data in configuration- and Fourier-space is generally consistent with that of simulations of DESI DR1, within a 2\% factor of galaxy bias. In Fourier-space, reducing the scale range to $k<0.2h$Mpc$^{-1}$ is necessary for reasonable agreement for BGS, LRG $0.8<z<1.1$, and QSO. The combination of galaxy bias factors and small-scale disagreement indicates that improvements can be made by updating the precise manner in which galaxies occupy dark matter halos in the simulations. For ELGs, reasonable agreement requires removing scales $k<0.02h$Mpc$^{-1}$, suggesting residual observational systematic contamination.

The catalogs we have presented are intended for cosmological analysis of 2-point functions on large-scales ($s\gtrsim 20 h^{-1}$Mpc, $k\lesssim 0.2 h$Mpc$^{-1}$). For higher-order and alternative clustering measurements that use similarly large scales that are significantly greater than those affected by fiber collisions ($\sim$0.05 degrees), we recommend using the same catalogs and using the information outlined above to account for any biases and/or systematic uncertainties in the measurements. To probe smaller scales, the effects of fiber assignment incompleteness must be corrected statistically, e.g., with PIP weights (see \cref{sec:fiberassigncomp}). DESI DR1 LSS catalogs that support PIP weights will be released as a separate version of the catalogs and are described in \cref{sec:pipcat}.

Despite the issues described above, observational systematic uncertainties are sub-dominant to the statistical uncertainties in the DESI 2024 cosmological analyses \cite{DESI2024.III.KP4,DESI2024.V.KP5}. While this study is being finalized, analysis of the nearly three years of data that will be released with Data Release 2 (DR2) has begun, which contains more than 20 million good extra-Galactic redshifts. For DR2 and beyond, DESI will continue to improve the LSS catalogs, with the aim of keeping observational systematic uncertainty sub-dominant while maximizing the signal-to-noise accessible for cosmological constraints.

\section*{Acknowledgements}

This material is based upon work supported by the U.S. Department of Energy (DOE), Office of Science, Office of High-Energy Physics, under Contract No. DE–AC02–05CH11231, and by the National Energy Research Scientific Computing Center, a DOE Office of Science User Facility under the same contract. Additional support for DESI was provided by the U.S. National Science Foundation (NSF), Division of Astronomical Sciences under Contract No. AST-0950945 to the NSF’s National Optical-Infrared Astronomy Research Laboratory; the Science and Technology Facilities Council of the United Kingdom; the Gordon and Betty Moore Foundation; the Heising-Simons Foundation; the French Alternative Energies and Atomic Energy Commission (CEA); the National Council of Humanities, Science and Technology of Mexico (CONAHCYT); the Ministry of Science, Innovation and Universities of Spain (MICIU/AEI/10.13039/501100011033), and by the DESI Member Institutions: \url{https://www.desi.lbl.gov/collaborating-institutions}.

The DESI Legacy Imaging Surveys consist of three individual and complementary projects: the Dark Energy Camera Legacy Survey (DECaLS), the Beijing-Arizona Sky Survey (BASS), and the Mayall z-band Legacy Survey (MzLS). DECaLS, BASS and MzLS together include data obtained, respectively, at the Blanco telescope, Cerro Tololo Inter-American Observatory, NSF’s NOIRLab; the Bok telescope, Steward Observatory, University of Arizona; and the Mayall telescope, Kitt Peak National Observatory, NOIRLab. NOIRLab is operated by the Association of Universities for Research in Astronomy (AURA) under a cooperative agreement with the National Science Foundation. Pipeline processing and analyses of the data were supported by NOIRLab and the Lawrence Berkeley National Laboratory. Legacy Surveys also uses data products from the Near-Earth Object Wide-field Infrared Survey Explorer (NEOWISE), a project of the Jet Propulsion Laboratory/California Institute of Technology, funded by the National Aeronautics and Space Administration. Legacy Surveys was supported by: the Director, Office of Science, Office of High Energy Physics of the U.S. Department of Energy; the National Energy Research Scientific Computing Center, a DOE Office of Science User Facility; the U.S. National Science Foundation, Division of Astronomical Sciences; the National Astronomical Observatories of China, the Chinese Academy of Sciences and the Chinese National Natural Science Foundation. LBNL is managed by the Regents of the University of California under contract to the U.S. Department of Energy. The complete acknowledgments can be found at \url{https://www.legacysurvey.org/}.

Any opinions, findings, and conclusions or recommendations expressed in this material are those of the author(s) and do not necessarily reflect the views of the U. S. National Science Foundation, the U. S. Department of Energy, or any of the listed funding agencies.

The authors are honored to be permitted to conduct scientific research on Iolkam Du’ag (Kitt Peak), a mountain with particular significance to the Tohono O’odham Nation. 

%%%%%%%%%%%%%%%%%%%%%%%%%%%%%%%%%%%%%%%%%%%%%%%%%%
\section*{Data Availability}

Data from the plots in this paper will be available on Zenodo as part of DESI's Data Management Plan.

%%%%%%%%%%%%%%%%%%%%%%%%%%%%%%%%%%%%%%%%%%%%%%%%%%
\bibliographystyle{JHEP}
\bibliography{DESI2024_updated25Sep,KP3_references}

\appendix
\section{Image Property Maps}
\label{sec:improp_maps}

The DESI LSS catalogs utilize property maps from a range of different sources to characterize systematics associated with the imaging used for DESI target selection, and to aid science analyses. These maps exist in HEALPixel \citep{Healpix} format which can be used directly for regression tests. But, for some analyses, the maps are also incorporated by determining the value of each map at the locations of points in the random catalogs described in Section 4.5 of \cite{TS.Pipeline.Myers.2023} (see also \cref{sec:target}). In \cref{tab:skymaps}, we list the full set of map names included with the LSS imaging properties product.\footnote{e.g. \url{https://github.com/desihub/LSS/blob/v1.2-DR1/py/LSS/imaging/sky_maps.py\#L64-L96}} These {\em map} names sometimes propagate into files associated with LSS analyses as {\em column} names.

Each map is associated with a particular file, and we now detail each map based on its associated filename, including how the maps were derived and the nature of the file's content. Each file is associated with a number listed in the ``File'' column of \cref{tab:skymaps}, and that table also summarizes some of the features of the maps.

\begin{table*}
\centering
\begin{tabular}{|l|l|c|c|c|c|}
\hline%\hline
Map/Column name (fig. label) & Map type & File & Resolution & Nested & Galactic \\
\hline
\texttt{HALPHA} &                 Halpha &  (1)  &     512 &   True &  True \\
\texttt{HALPHA_ERROR} &          Halpha &  (2)  &     512 &   True &  True \\
\texttt{HALPHA\_MASK} &           Halpha &  (3)  &     512 &   True &  True \\
\texttt{CALIB\_G} &          calibration &  (4)  &     128 &  False &  False \\
\texttt{CALIB\_R} &          calibration &  (5)  &     128 &  False &  False \\
\texttt{CALIB\_Z} &          calibration &  (6)  &     128 &  False &  False \\
\texttt{CALIB\_G\_MASK} &    calibration &  (4)  &     128 &  False &  False \\
\texttt{CALIB\_R\_MASK} &    calibration &  (5)  &     128 &  False &  False \\
\texttt{CALIB\_Z\_MASK} &    calibration &  (6)  &     128 &  False &  False \\
\texttt{EBV\_CHIANG\_SFDcorr} (`\texttt{EBVnoCIB}') &      EBV &  (7)  &    2048 &   True &  True \\
\texttt{EBV\_CHIANG\_LSS\_MASK} &    EBV &  (8)  &    2048 &   True &  True \\
\texttt{EBV\_MPF\_Mean\_FW15} &      EBV &  (9)  &    2048 &  False &  True \\
\texttt{EBV\_MPF\_Mean\_ZptCorr\_FW15} & EBV & (9) &  2048 &  False &  True \\
\texttt{EBV\_MPF\_Var\_FW15} &       EBV &  (9)  &    2048 &  False &  True \\
\texttt{EBV\_MPF\_VarCorr\_FW15} &   EBV &  (9)  &    2048 &  False &  True \\
\texttt{EBV\_MPF\_Mean\_FW6P1} &     EBV &  (9)  &    2048 &  False &  True \\
\texttt{EBV\_MPF\_Mean\_ZptCorr\_FW6P1} & EBV & (9) & 2048 &  False &  True \\
\texttt{EBV\_MPF\_Var\_FW6P1} &      EBV &  (9)  &    2048 &  False &  True \\
\texttt{EBV\_MPF\_VarCorr\_FW6P1} &  EBV &  (9)  &    2048 &  False &  True \\
\texttt{EBV\_SGF14} &                EBV & (10) &      512 &  False &  True \\
\texttt{EBV\_SGF14\_MASK} &          EBV & (10) &      512 &  False &  True \\
\texttt{BETA\_ML} &                  EBV & (11) &      256 &   True &  True \\
\texttt{BETA\_MEAN} &                EBV & (11) &      256 &   True &  True \\
\texttt{BETA\_RMS} &                 EBV & (11) &      256 &   True &  True \\
\texttt{HI} &                        NHI & (12) &     1024 &  False &  True \\
\texttt{KAPPA\_PLANCK} &           kappa & (13) &     2048 &  False &  True \\
\texttt{KAPPA\_PLANCK\_MASK} &     kappa & (14) &     2048 &  False &  True \\
\texttt{FRACAREA} &       pixweight-dark & (15) &      256 &   True &  False \\
\texttt{STARDENS} &             stardens & (16) &      512 &   True &  False \\
\texttt{ELG} &            pixweight-dark & (15)&       256 &   True &  False \\
\texttt{LRG} &            pixweight-dark & (15) &      256 &   True &  False \\
\texttt{QSO} &            pixweight-dark & (15) &      256 &   True &  False \\
\texttt{BGS\_ANY} &     pixweight-bright & (17) &      256 &   True &  False \\
\hline
\end{tabular}
\caption{\label{tab:skymaps} Sky maps potentially incorporated into the DESI LSS catalogs. Similar types of mask are collected with a key word (e.g. ``EBV" for a dust map). The ``Resolution" column refers to the nside at which the mask is stored using the HEALPix method \citep{Healpix}. The ``Nested" column is True (False) when the mask is stored in the nested (ring) HEALPix scheme. The ``Galactic" column is True (False) when the map is stored in Galactic (Equatorial) coordinates. The maps are detailed further in this appendix, labeled by the ``File" number.}
\end{table*}

\begin{itemize}
\item {\bf (1) Halpha\_fwhm06\_0512.fits; (2) Halpha\_error\_fwhm06\_0512.fits; (3) Halpha\_mask\_fwhm06\_0512.fits}: Maps of the intensity of H$\alpha$ emission at $6^\prime$ (FWHM) resolution, as originally compiled by \cite{Finkbeiner.2003},\footnote{See, e.g., the Legacy Archive for Microwave Background Data Analysis at \url{https://lambda.gsfc.nasa.gov/product/foreground/fg_halpha_get.html}} together with the associated error, and a mask indicating bad pixels.
\item {\bf (4) decam-ps1-0128-g.fits; (5) decam-ps1-0128-r.fits; decam-ps1-0128-z.fits}: $g-$, $r-$ and $z-$band systematic calibration residuals, in magnitudes, constructed by comparing LS stars to stars from Pan-STARRS1 (PS1) \cite{PS1.Chambers}, in magnitudes. The sense of the residuals is LS
minus PS1.
\item {\bf (7) Chiang23\_SFD\_corrected\_hp2048\_nest.fits}; and \\{\bf (8) Chiang23\_mask\_hp2048\_nest.fits}: Dust map from \cite{chiang_csfd}, and the associated mask of bad pixels.
\item {\bf (9) recon\_fw15\_final\_mult.fits}: Dust maps generated from a combination of stellar reddenings derived from PS1 and 2MASS \cite{2MASS.Skrutskie} photometry and Gaia EDR3 parallaxes \cite{Gaia.EDR3}, using the {\em Bayestar} stellar inference pipeline \cite{dust.green}. The stars are filtered so as to remove objects that are too nearby, and those that are possible extragalactic sources. Mean and variance maps are then generated for the part of the sky at Galactic Latitude $|b|>20^\circ$ that is within the PS1 footprint. Maps are generated both with a FWHM of $6.1^\prime$ and with a FWHM of $15^\prime$. See \cite{dust.mudur} for more details.
\item {\bf (10) ps1-ebv-4.5kpc.fits}: Dust map from \cite{Schlafly.2014}, and the associated mask of bad pixels.
\item {\bf (11) COM\_CompMap\_dust-commander\_0256\_R2.00.fits}: Thermal dust map information from \cite{Planck.Foregrounds} as served by the NASA/IPAC Infrared Science Archive.\footnote{\url{https://irsa.ipac.caltech.edu/data/Planck/release_2/all-sky-maps/foregrounds.html}} The included maps correspond to the dust emissivity index posterior maximum, mean and root mean square.

\item {\bf (12) NHI\_HPX.fits.gz}: All-sky H\,I column densities assembled by combining the Effelsberg-Bonn H\,I Survey and the third revision of the Galactic All-Sky Survey, as detailed in \cite{HI.Bekhti}.\footnote{See \url{cdsarc.u-strasbg.fr/viz-bin/qcat?J/A+A/594/A116\#/browse}}

\item {\bf (13) dat\_klm.fits; (14) mask.fits.gz}: Map of lensing convergence ($\hat{\kappa}_{LM}$) from Planck \cite{Planck.Lensing}, and the associated mask indicating bad pixels. The map values are, specifically, the mean-field-subtracted minimum-variance estimate from temperature and polarization.\footnote{See also \url{https://wiki.cosmos.esa.int/planck-legacy-archive/index.php/Lensing}} The map was cut to modes of $2 < l < 2048$ before being converted to the Healpix scheme.
\item {\bf (15) pixweight-1-dark.fits; (17) pixweight-1-bright.fits}: Maps of values of systematics and of the density (${\rm deg}^{-2}$) of targets in the dark- and bright-time portions of the DESI Main Survey. Derived from the \textsc{desitarget} random catalog code.\footnote{See \url{https://github.com/desihub/desitarget/blob/1.1.1/py/desitarget/randoms.py\#L1076-L1127}}
\item {\bf (16) stardens.fits}: Map of the density (${\rm deg}^{-2}$) of stars from Gaia DR2 \cite{GaiaDR2}, calculated using the \textsc{desitarget} \texttt{randoms.stellar_density()} function.\footnote{See \url{https://github.com/desihub/desitarget/blob/1.1.1/py/desitarget/randoms.py\#L942}} Limited to point-like sources in the range $12 \leq G < 17$.
\end{itemize}

To produce the \textsc{Healpix} maps used for DESI DR1 regression analysis, the values of the maps listed above are queried at each position in the random catalogs. These random catalogs already contained information from the Legacy Survey DR9 imaging data, including the PSF size and depth in each band.\footnote{As detailed here \url{https://www.Legacy Survey.org/dr9/files/\#randoms-1-fits}.} Columns that we use to build \textsc{Healpix} maps include
\begin{itemize}
    \item \texttt{PSFSIZE\_<band>} 
    \item \texttt{PSFDEPTH\_<band>} 
    \item \texttt{GALDEPTH\_<band>} .
\end{itemize}

After all of the desired information is matched to each random point, we create a set of maps at \textsc{Healpix} resolution \textsc{Nside}=256 in \textsc{Nested} format by averaging the values associated with each random point within each pixel. In this way, we can obtain the mean value in each pixel in a manner that is fully consistent with the footprint of the LSS sample. The random catalogs have small differences in their footprints due to differences in the applied veto masks. Thus, the maps used for regression analysis differ slightly for each tracer (even when the map type is the same). The only exception is for how we use the $E(B-V)$ maps determined based on DESI stars in \cite{KP3s14-Zhou}. When using these maps, we simply take the original \textsc{Nside}=256 maps and convert them from \textsc{Ring} to \textsc{Nested} format.

\section{Catalog Version Changes after Unblinding}
\label{app:catver}
The LSS catalogs that were first used for unblinded DR1 BAO measurements are version `v1'. Afterward, small bugs were found and fixed that produced subsequent versions of the LSS catalogs. The first was that initially, groups of overlapping tiles found in the randoms but not the data were given zero completeness for the parameter $f_{\rm tile}$, defined in \cref{sec:compdef}. Such groups of tiles tend to be small areas with a large number of overlapping tiles, which would thus likely have a completeness of 1 if any targets existed in the area. Versions `v1.2' and higher have $f_{\rm tile}$ set to 1 in these regions. The effect was greatest on the QSO sample, as it is the most sparse and thus the most likely to have targets absent from any group of overlapping tiles. After fixing the issue, we confirmed that the large-scale clustering of the (blinded) QSO sample was consistent whether we applied $f_{\rm tile}$ to the randoms or 1/$f_{\rm tile}$ to the data. In `v1', before the fix, the large-scale power of the QSO sample is significantly higher for the (fiducial) case where we apply $f_{\rm tile}$ to the randoms. Version v1.2 of the LSS catalogs for the DESI cosmology results in \cite{DESI2024.III.KP4,DESI2024.VI.KP7A}.

A second issue was discovered where the `template signal-to-noise ratio' \texttt{TSNR2} threshold cut (defined in \cref{sec:hardware_veto}) was applied only when making the clustering catalogs. This was not a logically consistent choice, as applying such a threshold is equivalent to including the threshold in the `good hardware' definition (and is part of this definition in the final catalogs). This logical inconsistency propagated to small differences in the angular distribution of data and randoms. It falsely increased the determination of the number of overlapping tiles (as all good hardware instances contribute to the number of overlapping tiles). 
 This was fixed in version v1.3 (and all subsequent versions) by applying the \texttt{TSNR2} threshold as part of the good hardware definition. 
 
 Before any Full-Shape analyses were unblinded, one further issue was found. A correlation was induced between different subsets of randoms due to the same random seed being used for the assignment of radial information to the random catalogs. This was fixed by explicitly setting a different random seed. Therefore, a version v1.4 was created, using a new stream-lined pipeline script. After v1.4 was frozen, it was noticed that the updated pipeline script had not included the application of the custom mask for the ELG sample (but it had been applied to v1.2). Thus, a v1.5 was created, which is used for the Full-Shape analysis \cite{DESI2024.V.KP5}. We publicly release both versions v1.2 and v1.5.

\section{Catalogs for small-scale clustering measurements (PIP weighted)}
\label{sec:pipcat}

As discussed in \cref{sec:comp}, the fiducial approach to mitigating fiber assignment incompleteness in the DESI 2024 cosmological analysis is to remove the angular scales within a DESI fiber patrol radius, $\theta <0.05$ degrees \cite{KP3s5-Pinon}. The fiducial DR1 LSS catalogs described throughout are built to be optimal for this application. Alternatively, one can correct for the effects of fiber assignment incompleteness at any scale via the combination of PIP weights and angular upweights \cite{2017Bianchi}. The LSS catalogs that support such measurements must be constructed differently and are released as v1.5pip. We detail what is different about them below.

As described in \cref{sec:altmtl}, we obtain a total of 129 realizations of the fiber assignment of the DESI DR1 data; one for the real data and 128 using the altmtl method described in \cite{KP3s7-Lasker}. Each target in the full catalogs has information on whether it was assigned or not in each of the 128 altmtl realizations stored in a bit array, $w^{(b)}$ and the total probability of assignment stored as $p_{\rm obs} = (1 + c) / (1 + 128)$, with $c=\mathrm{popcnt}(w^{(b)})$ the number of 1 bits in $w^{(b)}$, i.e. the number of realisations in which the target is assigned. Using these data, the completeness weights for the data, $w_{\rm comp}$ are determined as 1/$p_{\rm obs}$; this is the individual inverse probability (IIP). The process outlined in \cref{sec:FKP} is then followed to obtain $w_{\rm tot}$ and $w_{\rm FKP}$, except the weights for the randoms are not multiplied by $f_{\rm tile}$, as the IIP weights account for any source of fiber assignment incompleteness.

The process described above provides the LSS catalogs with the needed data to measure PIP-weighted 2-point clustering measurements.  
Then, each pair weight, $w^{PIP}_{ij}$ is obtained as  
\begin{equation}\label{eq:w_pip}
w^{PIP}_{ij} = w_{\rm comp,i} w_{\rm comp,j}  \frac{w^{\rm eff}_{ij}}{g_{\rm 128}(c_i,c_j)} \ ,
\end{equation}
where $g_{\rm 128}$ (defined fully in \cite{KP3s6-Bianchi}) represents the expectation value of $w^{PIP}$ in the limit of independent probabilities and $w^{\rm eff}_{ij} = (1 + 128) / (1 + \mathrm{popcnt}(w_{i}^{(b)} \& w_{j}^{(b)}))$. As described in detail in \cite{KP3s6-Bianchi}, this expression is more robust to scale-dependent artifacts induced by the combination of low probabilities and low number of realisations of the targeting.
The randoms are still weighted by $w_{\rm tot}$.

\begin{figure*}
    \centering
    \includegraphics[width=0.49\columnwidth]{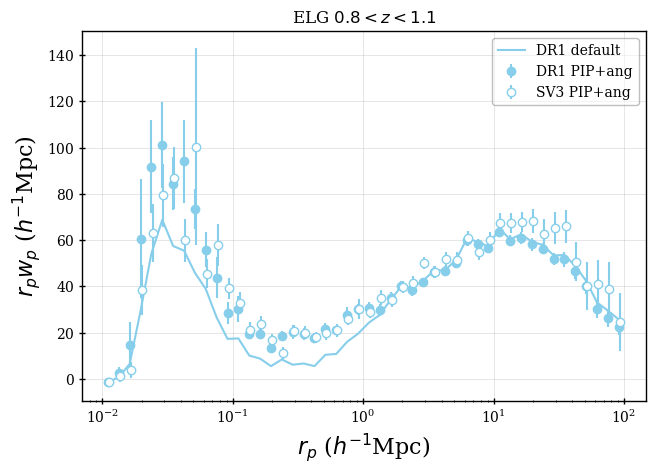}
    \includegraphics[width=0.49\columnwidth]{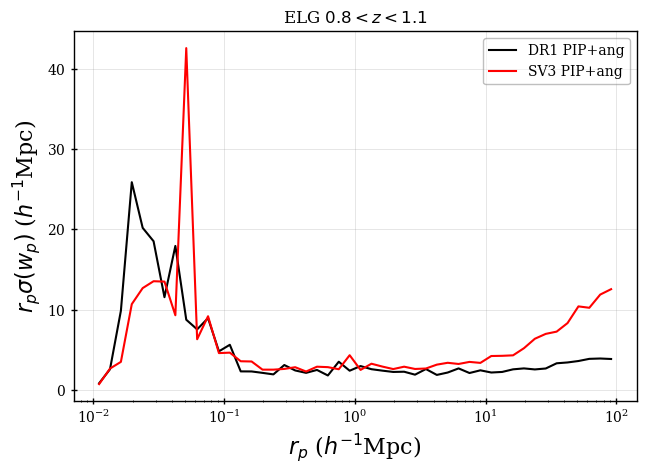}

\caption{For ELG data with $0.8<z<1.1$, the left-hand plot displays the projected clustering when using PIP and angular up-weighting comparing the results from EDR and DR1 data (open vs. filled symbols). The EDR points have their $r_p$ values slightly offset for legibility. We also compare to the results obtained using the default DR1 weights, which are known to not properly correct the measurements on small scales. The right-hand plot compares uncertainty recovered from 128 jack-knife realizations of the same DR1/EDR PIP and angular up-weighted projected clustering measurements.
}
    \label{fig:ELGwppipcomp}
\end{figure*}

For DESI DR1 LSS catalogs, angular up-weighting is additionally necessary. This is due to much of the area being covered by only 1 or 2 tiles, which leads to a large number of pairs that have 0 probability and thus will not be included in the PIP weights. The angular up-weighting uses the angular separation of pairs in the `full' LSS catalog, using all of the data (`parent' in \cite{2017Bianchi}), and also the selection of the data that was assigned a fiber and observed (`fibered' in \cite{2017Bianchi}). To do so consistently, we must include weights that account for the variation of $w_{\rm tot}/w_{\rm comp}$ and $w_{\rm FKP}$ with $n_{\rm tile}$. The mean value of these quantities in the data catalog is determined as a function of $n_{\rm tile}$ and is added as two new columns to the full LSS catalog.\footnote{They are \texttt{WEIGHT\_NTILE} and \texttt{WEIGHT\_FKP\_NTILE}.} These weights are then applied when obtaining the pair counts used for the angular up-weighting. In practice, angular up-weights $DD_{\mathrm{parent}}(\theta) / DD_{\mathrm{fibered}}(\theta)$ and $DR_{\mathrm{parent}}(\theta) / DR_{\mathrm{fibered}}(\theta)$ are pre-computed as a function of angular separation $\theta$ with a logarithmic binning (40 bins from $10^{-4}$ to $10^{0.5} \; \mathrm{deg}$). When running pair counts with angular up-weights, these weights are interpolated linearly at the cosine separation $\cos{\theta}$ between the galaxies of each pair.

To illustrate the impact of PIP and angular up-weighting on the measured clustering signal, we present the measured projected clustering applying PIP and angular up-weights to the DR1 v1.5pip catalogs for ELGs with $0.8<z<1.1$  in \cref{fig:ELGwppipcomp}. We compare the results to those obtained from the same weighting of EDR (SV3) \cite{abacusHODELG} and those obtained from the default catalogs (v1.5) and default weighting. The projected clustering $w_p(r_p)$ is defined by
\begin{equation}
    w_p(r_p) =  \int_{-r_{\pi,{\rm max}}}^{r_{\pi,{\rm max}}} \xi(r_p,r_\pi){\rm d}r_\pi,
\end{equation}
where $r_{\pi}$ is the line of sight separation $s\mu$ (with $s$ and $\mu$ defined in \cref{sec:xicalc}) and $r_p$ is the projected separation $\sqrt{1-\mu^2}s$. 
 One can observe that the PIP and angular up-weights are required to obtain small-scale clustering that is in rough agreement with the SV3 results. The sky area covered in SV3 was covered up to 13 times to minimize fiber assignment incompleteness, making it ideal for measuring small-scale clustering without major systematic concerns stemming from fiber assignment. The SV3 clustering at large scales is systematically greater than the DR1 clustering, which is likely due to some combination of correlated noise, uncorrected imaging systematics affecting the SV3 measurements, and variations in the intrinsic DESI ELG galaxy population with sky location. In the right-hand panel, we compare the uncertainty obtained from 128 jackknife realizations of the DR1 and SV3 clustering. One can observe that the uncertainty at small scales is similar for DR1 and SV3, but the improvement on large-scales is dramatic for DR1 compared to SV3. As DESI improves its coverage so that median number of overlapping tiles is greater than 4, the small-scale uncertainty will be dramatically improved.

\begin{figure*}
    \centering
    \includegraphics[width=0.49\columnwidth]{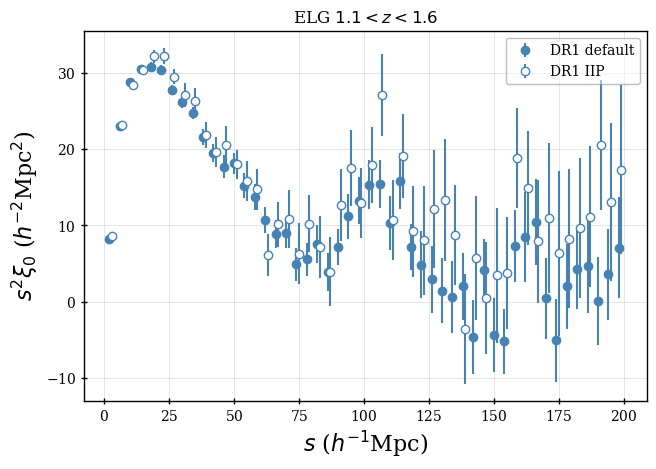}
    \includegraphics[width=0.49\columnwidth]{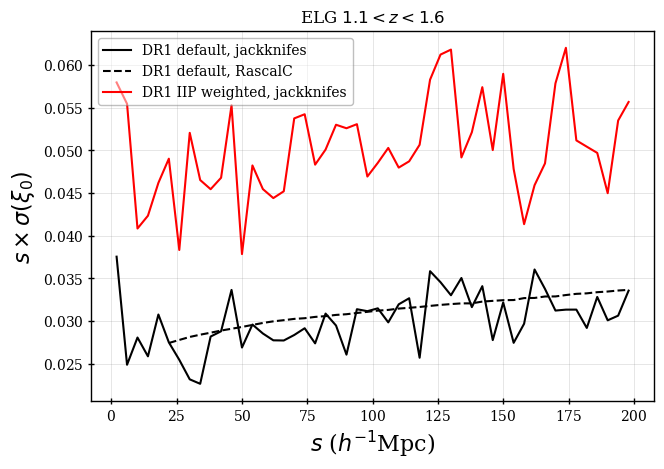}

\caption{For ELG data with $1.1<z<1.6$, the left-hand plot compares the monopole of the 2-point correlation function obtained from the DESI LSS catalogs used for 2024 cosmological analysis, to that obtained from the version of the LSS catalogs that uses pure individual inverse probability (IIP) weights for the completeness weight. The right-hand plot compares the same cases, but for the uncertainty determined from 60 jack-knife samples.}
    \label{fig:ELGpipcomp}
\end{figure*} 

We expect that the application of PIP and angular up-weighting to the v1.5pip LSS catalogs should provide clustering measurements that are unbiased by fiber assignment incompleteness. Based on comparisons of jackknife error estimates, we find that they are noisier on large scales than the results from the fiducial catalogs (v1.5). The effect is most extreme for the ELG sample, which we show in \cref{fig:ELGpipcomp}. In that figure, we compare the results obtained from the v1.5 catalogs with our default weighting ($w_{\rm tot}w_{\rm FKP}$ on both data and randoms) to that obtained from v1.5pip with the same weighting. The difference is therefore that the IIP weight is used for v1.5pip $w_{\rm comp}$ and there is no $f_{\rm tile}$ weighting applied to the randoms. The uncertainty estimated from 60 jackknife samples is approximately 60\% higher for v1.5pip. When applying PIP and angular up-weighting, the results (not shown) at large scales are nearly identical. The primary difference is in how $w_{\rm comp}$ is obtained. There is more variance in the $w_{\rm comp}$ obtained from the IIP than from $1/f_{\rm TLID}$. Further study of the treatment of fiber assignment incompleteness and the clustering estimators that can be applied can be found in \cite{KP3s6-Bianchi}.

\section{Color Scheme for Figures}
\label{app:colors}
%Of utmost importance was choosing a consistent color scheme for plotting results from different DESI DR1 tracers and redshift bins. In no sense was too much time spent on this.\footnote{In fact, at many times it was a necessary distraction from more mundane work. Though it drove some of us crazy...}
For the DESI 2024 cosmological analyses, we consistently applied a common color scheme associated with each tracer and redshift bin. This enabled easy comparisons of the results. Here, we describe the color scheme and some of the motivation behind the choices.

Certain choices were obvious: LRGs are red and ELGs are blue. The QSO sample was given green because SDSS \cite{ebosslss} did so for a reason not remembered; the particular shade\footnote{All colors in `' refer to named colors from \url{https://matplotlib.org/stable/gallery/color/named_colors.html}.} of `seagreen' was determined to be distinguishable when applying color-blindness filters using \url{https://colororacle.org/}. The Lyman $\alpha$ forest is observed at the lowest wavelengths of DESI spectra, so they were awarded purple.%\footnote{They deserved something nice like that.}

\begin{figure*}
    \centering    \includegraphics[width=0.9\columnwidth]{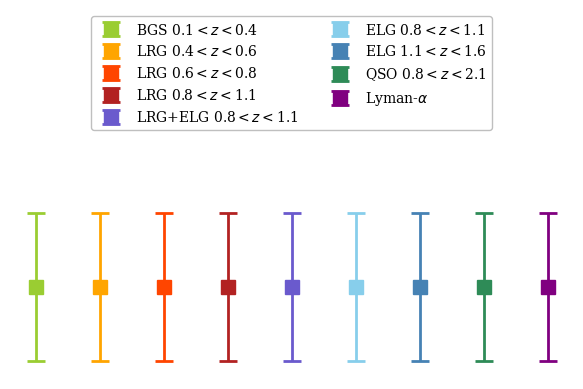} 
\caption{The color scheme used to display different tracers and redshifts bins in the DESI DR1 analysis.}
    \label{fig:colorkey}
\end{figure*}

The LRG redshift bins were given the progression orange, `orangered', and `firebrick'.%, which seemed natural. 
%In retrospect perhaps we should have used `orangered', red, `firebrick', but then red would not have been unique to the whole LRG redshift range. 
For the ELGs, we used `skyblue' and `steelblue' for $0.8<z<1.1$ and $1.1<z<1.6$, respectively. They make a handsome and distinguishable pair. The combination of LRG and ELG in the $0.8<z<1.1$ redshift range was given the color `slateblue', which is a quite lovely shade of purple that is distinguishable from any color used by the Lyman-$\alpha$ forest group. For BGS, chose `yellowgreen' to make sure it was especially distinguishable from the lowest LRG bin.
%This left BGS ... `darkgoldenrod' was nice and used initially, but it was too close to orange. We changed to `yellowgreen', which will not be confused with orange. 
\cref{fig:colorkey} shows the colors of all tracers and redshift bins used in the DESI DR1 analyses.

%The process was completed by consensus within the full collaboration.\footnote{People made changes to \url{https://github.com/cosmodesi/desi-y1-kp/blob/main/desi_y1_plotting/kp3.py} and opinions were sought on Slack. The Lyman-$\alpha$ forest group was consulted to make sure that they liked the color purple.}

\section{Glossary}

A glossary of DESI quantities and jargon is available at \url{https://data.desi.lbl.gov/doc/glossary/} and \cite{KP3s15-Ross}. Below, we repeat some of the entries that are most relevant for this work and provide additional entries.

\noindent\textbullet~ {\bf assignment completeness ($C_{\rm assign}$)}: 
The fiber assignment completeness for any arbitrary selection of targets is the number of those targets assigned to a fiber, divided by the total number of those targets.

\noindent\textbullet~ {\bf BGS}: `Bright Galaxy Sample'; Galaxies targeted during bright time (see below); the sample is primarily flux-limited \cite{BGStarget}.

\noindent\textbullet~ {\bf bright time}: DESI observations taken during `bright' conditions, as defined in \cite{surveyops}. BGS are the only LSS targets during bright time.

\noindent\textbullet~ {\bf dark time}: DESI observations taken during `dark' conditions, as defined in \cite{surveyops}. ELG, LRG, QSO are observed during dark time.

\noindent\textbullet~ \textbf{\textsc{desitarget}}: The code package used to select targets for DESI spectroscopic observation and change their status based on their observation history; \cite{DESItarget} \url{https://github.com/desihub/desitarget}.

\noindent\textbullet~ {\bf ELG}: `Emission Line Galaxy'; a class of DESI targets (see below) selected with the expectation they will yield a detection of OII flux with redshift $z$ between 0.6 and 1.6. The selection is defined in \cite{ELGtarget}.

\noindent\textbullet~ \textbf{fiber}: An individual fiber optic from a positioner on the focal plane to a spectrograph. Fibers are numbered sequentially from 0 to 4999, corresponding to value of \texttt{FIBER} in the catalogs.

\noindent\textbullet~ \textbf{fiber positioner}: A two-arm moveable robot holding a DESI fiber on the focal plane.

\noindent\textbullet~ \textbf{\textsc{fiberassign}}: The code package used to determine which targets can be assigned to which fiber on the DESI focal plane; \cite{FBA.Raichoor.2024}, \url{https://github.com/desihub/fiberassign}.

\noindent\textbullet~ {\bf Legacy Surveys (LS)}: The program that delivered the photometric information used to select targets for DESI spectroscopy, via their Data Release 9 (DR9); \cite{LS.Overview.Dey.2019,LS.dr9.Schegel.2024}, \url{https://www.Legacy Survey.org/dr9/}.

\noindent\textbullet~ \textbf{\texttt{LOCATION}}: The identifier corresponding to a particular fiber positioner. Each \texttt{LOCATION} value has a one-to-one mapping to each \texttt{FIBER} value.

\noindent\textbullet~ {\bf LRG}: `Luminous Red Galaxy'; a class of DESI targets distinguished by their red colours, resulting from a strong 4000 angstrom break.

\noindent\textbullet~ {\bf MTL}: The `Merged Target Ledger' contains all of the information on how the state of a target has changed. It is updated through \textsc{desitarget} (see above) and controls its priority in \textsc{fiberassign} (see above). Primarily, a target will go from unobserved (and thus high \texttt{PRIORITY}; see below) to observed (and thus low \texttt{PRIORITY}). 

\noindent\textbullet~ {\bf \texttt{PRIORITY}}: The quantity given to targets to determine the relative preference for assigning a fiber. The initial \texttt{PRIORITY} are determined based on the target type and the values are reduced after a successful observation.

\noindent\textbullet~ {\bf QSO}: Technically `Quasi-Stellar Object', but we use it synonymously with `quasar'; a class of DESI targets likely to be quasars \cite{QSOtarget}. Those with redshifts $>$ 2.1 are 'Lyman-$\alpha$' quasars, which are at high enough redshift to allow measurement of 'Lyman-$\alpha$' forest absorption. 

\noindent\textbullet~ {\bf random}: Object with celestial coordinates randomly selected at a uniform density from a specified region on the sky.

\noindent\textbullet~ {\bf target}: Object selected via photometry for DESI spectroscopic followup by \textsc{desitarget} (see above), but not necessarily observed (yet) by DESI. Each has a unique \texttt{TARGETID}. Similarly, randoms (see above) that only occupy sky locations where there was Legacy Survey (see above) DR9 imaging were produced by \textsc{desitarget} and have unique \texttt{TARGETID}.

\noindent\textbullet~ {\bf tile}: A single DESI pointing on the sky with assignments of which fibers should observe which targets; each has a unique \texttt{TILEID}.

\noindent\textbullet~ \texttt{TILES}: A string listing the tiles that the target appeared on, using the \texttt{TILEID}s sorted in ascending order and separated by `-'. Each unique \texttt{TILES} represents a unique group of overlapping tiles.

\noindent\textbullet~ {\bf \texttt{TILELOCID}}: The identifier we use to match to information associated with a particular tile and fiber, defined as 10000\texttt{TILEID}+\texttt{LOCATION}.

\noindent\textbullet~ {\bf \texttt{TSNR2}}: Template Signal-to-Noise Squared. A signal-to-noise metric weighted by what wavelengths matter most for determining the redshift of DESI targets, given their magnitude and redshift distributions. This depends upon target class, e.g. Lyman-alpha QSO TSNR2 more heavily weights blue wavelengths, while ELG TSNR2 more heavily weights redder wavelengths which cover the emission lines for the DESI redshifts of interest. TSNR2 depends upon the noise properties of individual spectra, but not the signal properties of the target. It is fully defined in \cite{DESIpipe}.

% Author list file generated with: mkauthlist 1.3.0+14.gcc6daf1.dirty 
%% Affiliations file. load \usepackage{hanging}. Use \input to call it after \appendix

\section{Author Affiliations}
\label{sec:affiliations}

\noindent \hangindent=.5cm $^{1}${Instituto de F\'{\i}sica Te\'{o}rica (IFT) UAM/CSIC, Universidad Aut\'{o}noma de Madrid, Cantoblanco, E-28049, Madrid, Spain}

\noindent \hangindent=.5cm $^{2}${Lawrence Berkeley National Laboratory, 1 Cyclotron Road, Berkeley, CA 94720, USA}

\noindent \hangindent=.5cm $^{3}${Physics Dept., Boston University, 590 Commonwealth Avenue, Boston, MA 02215, USA}

\noindent \hangindent=.5cm $^{4}${Tata Institute of Fundamental Research, Homi Bhabha Road, Mumbai 400005, India}

\noindent \hangindent=.5cm $^{5}${Centre for Extragalactic Astronomy, Department of Physics, Durham University, South Road, Durham, DH1 3LE, UK}

\noindent \hangindent=.5cm $^{6}${Institute for Computational Cosmology, Department of Physics, Durham University, South Road, Durham DH1 3LE, UK}

\noindent \hangindent=.5cm $^{7}${Department of Physics, University of Michigan, Ann Arbor, MI 48109, USA}

\noindent \hangindent=.5cm $^{8}${Leinweber Center for Theoretical Physics, University of Michigan, 450 Church Street, Ann Arbor, Michigan 48109-1040, USA}

\noindent \hangindent=.5cm $^{9}${IRFU, CEA, Universit\'{e} Paris-Saclay, F-91191 Gif-sur-Yvette, France}

\noindent \hangindent=.5cm $^{10}${Institut de F\'{i}sica d’Altes Energies (IFAE), The Barcelona Institute of Science and Technology, Campus UAB, 08193 Bellaterra Barcelona, Spain}

\noindent \hangindent=.5cm $^{11}${Instituto de Ciencias F\'{\i}sicas, Universidad Aut\'onoma de M\'exico, Cuernavaca, Morelos, 62210, (M\'exico)}

\noindent \hangindent=.5cm $^{12}${Instituto Avanzado de Cosmolog\'{\i}a A.~C., San Marcos 11 - Atenas 202. Magdalena Contreras, 10720. Ciudad de M\'{e}xico, M\'{e}xico}

\noindent \hangindent=.5cm $^{13}${Physics Department, Yale University, P.O. Box 208120, New Haven, CT 06511, USA}

\noindent \hangindent=.5cm $^{14}${Department of Physics and Astronomy, University of California, Irvine, 92697, USA}

\noindent \hangindent=.5cm $^{15}${Department of Physics, Kansas State University, 116 Cardwell Hall, Manhattan, KS 66506, USA}

\noindent \hangindent=.5cm $^{16}${Department of Physics \& Astronomy, University of Rochester, 206 Bausch and Lomb Hall, P.O. Box 270171, Rochester, NY 14627-0171, USA}

\noindent \hangindent=.5cm $^{17}${Institute for Astronomy, University of Edinburgh, Royal Observatory, Blackford Hill, Edinburgh EH9 3HJ, UK}

\noindent \hangindent=.5cm $^{18}${Dipartimento di Fisica ``Aldo Pontremoli'', Universit\`a degli Studi di Milano, Via Celoria 16, I-20133 Milano, Italy}

\noindent \hangindent=.5cm $^{19}${Centre for Astrophysics \& Supercomputing, Swinburne University of Technology, P.O. Box 218, Hawthorn, VIC 3122, Australia}

\noindent \hangindent=.5cm $^{20}${NSF NOIRLab, 950 N. Cherry Ave., Tucson, AZ 85719, USA}

\noindent \hangindent=.5cm $^{21}${Department of Physics \& Astronomy, University College London, Gower Street, London, WC1E 6BT, UK}

\noindent \hangindent=.5cm $^{22}${Department of Astronomy and Astrophysics, University of Chicago, 5640 South Ellis Avenue, Chicago, IL 60637, USA}

\noindent \hangindent=.5cm $^{23}${Fermi National Accelerator Laboratory, PO Box 500, Batavia, IL 60510, USA}

\noindent \hangindent=.5cm $^{24}${Korea Astronomy and Space Science Institute, 776, Daedeokdae-ro, Yuseong-gu, Daejeon 34055, Republic of Korea}

\noindent \hangindent=.5cm $^{25}${Institute of Cosmology and Gravitation, University of Portsmouth, Dennis Sciama Building, Portsmouth, PO1 3FX, UK}

\noindent \hangindent=.5cm $^{26}${Departamento de Astrof\'{\i}sica, Universidad de La Laguna (ULL), E-38206, La Laguna, Tenerife, Spain}

\noindent \hangindent=.5cm $^{27}${Instituto de Astrof\'{\i}sica de Canarias, C/ V\'{\i}a L\'{a}ctea, s/n, E-38205 La Laguna, Tenerife, Spain}

\noindent \hangindent=.5cm $^{28}${Department of Physics and Astronomy, University of Sussex, Brighton BN1 9QH, U.K}

\noindent \hangindent=.5cm $^{29}${Departamento de F\'{i}sica, Instituto Nacional de Investigaciones Nucleares, Carreterra M\'{e}xico-Toluca S/N, La Marquesa,  Ocoyoacac, Edo. de M\'{e}xico C.P. 52750,  M\'{e}xico}

\noindent \hangindent=.5cm $^{30}${Institute for Advanced Study, 1 Einstein Drive, Princeton, NJ 08540, USA}

\noindent \hangindent=.5cm $^{31}${Center for Cosmology and AstroParticle Physics, The Ohio State University, 191 West Woodruff Avenue, Columbus, OH 43210, USA}

\noindent \hangindent=.5cm $^{32}${NASA Einstein Fellow}

\noindent \hangindent=.5cm $^{33}${School of Mathematics and Physics, University of Queensland, 4072, Australia}

\noindent \hangindent=.5cm $^{34}${Department of Physics and Astronomy, The University of Utah, 115 South 1400 East, Salt Lake City, UT 84112, USA}

\noindent \hangindent=.5cm $^{35}${Instituto de F\'{\i}sica, Universidad Nacional Aut\'{o}noma de M\'{e}xico,  Cd. de M\'{e}xico  C.P. 04510,  M\'{e}xico}

\noindent \hangindent=.5cm $^{36}${CIEMAT, Avenida Complutense 40, E-28040 Madrid, Spain}

\noindent \hangindent=.5cm $^{37}${Department of Physics \& Astronomy and Pittsburgh Particle Physics, Astrophysics, and Cosmology Center (PITT PACC), University of Pittsburgh, 3941 O'Hara Street, Pittsburgh, PA 15260, USA}

\noindent \hangindent=.5cm $^{38}${Department of Astronomy, School of Physics and Astronomy, Shanghai Jiao Tong University, Shanghai 200240, China}

\noindent \hangindent=.5cm $^{39}${Space Sciences Laboratory, University of California, Berkeley, 7 Gauss Way, Berkeley, CA  94720, USA}

\noindent \hangindent=.5cm $^{40}${University of California, Berkeley, 110 Sproul Hall \#5800 Berkeley, CA 94720, USA}

\noindent \hangindent=.5cm $^{41}${Universities Space Research Association, NASA Ames Research Centre}

\noindent \hangindent=.5cm $^{42}${Center for Astrophysics $|$ Harvard \& Smithsonian, 60 Garden Street, Cambridge, MA 02138, USA}

\noindent \hangindent=.5cm $^{43}${Department of Physics, The Ohio State University, 191 West Woodruff Avenue, Columbus, OH 43210, USA}

\noindent \hangindent=.5cm $^{44}${The Ohio State University, Columbus, 43210 OH, USA}

\noindent \hangindent=.5cm $^{45}${Kavli Institute for Particle Astrophysics and Cosmology, Stanford University, Menlo Park, CA 94305, USA}

\noindent \hangindent=.5cm $^{46}${SLAC National Accelerator Laboratory, Menlo Park, CA 94305, USA}

\noindent \hangindent=.5cm $^{47}${Instituto de Astrof\'{i}sica de Andaluc\'{i}a (CSIC), Glorieta de la Astronom\'{i}a, s/n, E-18008 Granada, Spain}

\noindent \hangindent=.5cm $^{48}${Institute of Physics, Laboratory of Astrophysics, \'{E}cole Polytechnique F\'{e}d\'{e}rale de Lausanne (EPFL), Observatoire de Sauverny, Chemin Pegasi 51, CH-1290 Versoix, Switzerland}

\noindent \hangindent=.5cm $^{49}${Departamento de F\'isica, Universidad de los Andes, Cra. 1 No. 18A-10, Edificio Ip, CP 111711, Bogot\'a, Colombia}

\noindent \hangindent=.5cm $^{50}${Observatorio Astron\'omico, Universidad de los Andes, Cra. 1 No. 18A-10, Edificio H, CP 111711 Bogot\'a, Colombia}

\noindent \hangindent=.5cm $^{51}${Department of Physics, The University of Texas at Dallas, Richardson, TX 75080, USA}

\noindent \hangindent=.5cm $^{52}${Institut d'Estudis Espacials de Catalunya (IEEC), 08034 Barcelona, Spain}

\noindent \hangindent=.5cm $^{53}${Institute of Space Sciences, ICE-CSIC, Campus UAB, Carrer de Can Magrans s/n, 08913 Bellaterra, Barcelona, Spain}

\noindent \hangindent=.5cm $^{54}${Departament de F\'{\i}sica Qu\`{a}ntica i Astrof\'{\i}sica, Universitat de Barcelona, Mart\'{\i} i Franqu\`{e}s 1, E08028 Barcelona, Spain}

\noindent \hangindent=.5cm $^{55}${Institut de Ci\`encies del Cosmos (ICCUB), Universitat de Barcelona (UB), c. Mart\'i i Franqu\`es, 1, 08028 Barcelona, Spain.}

\noindent \hangindent=.5cm $^{56}${Consejo Nacional de Ciencia y Tecnolog\'{\i}a, Av. Insurgentes Sur 1582. Colonia Cr\'{e}dito Constructor, Del. Benito Ju\'{a}rez C.P. 03940, M\'{e}xico D.F. M\'{e}xico}

\noindent \hangindent=.5cm $^{57}${Departamento de F\'{i}sica, Universidad de Guanajuato - DCI, C.P. 37150, Leon, Guanajuato, M\'{e}xico}

\noindent \hangindent=.5cm $^{58}${Centro de Investigaci\'{o}n Avanzada en F\'{\i}sica Fundamental (CIAFF), Facultad de Ciencias, Universidad Aut\'{o}noma de Madrid, ES-28049 Madrid, Spain}

\noindent \hangindent=.5cm $^{59}${Excellence Cluster ORIGINS, Boltzmannstrasse 2, D-85748 Garching, Germany}

\noindent \hangindent=.5cm $^{60}${University Observatory, Faculty of Physics, Ludwig-Maximilians-Universit\"{a}t, Scheinerstr. 1, 81677 M\"{u}nchen, Germany}

\noindent \hangindent=.5cm $^{61}${Department of Astrophysical Sciences, Princeton University, Princeton NJ 08544, USA}

\noindent \hangindent=.5cm $^{62}${Institut d'Astrophysique de Paris. 98 bis boulevard Arago. 75014 Paris, France}

\noindent \hangindent=.5cm $^{63}${Department of Astronomy, University of Florida, 211 Bryant Space Science Center, Gainesville, FL 32611, USA}

\noindent \hangindent=.5cm $^{64}${Institute for Fundamental Physics of the Universe, via Beirut 2, 34151 Trieste, Italy}

\noindent \hangindent=.5cm $^{65}${International School for Advanced Studies, Via Bonomea 265, 34136 Trieste, Italy}

\noindent \hangindent=.5cm $^{66}${Kavli Institute for Cosmology, University of Cambridge, Madingley Road, Cambridge CB3 0HA, UK}

\noindent \hangindent=.5cm $^{67}${Department of Astronomy, The Ohio State University, 4055 McPherson Laboratory, 140 W 18th Avenue, Columbus, OH 43210, USA}

\noindent \hangindent=.5cm $^{68}${Department of Physics, Southern Methodist University, 3215 Daniel Avenue, Dallas, TX 75275, USA}

\noindent \hangindent=.5cm $^{69}${The Ohio State University, Columbus, 43210 OH, USA"}

\noindent \hangindent=.5cm $^{70}${Department of Physics and Astronomy, University of Waterloo, 200 University Ave W, Waterloo, ON N2L 3G1, Canada}

\noindent \hangindent=.5cm $^{71}${Perimeter Institute for Theoretical Physics, 31 Caroline St. North, Waterloo, ON N2L 2Y5, Canada}

\noindent \hangindent=.5cm $^{72}${Waterloo Centre for Astrophysics, University of Waterloo, 200 University Ave W, Waterloo, ON N2L 3G1, Canada}

\noindent \hangindent=.5cm $^{73}${Graduate Institute of Astrophysics and Department of Physics, National Taiwan University, No. 1, Sec. 4, Roosevelt Rd., Taipei 10617, Taiwan}

\noindent \hangindent=.5cm $^{74}${Schmidt Sciences, 155 W 23rd St, New York, NY 10011, USA}

\noindent \hangindent=.5cm $^{75}${Sorbonne Universit\'{e}, CNRS/IN2P3, Laboratoire de Physique Nucl\'{e}aire et de Hautes Energies (LPNHE), FR-75005 Paris, France}

\noindent \hangindent=.5cm $^{76}${Department of Astronomy and Astrophysics, UCO/Lick Observatory, University of California, 1156 High Street, Santa Cruz, CA 95064, USA}

\noindent \hangindent=.5cm $^{77}${Department of Astronomy and Astrophysics, University of California, Santa Cruz, 1156 High Street, Santa Cruz, CA 95065, USA}

\noindent \hangindent=.5cm $^{78}${Department of Astronomy \& Astrophysics, University of Toronto, Toronto, ON M5S 3H4, Canada}

\noindent \hangindent=.5cm $^{79}${University of Science and Technology, 217 Gajeong-ro, Yuseong-gu, Daejeon 34113, Republic of Korea}

\noindent \hangindent=.5cm $^{80}${Departament de F\'{i}sica, Serra H\'{u}nter, Universitat Aut\`{o}noma de Barcelona, 08193 Bellaterra (Barcelona), Spain}

\noindent \hangindent=.5cm $^{81}${Laboratoire de Physique Subatomique et de Cosmologie, 53 Avenue des Martyrs, 38000 Grenoble, France}

\noindent \hangindent=.5cm $^{82}${Instituci\'{o} Catalana de Recerca i Estudis Avan\c{c}ats, Passeig de Llu\'{\i}s Companys, 23, 08010 Barcelona, Spain}

\noindent \hangindent=.5cm $^{83}${Max Planck Institute for Extraterrestrial Physics, Gie\ss enbachstra\ss e 1, 85748 Garching, Germany}

\noindent \hangindent=.5cm $^{84}${Department of Physics and Astronomy, Siena College, 515 Loudon Road, Loudonville, NY 12211, USA}

\noindent \hangindent=.5cm $^{85}${Department of Physics \& Astronomy, University  of Wyoming, 1000 E. University, Dept.~3905, Laramie, WY 82071, USA}

\noindent \hangindent=.5cm $^{86}${National Astronomical Observatories, Chinese Academy of Sciences, A20 Datun Rd., Chaoyang District, Beijing, 100012, P.R. China}

\noindent \hangindent=.5cm $^{87}${Steward Observatory, University of Arizona, 933 N, Cherry Ave, Tucson, AZ 85721, USA}

\noindent \hangindent=.5cm $^{88}${Aix Marseille Univ, CNRS, CNES, LAM, Marseille, France}

\noindent \hangindent=.5cm $^{89}${Departament de F\'isica, EEBE, Universitat Polit\`ecnica de Catalunya, c/Eduard Maristany 10, 08930 Barcelona, Spain}

\noindent \hangindent=.5cm $^{90}${Aix Marseille Univ, CNRS/IN2P3, CPPM, Marseille, France}

\noindent \hangindent=.5cm $^{91}${University of California Observatories, 1156 High Street, Sana Cruz, CA 95065, USA}

\noindent \hangindent=.5cm $^{92}${Department of Physics \& Astronomy, Ohio University, Athens, OH 45701, USA}

\noindent \hangindent=.5cm $^{93}${Department of Physics and Astronomy, Sejong University, Seoul, 143-747, Korea}

\noindent \hangindent=.5cm $^{94}${Abastumani Astrophysical Observatory, Tbilisi, GE-0179, Georgia}

\noindent \hangindent=.5cm $^{95}${Faculty of Natural Sciences and Medicine, Ilia State University, 0194 Tbilisi, Georgia}

\noindent \hangindent=.5cm $^{96}${Space Telescope Science Institute, 3700 San Martin Drive, Baltimore, MD 21218, USA}

\noindent \hangindent=.5cm $^{97}${Centre for Advanced Instrumentation, Department of Physics, Durham University, South Road, Durham DH1 3LE, UK}

\noindent \hangindent=.5cm $^{98}${Physics Department, Brookhaven National Laboratory, Upton, NY 11973, USA}

\noindent \hangindent=.5cm $^{99}${Beihang University, Beijing 100191, China}

\noindent \hangindent=.5cm $^{100}${Department of Astronomy, Tsinghua University, 30 Shuangqing Road, Haidian District, Beijing, China, 100190}

\noindent \hangindent=.5cm $^{101}${Physics Department, Stanford University, Stanford, CA 93405, USA}

\noindent \hangindent=.5cm $^{102}${Department of Physics, University of California, Berkeley, 366 LeConte Hall MC 7300, Berkeley, CA 94720-7300, USA}

\noindent \hangindent=.5cm $^{103}${Institute of Physics, Laboratory of Astrophysics, \'{E}cole Polytechnique F\'{e}d\'{e}rale de Lausanne (EPFL), Observatoire de Sauverny, CH-1290 Versoix, Switzerland}

\end{document}